\documentclass[12pt]{article}

\usepackage[numbers,sort&compress]{natbib}
\usepackage{amsmath}
\usepackage{amsthm,amscd,amsfonts}
\usepackage{amssymb, upref, color}
\usepackage{amsmath,amssymb,amsthm}
\usepackage{color}
\usepackage[colorlinks]{hyperref}
\usepackage{graphicx}
\usepackage{epstopdf}
\usepackage{latexsym,bm}
\usepackage{subfigure}
\usepackage{float}
\usepackage{booktabs}
\usepackage{dutchcal}

\hypersetup{
	colorlinks=true, 
	linkcolor=blue, 
	citecolor=blue, 
	urlcolor=blue 
}

\numberwithin{equation}{section}

\theoremstyle{plain}
\newtheorem{exam}{Example}[section]
\newtheorem{theorem}[exam]{Theorem}

\begin{document}
	
	\title{Pattern formation and global analysis of a systematically reduced plant model in dryland environment\footnote{Y. Xia was supported by the Zhejiang Provincial Natural Science Foundation of China (No.LZ24A010006) and  NNSFC (No.11931016). 
J. Yu was supportted by the NNSFC (No.12331017).}}
	\author{Yonghui Xia$^a$ \,\,\,\,\,
		Jianglong Xiao$^b$ \,\,\,\,\, Jianshe Yu$^c$\footnote{Corresponding author. Jianshe Yu, jsyu@gzhu.edu.cn} \\
{\small \textit{$^a$School of Mathematics, Foshan University, Foshan, 528000, China }}\\
		{\small \textit{$^b$School of Mathematical Sciences,  Zhejiang Normal University,  Jinhua, 321004, China}}\\
		{\small \textit{$^c$College of Mathematics and Information Sciences, Guangzhou University, Guangzhou, 510006, China}}\\
		{\small Email: xiadoc@163.com;yhxia@zjnu.cn; jianglongxiao@zjnu.edu.cn; jsyu@gzhu.edu.cn}
	}
	\date{}
	
	\maketitle

		\begin{abstract}
	This paper delves into a systematically	reduced plant system proposed by Ja\"{\i}bi et al.  \cite[Phys. D, 2020]{Physica D} in arid area. They used the method of geometric singular perturbation to study the existence of abundant orbits. Instead, we deliberate the stability and distributed patterns of this system. For a non-diffusive scenario for the model, we scrutinize the local and global stability of equilibria and derive conditions for the existence or non-existence of the limit cycle. The bifurcation behaviors are also explored. For the spatial model, we investigate Hopf, Turing, Hopf-Turing, Turing-Turing bifurcations. Specially, 
	the evolution process from periodic solutions to spatially nonconstant steady states is observed near the Hopf-Turing bifurcation point.
	And mixed nonconstant steady states near the Turing-Turing bifurcation point are observed. Furthermore, it's found that there exist gap, spot, stripe and mixed patterns. The seed-dispersal rate enables the transformation of pattern structures. Reasonable control of system parameters may prevent desertification from occurring.\\
	{\bf Keywords}: Turing bifurcation; Hopf bifurcation; Plant pattern; Stability
	\end{abstract}
	{\bf MSC 2020} 92C15; 92D40; 34C23; 35B32 
	
	\section{Introduction} 
	This paper explores the space-time dynamics of the following plant system proposed by Ja\"{\i}bi et al. \cite{Physica D}.
	\begin{equation}\label{original diffusion system}
		\begin{cases}
				(\partial_{\tau}-d_{B}\Delta)B(x,t)=\Theta BW(1-\frac{B}{K})(1+\Lambda B)-NB, \ \ \ \ \ \ \ \ \ \ \ \ \ \ \  \ \ \ \ \ \ \ \ x\in (0,\mathcal{l}\pi),\\
				(\partial_{\tau}-d_{W}\Delta)W(x,t)=P-QW(1-\frac{RB}{K})-MBW(1+\Lambda B), \ \ \ \ \ \ \ \ \ \ \ \ \ x\in (0,\mathcal{l}\pi),\\
				B_{\textbf{n}}=W_{\textbf{n}}=0, \ \ \ \ \ \ \ \ \ \ \ \ \ \ \ \ \ \ \ \ \ \ \ \ \ \ \ \ \ \ \ \ \ \ \ \ \ \ \ \ \ \ \ \ \ \ \ \ \ \  \ \ \ \ \ \ \ \ \ \ \ \ \ \ \ \ \ \ \ \ \ \ \ t \geq 0,x=0,\mathcal{l}\pi,
		\end{cases}
	\end{equation}
	where $B(x,t)$ and $W(x,t)$ represent, respectively, plant biomass and soil-water density. And the biological explanations of the parameters are shown in Table \ref{parameters}.\\
	\indent Desertification is an ecological issue characterized by the degradation of land in arid, semi-arid, and dry sub-humid areas.  This process involves the loss of vegetation, soil erosion, and the decline in biodiversity. Desertification poses significant environmental, economic, and social challenges, including reduced agricultural productivity, water scarcity, displacement of communities, and loss of livelihoods. Fortunately, vegetation conservation stands as an effective measure in combating desertification. Vegetation plays a crucial role in maintaining soil structure stability, reducing soil erosion, enhancing soil-water retention capacity, and promoting biodiversity \cite{Abera et al Glob, Chen et al PLOS}. Mathematically, modeling the relationship between vegetation and water in arid regions, and studying their dynamic behaviors including plant patterns is essential. This is beneficial to understanding plant patterns and implementing effective conservation strategies.\\
	\indent Over the past few decades, numerous plant models have been proposed and continually improved to explore plant dynamics. The earliest plant-water model, developed by Klausmeier \cite{Klausmeier 1999}, described the interaction between plant biomass and water in semi-arid regions. The pivotal role of nonlinear mechanisms in shaping the spatial structure of plant communities was elucidated \cite{Klausmeier 1999}. By introducing a system that takes into account soil-water, 	Hardeberg et al. \cite{Hardeberg et al. 2001} presented a novel description to elucidate desertification phenomena and classify droughts. Later, Rietkerk and colleagues \cite{Rietkerk 2002} partitioned water into soil moisture and surface water, and established a three-component model to investigate the positive feedback between water infiltration and vegetation density. Gilad et al. \cite{Gilad et al. 2007} presented a three-component model for arid regions, which seizes a variety of pattern structure of plants. One notable divergence between two models in  \cite{Rietkerk 2002} and \cite{Gilad et al. 2007}  is that the latter incorporates lateral root water conduction as an additional mechanism for water transport. Based on Gilad et al. \cite{Gilad et al. 2007},  Zelnik et al. \cite{Zelnik et al. 2015} used a simplified version of the model to explore the dynamics in fairy circles. And their findings indicated that the formation and decay processes of fairy circles correspond to spatially constrained transitions between distinct stable states. 
    The level of research attention is greater for the Klausmeier model and its extended versions \cite{Lejeune 2004,Gowda et al. 2014,Dijkstra 2011,Zelnik et al. 2013}, including some delayed models \cite{Guo Gaihui CSF 2024,Sun CSF 2023,Yuan Sanling 2023-JMB}. \\
			\begin{table}[H]
				\caption{Parameter descriptions of system \eqref{original diffusion system}}
				\centering
				\begin{tabular}{cc}
					\toprule
				Parameter &   Description  \\
					\midrule
					K& Maximal standing biomass\\
					M&  Water-uptake rate \\
					N& Plant mortality rate\\
					P&Precipitation rate\\
					Q& Evaporation rate\\
				    R&Reduction evaporation rate because of shading,\\
					$\Lambda$& Root-to-shoot ratio\\
					$\Theta$& Biomass growth rate coefficient\\
					$d_{B}$&  Seed-dispersal coefficient\\
					$d_{W}$& Soil-water diffusivity \\
					\bottomrule
				\end{tabular}
			\end{table}\label{parameters}
	\indent To better elucidate the impact of parameter perturbations on the formation of plant patterns, Sun et al. \cite{Sun 2022-JMB} conducted a thorough analysis of the model proposed by Zelnik et al. \cite{Zelnik et al. 2015}. They found that the rate of water diffusion suppresses plant growth, while shading parameters facilitate an increase in plant biomass. By controlling properly parameters, the uniform state gradually transitions to gap patterns. In effect, a wide variety of patterns such as spot, strip and gap patterns are observed in plant system \cite{Borgogno et al. 2009,Hardeberg et al. 2001,Rietkerk 2002,Rietkerk and Koppel 2008,Sun 2022-JMB}. Turing bifurcation is indeed a significant mathematical concept in understanding pattern formation, especially in fields like nonlinear dynamics and mathematical biology. It has been used to explain various observed patterns in nature, such as animal coat patterns, chemical patterns, and even patterns in biological systems like the development of animal skin or the organization of cells in embryos. This has prompted scholars to investigate pattern formation across various systems, such as \cite{Dai BX 2020,Song Yang Sun AMM,Cai YL 2018,Vanag Phys. Chem. Chem. Phys.,Peng YH 2016AMC}. 
	
\subsection*{Motivation} 
	Based on Zelnik et al. \cite{Zelnik et al. 2015}, Ja\"{\i}bi et al. \cite{Physica D} reduced systematically the model under one space dimension and obtain \eqref{original diffusion system}, which enables a more direct connection between ecological mechanisms and observations. They employed the geometric singular perturbation means to investigate localized patterns and showed the existence of numerous heteroclinic, homoclinic and periodic orbits. However, no one has studied the space-time dynamics of \eqref{original diffusion system} so far. The mechanisms and patterns of spatial diffusion are particularly important for the study of biodynamics \cite{An,Huang Jicai JBD,Huang Jicai JDE,Jiang Weihua2020,Rui Peng,Lou4,Lou1,Lou2,Song J. Nonlinear Sci.,Wu Shiliang,Wu Zhao,Xiao yanni JMA,Yi JDE,Xue Song Wang,Zhou Xiao,Linyi}. In view of this, it is necessary to study bifurcation behaviors and the effect of parameter perturbation on patterned vegetation distribution for \eqref{original diffusion system}. \\
\indent Making conversions 
\begin{flalign*}
	 &u=\sqrt{\frac{P\Lambda\Theta}{KN^{2}}}B, \ v=\frac{N}{P}W, \ t=N\tau, \ a=\frac{P\Theta}{N^{2}}, \ b=\frac{Q}{N}, \\
	 &c=\frac{MK-RQ}{\sqrt{\Lambda\Theta PK}}, \ d=\frac{MNK}{P\Theta}, \
	 m=\frac{BK-1}{B}\sqrt{\frac{\Lambda\Theta P}{KN^{2}}}, \ d_{1}=\frac{d_{B}}{N}, \ d_{2}=\frac{d_{W}}{N},
\end{flalign*}
then \eqref{original diffusion system} becomes 
	\begin{equation}\label{diffusion system}
		\begin{cases}
		    (\partial_{t}-d_{1}\Delta)u(x,t)=(av-1)u+mu^{2}v-u^{3}v,\\
			(\partial_{t}-d_{2}\Delta)v(x,t)=1-(b+cu+du^{2})v,
		\end{cases}
	\end{equation}
	where $a, \ b, \ d \textgreater0$ and $m, \ c \in \mathbb{R}$.\\
\indent The remainder of this paper unfolds as follows. Section 2 deliberates the local and global stability of equilibria, analyzes the conditions for the existence and non-existence of limit cycles, and explores potential bifurcations that system \eqref{local system} undergoes. Section 3 analyzes several bifurcations of codimension 2 for system \eqref{diffusion system}. Section 4 firstly validates the complex theoretical analysis through numerical computation and provide simulation results. Additionally, abundant pattern structures of vegetation are observed. Eventually, Section 5 concludes this paper.
	
\section{Analysis for the local system}
This section deals with the global dynamics of the following temporal system corresponding to \eqref{diffusion system}.
	\begin{equation}\label{local system}
		\begin{cases}
			\frac{\text{d}u}{\text{d}t}=(av-1)u+mu^{2}v-u^{3}v,\\
		\frac{\text{d}v}{\text{d}t}=1-(b+cu+du^{2})v.
		\end{cases}
	\end{equation}
	
	Apparently, the system admits only one boundary equilibrium $E_{0}(0,\frac{1}{b})$. By computation, the positive equilibrium lies in the following equations:
	\begin{equation}
		\begin{cases}
			(a+mu-u^{2})v=1,\\
			(b+cu+du^{2})v=1.
		\end{cases}
	\end{equation} 
	Ulteriorly, we have 
	\begin{equation}\label{positive equilibrium equation}
		(1+d)u^{2}+(c-m)u+b-a=0.
	\end{equation}
	By analysing \eqref{positive equilibrium equation}, the following outcomes are gained.
	\begin{theorem}
		System \eqref{local system} has one boundary equilibrium $E_{0}(0,\frac{1}{b})$. Letting $\Delta=(c-m)^{2}-4(1+d)(b-a)$, then \\
		(i) if $b\textless a$, then there is a positive equilibrium $E_{10}(u_{10},v_{10})$, where $u_{10}=\frac{m-c+\sqrt{\Delta}}{2(1+d)}$ and $v_{10}=\frac{1}{a+mu_{10}-u^{2}_{10}}$;\\
		(ii) if $b=a$ and $m\textgreater c$, then there is a positive equilibrium $E_{11}(u_{11},v_{11})$, where $u_{11}=\frac{m-c}{1+d}$ and $v_{10}=\frac{1}{a+mu_{11}-u^{2}_{11}}$;\\
		(iii) if $a \textless b \textless a+\frac{(c-m)^{2}}{4(1+d)}$ and $m\textgreater c$, then there are two positive equilibria $E_{10}(u_{10},v_{10})$ and $E_{12}(u_{12},v_{12})$, where $u_{12}=\frac{m-c-\sqrt{\Delta}}{2(1+d)}$ and $v_{12}=\frac{1}{a+mu_{12}-u^{2}_{12}}$;\\
		$(iv)$ if $b=a+\frac{(c-m)^{2}}{4(1+d)}$ and $m\textgreater c$, there is a positive equilibrium $E_{13}(u_{13},v_{13})$, where $u_{13}=\frac{m-c}{2(1+d)}$ and $v_{13}=\frac{1}{a+mu_{13}-u^{2}_{13}}$;\\
		$(v)$ if $b \textgreater a+\frac{(c-m)^{2}}{4(1+d)}$, then there is no positive equilibrium.
	\end{theorem}
The Jacobian matrix of \eqref{local system} at $E(u,v)$ reads 
\begin{equation}\label{Jacobian matrix}
	J(E(u,v))=
	\left(
	\begin{array}{cc}
		av-1+2muv-3u^{2}v   &   au+mu^{2}-u^{3}   \\
		-(c+2du)v    &   -(b+cu+du^{2})
	\end{array}
	\right).
\end{equation}
	Inserting $E_{0}(0,\frac{1}{b})$ and $E_{1j}(u_{1j},v_{1j}) \ (j=0,1,2,3)$ into \eqref{Jacobian matrix} yields 
	\begin{equation}
	J(E_{0})=
	\left(
	\begin{array}{cc}
		\frac{a}{b}-1   &   0   \\
		-\frac{c}{b}    &   -b
	\end{array}
	\right) \text{and} \
	J(E_{1j})=
	\left(
	\begin{array}{cc}
		  u_{1j}v_{1j}(m-2u_{1j}) &     \frac{u_{1j}}{v_{1j}} \\
		    -(c+2du_{1j})v_{1j}   &     -\frac{1}{v_{1j}}
	\end{array}
	\right), 
\end{equation}
respectively. 
	The two eigenvalues of $J(E_{0})$ are $\frac{a}{b}-1$ and $-b$.  Hence, the following outcomes are gained. 
	\begin{theorem}
		 $E_{0}$ is a saddle for $a \textgreater b$ and is a stable node for $a \textless b$. If $b=a$, then \\
		(i) if $m = c$, then $E_{0}$ is a degenerate stable node;\\
		(ii) if $m \neq c$, then $E_{0}$ is an attracting saddle-node that contains a stable parabolic sector in $\mathbb{R}^{+}_{2}$.
	\end{theorem}
		\begin{figure}[H]
		\centering
		\begin{minipage}[b]{.45\linewidth}
			\centering
			\subfigure[]
			{
				\includegraphics[scale=0.28]{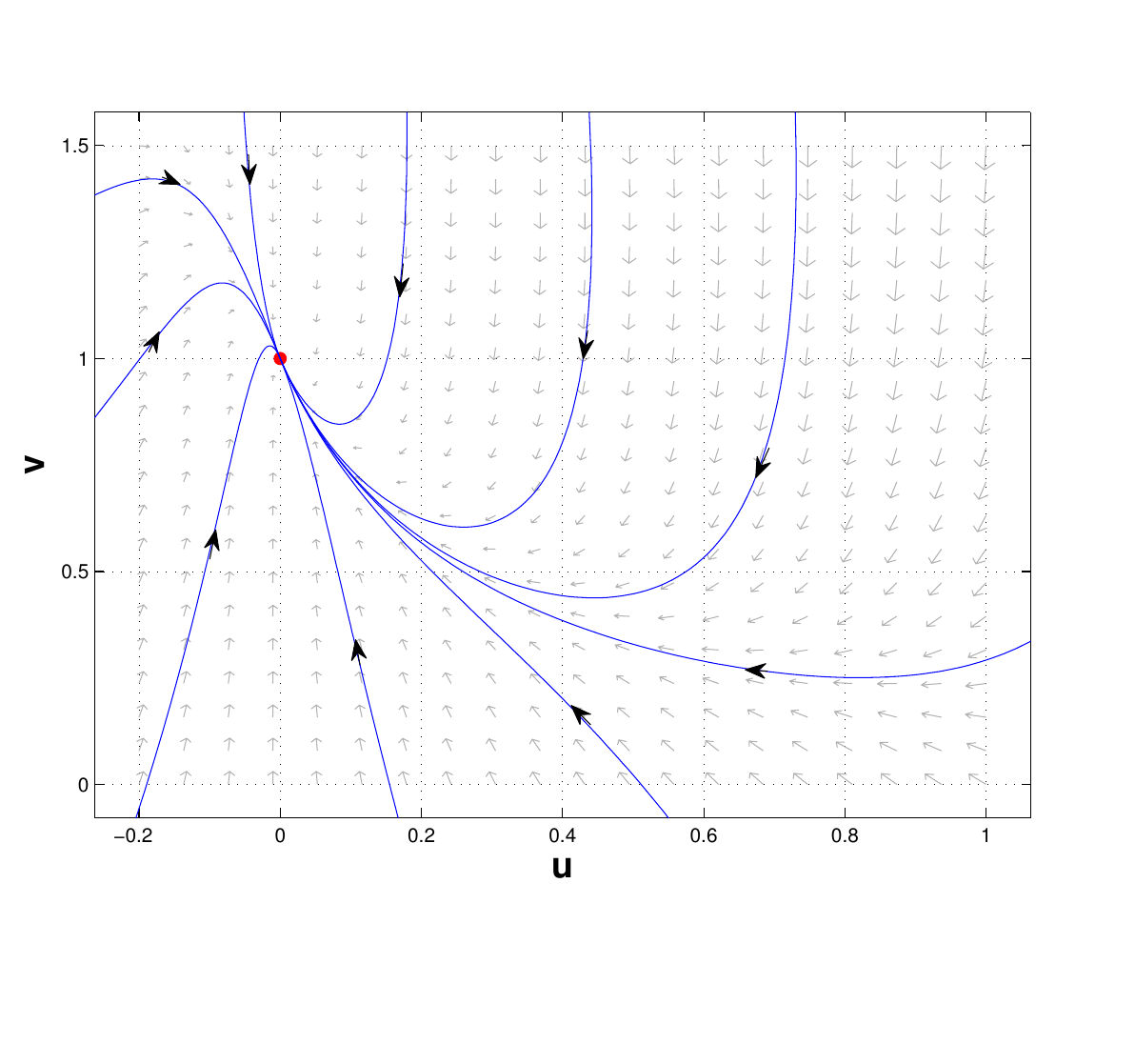}
			}
		\end{minipage}
		\begin{minipage}[b]{.45\linewidth}
			\subfigure[]
			{
				\includegraphics[scale=0.28]{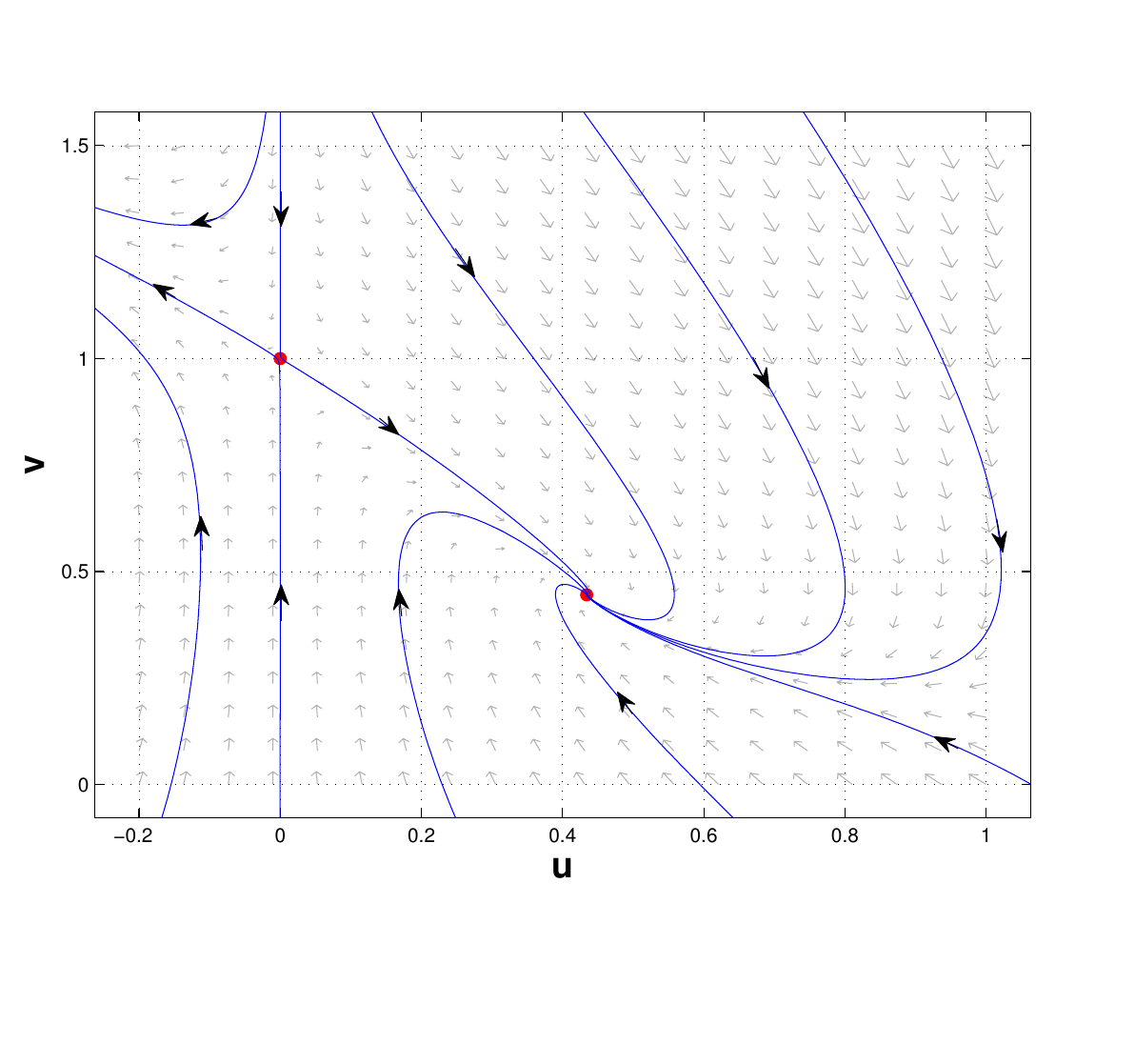}
			}
		\end{minipage}
		\begin{minipage}[b]{.45\linewidth}
			\centering
			\subfigure[]
			{
				\includegraphics[scale=0.28]{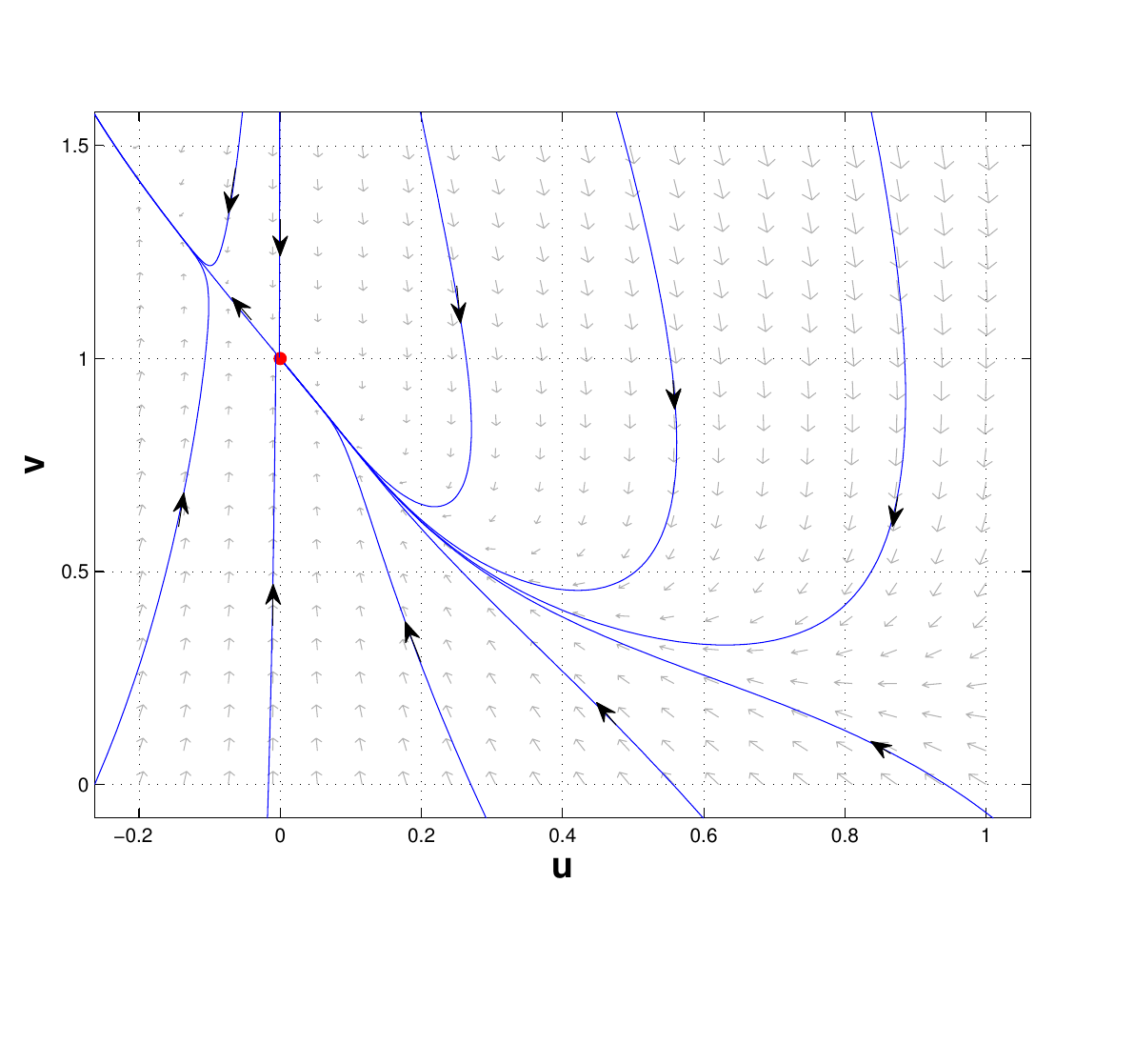}
			}
		\end{minipage}
		\begin{minipage}[b]{.45\linewidth}
			\subfigure[]
			{
				\includegraphics[scale=0.28]{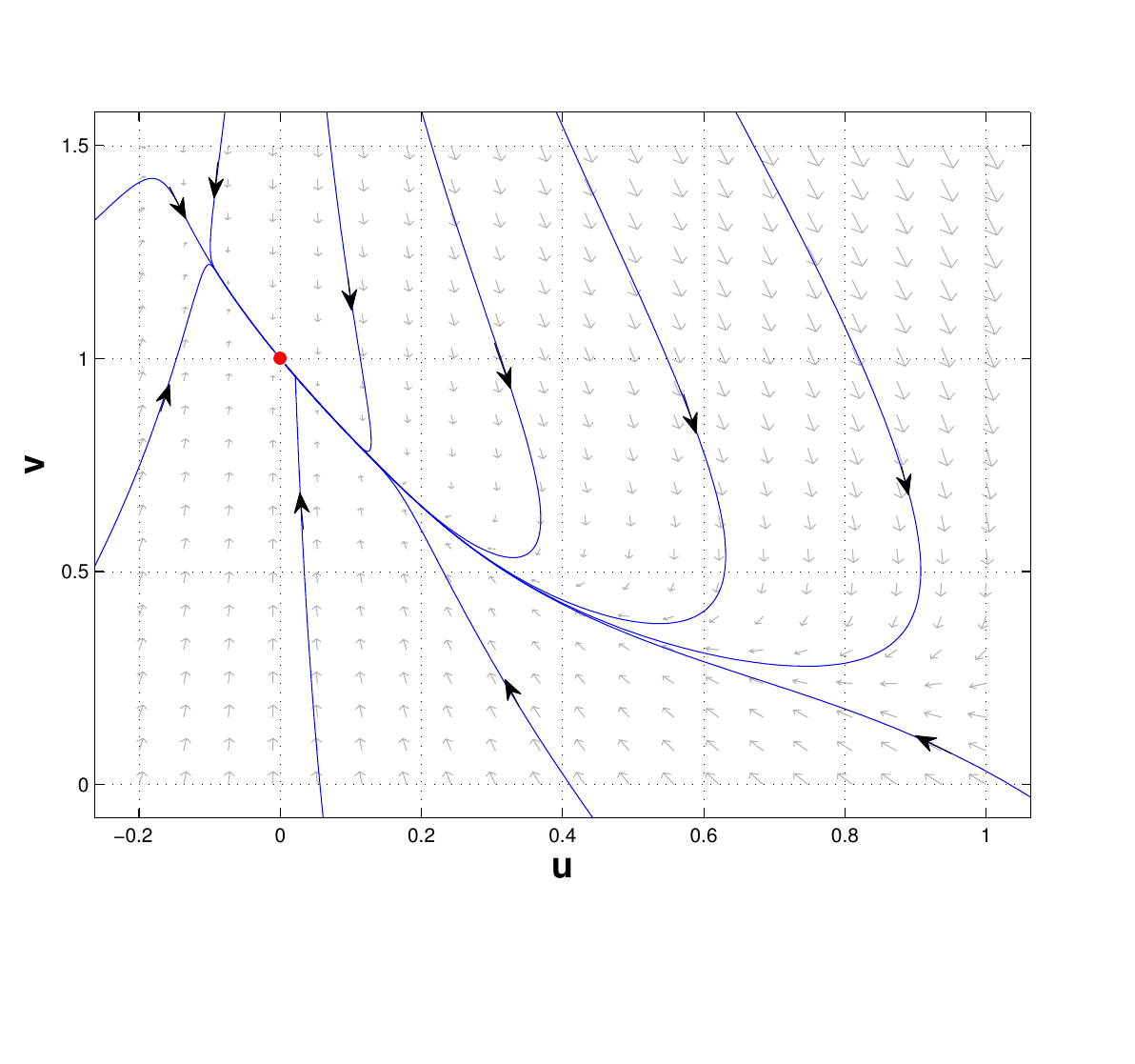}
			}
		\end{minipage}
		\caption{Types of equilibrium $E_{0}(0,1)$. Take $(b,c,d)=(1,2,2)$ and vary the value of parameters $a$ and $m$. $(a)$ a stable node for $m=1$ and $a=\frac{1}{2}\textless b$; $(b)$ a saddle for $m=1$ and $a=2\textgreater b$; $(c)$ an attracting saddle-node for $m=1$ and $a=b=1$. $(d)$ a degenerate stable node for $m=c=2$ and $a=b=1$.}\label{E0}
	\end{figure}
	
	\noindent\textbf{Proof.} The two cases of $b \neq a$ are evident, thus we directly prove the case of $b=a$. Utilizing the following conversions successively:
	\begin{equation*}
	U=u, V=v-\frac{1}{b}; \ U=u, V=-\frac{c}{b}u-bv,
	\end{equation*} 
	then \eqref{local system} becomes 
	\begin{equation}\label{E_1 conversion 1}
		\begin{cases}
			\frac{\text{d}U}{\text{d}t}=\frac{m-c}{a}U^{2}-UV-\frac{m}{a}U^{2}V-\frac{a+mc}{a^{2}}U^{3}+O(|(U,V)|^{3}), \\
			\frac{\text{d}V}{\text{d}t}=-bV+\frac{a(ad-c^{2})+c(c-m)}{a^{2}}U^{2}+\frac{c(1-a)}{a}UV+\frac{mc-a^{2}d}{a^{2}}U^{2}V-\frac{c(a^{2}d-cm-a)}{a^{3}}U^{3}+O(|(U,V)|^{3}).
		\end{cases}
	\end{equation}
	Readjusting $t=-a\tau$, then \eqref{E_1 conversion 1} becomes 
	\begin{equation}
		\begin{cases}
			\frac{\text{d}U}{\text{d}\tau}=\frac{c-m}{a^{2}}U^{2}+\frac
			{1}{a}UV+\frac{m}{a^{2}}U^{2}V+\frac{a+mc}{a^{3}}U^{3}+O(|(U,V)|^{3}), \\
			\frac{\text{d}V}{\text{d}\tau}=V-\frac{a(ad-c^{2})+c(c-m)}{a^{3}}U^{2}-\frac{c(1-a)}{a^{2}}UV-\frac{mc-a^{2}d}{a^{3}}U^{2}V+\frac{c(a^{2}d-cm-a)}{a^{4}}U^{3}+O(|(U,V)|^{3}).
		\end{cases}
	\end{equation}
	Plugging $V=\frac{a(ad-c^{2})+c(c-m)}{a^{3}}U^{2}+\cdots$ into 
	$\frac{\text{d}U}{\text{d}\tau}$ yields $\frac{\text{d}U}{\text{d}\tau}=\frac{c-m}{a^{2}}U^{2}+\frac{1+d}{a^{2}}U^{3}+\cdots$.
	Utilizing Theorem 7.1 of Chapter 2 in \cite{Zhang zhifen}, we assert that $(i)$ and $(ii)$ hold.
	
	In order to analyze the stability of the positive equilibria, it's necessary to judge the symbols of the following formulas in detail, namely 
	\begin{equation}
		\text{Tr}J(E_{1j})= u_{1j}v_{1j}(m-2u_{1j})-\frac{1}{v_{1j}}, \  \text{Det}J(E_{1j})=2(1+d)u_{1j}\left(u_{1j}-\frac{m-c}{2(1+d)}\right).
	\end{equation}
	Thus, the following series of outcomes are obtained.
	\begin{theorem}
		$E_{12}$ is a hyperbolic saddle.
	\end{theorem}
	\noindent\textbf{Proof.} Direct calculation displays $\text{Det}J(E_{12})=-u_{12}\sqrt{\Delta}\textless 0$. Then the conclusion is clear.

	Next we focus on $E_{11}$. Direct calculation reveals $\text{Det}J(E_{11})=\frac{(m-c)^{2}}{1+d} \textgreater 0,  \text{Tr}J(E_{11})=\frac{u_{11}(m-2u_{11})-(a+mu_{11}-u^{2}_{11})^{2}}{a+mu_{11}-u^{2}_{11}}$. And $u_{11}(m-2u_{11})-(a+mu_{11}-u^{2}_{11})^{2}=\frac{\mathcal{A}a^{2}+\mathcal{B}a+\mathcal{C}}{(1+d)^{4}}$, where  
	\begin{flalign*}
		\mathcal{A}=-(1+d)^{4}\textless 0, \ \mathcal{B}=2(1+d)^{2}(c-m)(c+md), \ \mathcal{C}=(m-c)g(d),
	\end{flalign*}
	with 
	\begin{equation}\label{g(d)}
		g(d)=md^{3}+(cm^{2}-m^{3}+2c+m)d^{2}+(2mc^{2}-2cm^{2}+4c-m)d+c^{3}-mc^{2}+2c-m.
	\end{equation}
	By computation, we gain $\mathcal{B}^{2}-4\mathcal{A}\mathcal{C}=4(m-c)(md+2c-m)(1+d)^{6}$. For $m(d-1)+2c \textless 0$, we have $\mathcal{B}^{2}-4\mathcal{A}\mathcal{C} \textless 0$, i.e., $\mathcal{A}a^{2}+\mathcal{B}a+\mathcal{C} \textless 0$, which means $\text{Tr}J(E_{11}) \textless 0$. That is, $E_{11}$ is stable if $m(d-1)+2c \textless 0$.
	For $m(d-1)+2c=0$, i.e., $(md+c)+(c-m)=0$, we have $md+c \textgreater 0$ and then $-\frac{\mathcal{B}}{2\mathcal{A}}\textless 0$, which means $\text{Tr}J(E_{11}) \textless 0$. Thus, $E_{11}$ is stable if $m(d-1)+2c=0$. For $m(d-1)+2c \textgreater 0$, we have $md+c \textgreater 0$, i.e., $-\frac{\mathcal{B}}{2\mathcal{A}}\textless 0$. In this case, we gain $\mathcal{A}a^{2}+\mathcal{B}a+\mathcal{C} \textless 0$ for 
	$ g(d)\leq 0$ and Eq. $\mathcal{A}a^{2}+\mathcal{B}a+\mathcal{C}=0$ admits a positive root $a^{*}$ for $g(d) \textgreater 0$, where 
	\begin{equation}\label{local Hopf point}
		a^{*}=\frac{(c-m)(md+c)+(1+d)\sqrt{(m-c)(md+2c-m)}}{(1+d)^{2}}.
	\end{equation}
	That is, if $m(d-1)+2c \textgreater 0$ and $g(d) \textgreater 0$, then 
	\begin{equation}
		\text{Tr}J(E_{11})
		\begin{cases}
			\textgreater 0 \ \ \ \text{for} \ a \textless a^{*}, \\
			= 0 \ \ \text{for} \  a = a^{*}, \\
			\textless 0 \ \ \ \text{for} \ a \textgreater a^{*},
		\end{cases}
	\end{equation}
	which explains the local stability range of $E_{11}$ in this situation. Next we verify the transversal condition. 
	Due to $\frac{\partial u_{11}}{\partial a}=0, \ \frac{\partial v_{11}}{\partial a}=-v^{2}_{11}\left( 1+(m-2u_{11})\frac{\partial u_{11}}{\partial a} \right)$ and $mu_{11}-2u^{2}_{11}=\frac{1}{v^{2}_{11}}$ at $a = a^{*}$, then we have
	\begin{flalign*}
		\frac{\partial \text{Tr}J(E_{11})}{\partial a}\bigg|_{a=a^{*}}&=\left( (mv_{11}-4u_{11}v_{11})\frac{\partial u_{11}}{\partial a}+(mu_{11}-2u^{2}_{11}+\frac{1}{v^{2}_{11}})\frac{\partial v_{11}}{\partial a}\right)\bigg|_{a = a^{*}}\\
		&=\frac{2}{v^{2}_{11}}\frac{\partial v_{11}}{\partial a}=-2.
	\end{flalign*}
	This explains the appearance of Hopf bifurcation at $a = a^{*}$. \\
	\indent From the above discussion, we gain the following outcome.
	\begin{figure}[H]
		\centering
		\begin{minipage}[b]{.45\linewidth}
				\centering
				\subfigure[]
				{
						\includegraphics[scale=0.28]{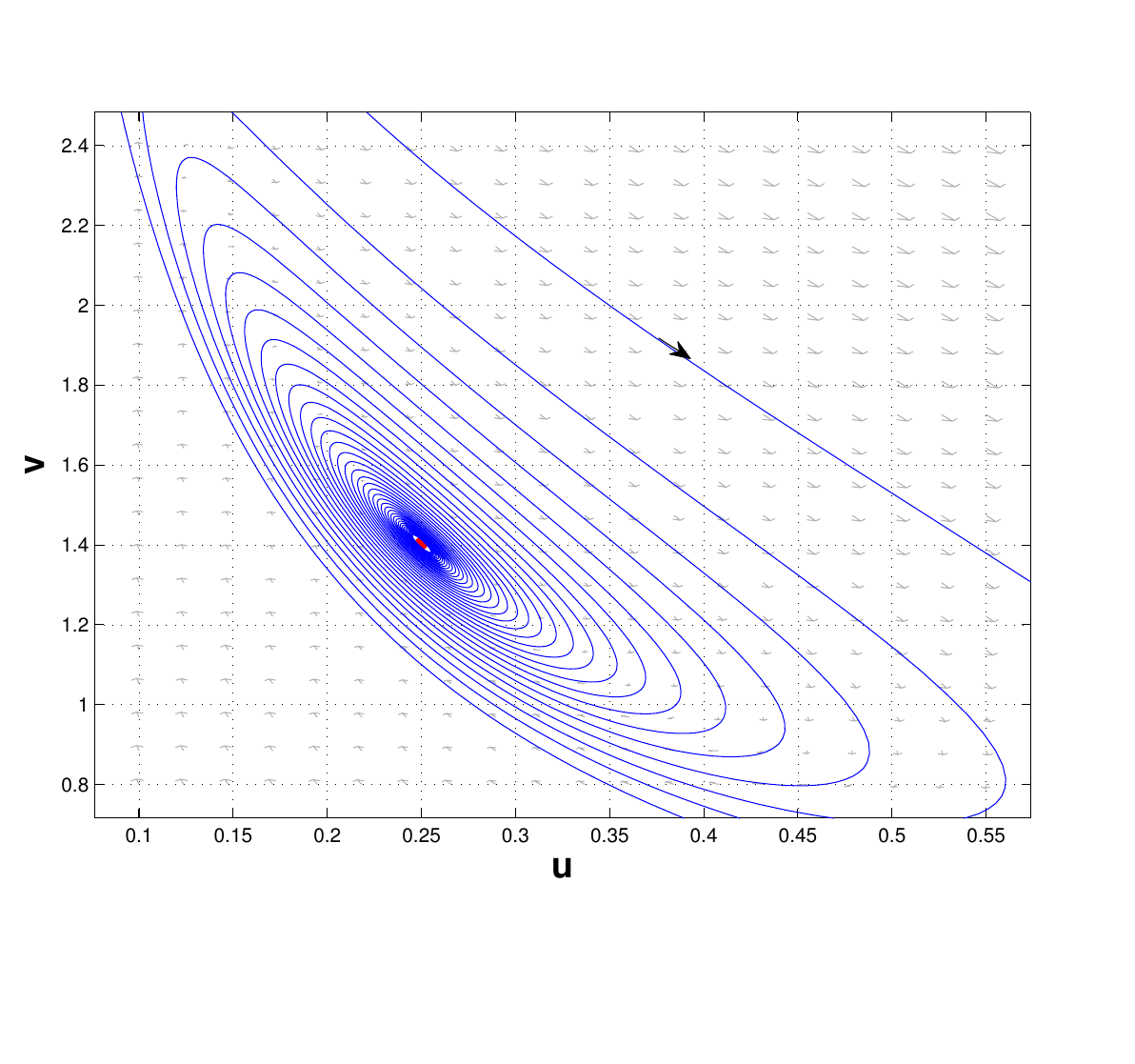}
					}
			\end{minipage}
		\begin{minipage}[b]{.45\linewidth}
				\subfigure[]
				{
						\includegraphics[scale=0.28]{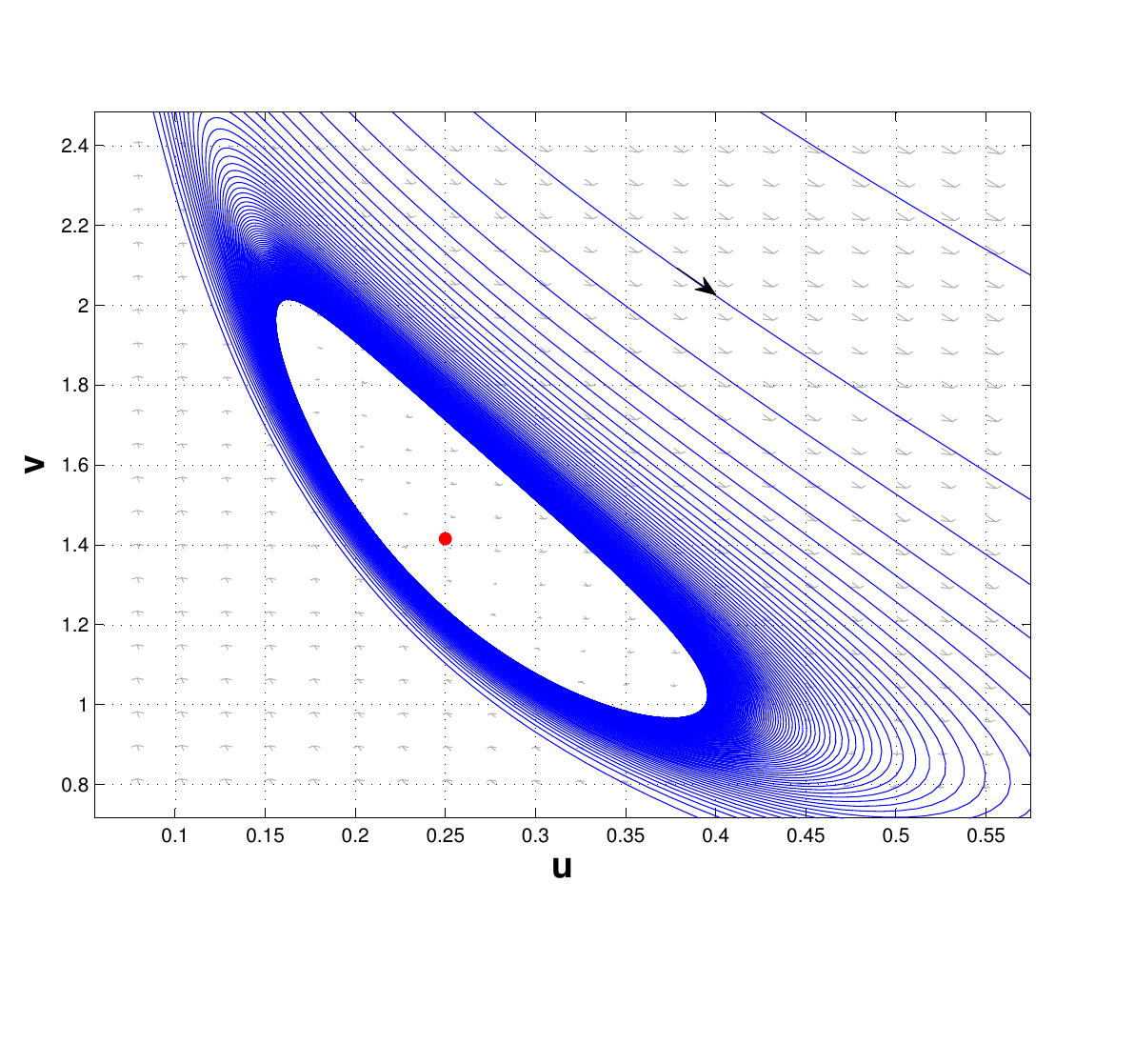}
					}
			\end{minipage}
		\caption{Taking $(b,c,d,m)=(\frac{\sqrt{2}}{2}-\frac{9}{16},2,1,\frac{5}{2})$, then we have  $g(d)=\frac{47}{8}$ and $a^{*}=\frac{\sqrt{2}}{2}-\frac{9}{16}$. $(a)$ For $a=0.15\textgreater a^{*}$, $E_{11}(\frac{1}{4},\sqrt{2})$ is locally stable. $(b)$ For $a=0.144 \textless a^{*}$, a limit cycle appears.}\label{E11 limit cycle}
	\end{figure}
	\begin{theorem}\label{E11 Hopf}
		If $m(d-1)+2c\leq 0$, then $E_{11}$ is stable. Furthermore, if  $m(d-1)+2c \textgreater 0$, then \\
		(i) if $g(d) \leq 0$, then $E_{11}$ is stable;\\
		(ii) if $g(d) \textgreater 0$, then $E_{11}$ is stable  for $a \textgreater a^{*}$ and is unstable for $a \textless a^{*}$ and Hopf bifurcation emerges at $a = a^{*}$.
	\end{theorem}
	
	Next we focus on $E_{13}$. Direct calculation reveals $\text{Det}J(E_{13})= 0,  \text{Tr}J(E_{13})=\frac{u_{13}(m-2u_{13})-(a+mu_{13}-u^{2}_{13})^{2}}{a+mu_{13}-u^{2}_{13}}$, which means that $E_{13}$ becomes a degenerate equilibrium. 
	Utilizing the following conversions successively:
	\begin{equation*}
		U=u-u_{13}, V=v-v_{13}; \ U=u-\frac{m_{10}}{n_{10}}v, V=v,
	\end{equation*} 
	then \eqref{local system} becomes 
	\begin{equation}\label{E_13 conversion 1}
		\begin{cases}
			\frac{\text{d}U}{\text{d}t}=p_{20}U^{2}+p_{11}UV+p_{02}V^{2}+O(|(U,V)|^{2}), \\
			\frac{\text{d}V}{\text{d}t}=q_{10}U+q_{01}V+q_{20}U^{2}+q_{11}UV+q_{02}V^{2}+O(|(U,V)|^{2}),
		\end{cases}
	\end{equation}
	where 
	\begin{flalign*}
		&p_{20}=m_{20}-\frac{m_{10}n_{20}}{n_{10}}, \ p_{11}=\frac{2m_{10}}{n_{10}}\left(m_{20}-\frac{m_{10}n_{20}}{n_{10}}\right)+m_{11}-\frac{m_{10}n_{11}}{n_{10}}, \\ &p_{02}=\frac{m^{2}_{10}}{n^{2}_{10}}\left(m_{20}-\frac{m_{10}n_{20}}{n_{10}}\right)+\frac{m_{10}}{n_{10}}\left( m_{11}-\frac{m_{10}n_{11}}{n_{10}}\right), \ q_{10}=n_{10}, \ q_{01}=m_{10}+n_{01}, \\
		&q_{20}=n_{20}, \ q_{11}=n_{11}+\frac{2n_{20}m_{10}}{n_{10}}, \ q_{02}=\frac{n_{20}m^{2}_{10}}{n^{2}_{10}}+\frac{n_{11}m_{10}}{n_{10}},
	\end{flalign*}
	with 
		\begin{flalign*}
	&m_{10}=av_{13}-1+2mu_{13}v_{13}-3u^{2}_{13}v_{13}, \  m_{11}=a+2mu_{13}-3u^{2}_{13}, \ m_{20}=mv_{13}-3u_{13}v_{13}, \\
	&n_{10}=-(c+2du_{13})v_{13}, \ n_{01}=-(b+cu_{13}+du^{2}_{13}), \ n_{11}=-(c+2du_{13}), \ n_{20}=-dv_{13}.
		\end{flalign*}
	Making conversion $U_{1}=U, \ V_{1}=q_{10}U+q_{01}V$, then \eqref{E_13 conversion 1} becomes 
	\begin{equation}\label{E_13 conversion 2}
		\begin{cases}
			\frac{\text{d}U_{1}}{\text{d}t}=\left(p_{20}+\frac{p_{02}q^{2}_{10}}{q^{2}_{01}}-\frac{p_{11}q_{10}}{q_{01}}\right)U^{2}_{1}+\left( \frac{p_{11}}{q_{01}}-\frac{2p_{02}q_{10}}{q^{2}_{01}}\right)U_{1}V_{1}+\frac{p_{02}}{q^{2}_{01}}V^{2}_{1}+O(|(U_{1},V_{1})|^{3}), \\
			\frac{\text{d}V_{1}}{\text{d}t}=V_{1}+\left(q_{20}-\frac{q_{11}q_{10}}{q_{01}}+\frac{q_{02}q^{2}_{10}}{q^{2}_{01}} \right)U^{2}_{1}+\left(\frac{q_{11}}{q_{01}}-\frac{2q_{02}q_{10}}{q^{2}_{01}}\right)U_{1}V_{1}+\frac{q_{02}}{q^{2}_{01}}V^{2}_{1}+O(|(U_{1},V_{1})|^{3}).
		\end{cases}
	\end{equation}
	Inserting $V_{1}=\left(q_{20}-\frac{q_{11}q_{10}}{q_{01}}+\frac{q_{02}q^{2}_{10}}{q^{2}_{01}} \right)U^{2}_{1}+\cdots$ into $\frac{\text{d}U_{1}}{\text{d}t}$ obtains the coefficient of $U^{2}_{1}$ as follows:
	\begin{equation}
     \delta=p_{20}-\frac{p_{11}q_{10}}{q_{01}}+\frac{p_{02}q^{2}_{10}}{q^{2}_{01}}-\left( \frac{p_{11}}{q_{01}}-\frac{2p_{02}q_{10}}{q^{2}_{01}} \right)\left( q_{20}-\frac{q_{11}q_{10}}{q_{01}}+\frac{q_{02}q^{2}_{10}}{q^{2}_{01}}\right).
	\end{equation}
		Utilizing Theorem 7.1 of Chapter 2 in \cite{Zhang zhifen}, we assert that $E_{13}$ is a saddle-node for $\delta\neq0$. \\
		\indent Next its attraction and repulsion are further discussed.
	Clearly,  $u_{13}(m-2u_{13})-(a+mu_{13}-u^{2}_{13})^{2}=\frac{\widehat{\mathcal{A}}a^{2}+\widehat{\mathcal{B}}a+\widehat{\mathcal{C}}}{16(1+d)^{4}}$, where  
	\begin{flalign*}
		\widehat{\mathcal{A}}=-16(1+d)^{4}\textless 0, \ \widehat{\mathcal{B}}=8(1+d)^{2}(c-m)(c+m+2md), \ \widehat{\mathcal{C}}=(m-c)h(d),
	\end{flalign*}
	with 
	\begin{equation}\label{h(d)}
		h(d)=8md^{3}+4(cm^{2}-m^{3}+2c+4m)d^{2}+4(mc^{2}-m^{3}+4c+2m)d+c^{3}+mc^{2}-cm^{2}-m^{3}+8c.
	\end{equation}
	By computation, we gain $\widehat{\mathcal{B}}^{2}-4\widehat{\mathcal{A}}\widehat{\mathcal{C}}=512(m-c)(md+c)(1+d)^{6}$. For $md+c \textless 0$, we have $\widehat{\mathcal{B}}^{2}-4\widehat{\mathcal{A}}\widehat{\mathcal{C}} \textless 0$, i.e., $\widehat{\mathcal{A}}a^{2}+\widehat{\mathcal{B}}a+\widehat{\mathcal{C}} \textless 0$, which means $\text{Tr}J(E_{13}) \textless 0$. That is, $E_{13}$ becomes an attracting saddle-node if $m+c \textless 0$.
	For $md+c=0$, we have $m \textgreater c=-md$ and then $m\textgreater 0$, which suggests $m+c+2md=(c+md)+m(1+d)\textgreater0$ and $-\frac{\widehat{\mathcal{B}}}{2\widehat{\mathcal{A}}}\textless 0$, which means $\text{Tr}J(E_{11}) \textless 0$. Thus, $E_{11}$ becomes an attracting saddle-node if $md+c=0$. For $md+c \textgreater 0$, we have $m \textgreater c\textgreater-md$ and $m\textgreater 0$, then $m+c+2md=(c+md)+m(1+d)\textgreater0$ and $-\frac{\widehat{\mathcal{B}}}{2\widehat{\mathcal{A}}}\textless 0$. In this case, we gain $\widehat{\mathcal{A}}a^{2}+\widehat{\mathcal{B}}a+\widehat{\mathcal{C}}=0$ for $ h(d)\leq 0$ and Eq. $\widehat{\mathcal{A}}a^{2}+\widehat{\mathcal{B}}a+\widehat{\mathcal{C}}=0$ admits a positive root $a_{*}$ for $h(d) \textgreater 0$, where 
\begin{equation}\label{saddle-node point}
	a_{*}=\frac{(c-m)(m+c+2md)+2(1+d)\sqrt{2(m-c)(c+md)}}{4(1+d)^{2}}.
\end{equation}
	That is, if $c+md \textgreater 0$ and $h(d)\textgreater 0$, then 
		\begin{equation}
		\text{Tr}J(E_{13})
		\begin{cases}
			\textgreater 0 \ \ \ \text{for} \ a \textless a_{*}, \\
			= 0 \ \ \text{for} \  a = a_{*}, \\
			\textless 0 \ \ \ \text{for} \ a \textgreater a_{*},
		\end{cases}
	\end{equation}
	which explains the attraction and repulsion of saddle-node $E_{13}$ in this case. \\
	\indent In particular, $E_{13}$ becomes a cusp for $a = a_{*}$ (i.e. $q_{01}=0$). To analyze the codimension at $a = a_{*}$, we utilize conversion $U_{2}=V, \ V_{2}=U$ and readjust $t=q_{10}\tau$,  then 
	\begin{equation}\label{E_13 conversion 3}
	\begin{cases}
		\frac{\text{d}U_{2}}{\text{d}t}=V_{2}+\frac{q_{11}}{q_{10}}U_{2}V_{2}+\frac{q_{02}}{q_{10}}U^{2}_{2}+\frac{q_{02}}{q_{10}}V^{2}_{2}+O(|(U_{2},V_{2})|^{3}), \\
		\frac{\text{d}V_{2}}{\text{d}t}=\frac{p_{02}}{q_{10}}U^{2}_{2}+\frac{p_{11}}{q_{10}}U_{2}V_{2}+\frac{p_{20}}{q_{10}}V^{2}_{2}+O(|(U_{2},V_{2})|^{3}).
	\end{cases}
\end{equation}
From \cite{Perko}, the following equivalent system of \eqref{E_13 conversion 3} around $(0,0)$ is gained.
\begin{equation}
	\begin{cases}
			\frac{\text{d}U_{2}}{\text{d}\tau}=V_{2}+O(|(U_{2},V_{2})|^{2}), \\
			\frac{\text{d}V_{2}}{\text{d}\tau}=\frac{p_{02}}{q_{10}}U^{2}_{2}+\frac{2q_{02}+q_{11}}{q10}U_{2}V_{2}+O(|(U_{2},V_{2})|^{2}).
	\end{cases}
\end{equation}
Skipping the tedious calculation, we directly show 
	\begin{equation}\label{cusp point}
			\begin{aligned}
					&\frac{2q_{02}+q_{11}}{q_{10}}=0 \Longleftrightarrow m_{10}(2m_{20}+n_{11})+m_{11}n_{10}=0 \Longleftrightarrow \\
			&a=a_{\#} \triangleq \frac{1}{4(c+md)(1+d)^{2}}\left( 2md^{3}+(2cm^{2}-2m^{3}+4c+2m)d^{2}+(3mc^{2}-2cm^{2}-m^{3}\right.\\
			&+\left.8c-2m)d+c^{3}-cm^{2}+4c-2m\right).
		\end{aligned}
	\end{equation}
Thus, the codimension at $a = a_{*}$ is apparent. \\
\indent From the above discussion, we gain the folllowing outcome.
	\begin{theorem}
	If $c+md\leq 0$, then $E_{13}$ is an attracting saddle-node. Furthermore, if  $c+md \textgreater 0$, then \\
	(\uppercase\expandafter{\romannumeral1}) if $h(d) \leq 0$, then an attracting saddle-node;\\
	(\uppercase\expandafter{\romannumeral2}) if $h(d) \textgreater 0$, then $E_{13}$ is an attracting saddle-node  for $a \textgreater a_{*}$ and is a repelling saddle-node for $a \textless a_{*}$. Besides, if $a=a_{*}$, then \\
	\indent (i) if $a_{*} \neq a_{\#}$, then $E_{13}$ is a cusp of codimension 2;\\
	\indent (ii) if $a_{*}=a_{\#}$, then $E_{13}$ is a cusp of at least codimension 3.
\end{theorem}

\begin{figure}[H]
	\centering
	\begin{minipage}[b]{.45\linewidth}
		\centering
		\subfigure[]
		{
			\includegraphics[scale=0.28]{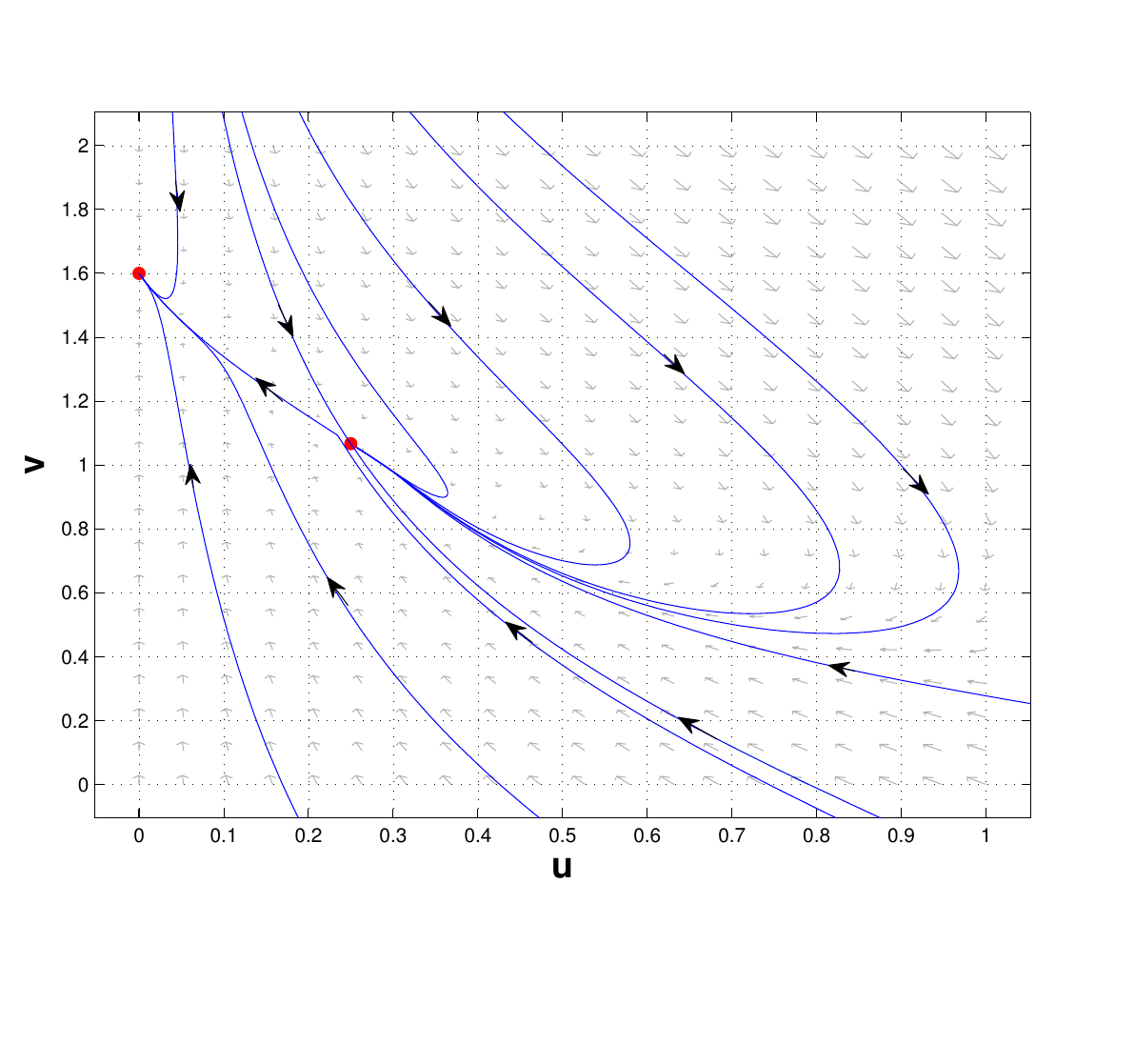}
		}
	\end{minipage}
	\begin{minipage}[b]{.45\linewidth}
		\subfigure[]
		{
			\includegraphics[scale=0.28]{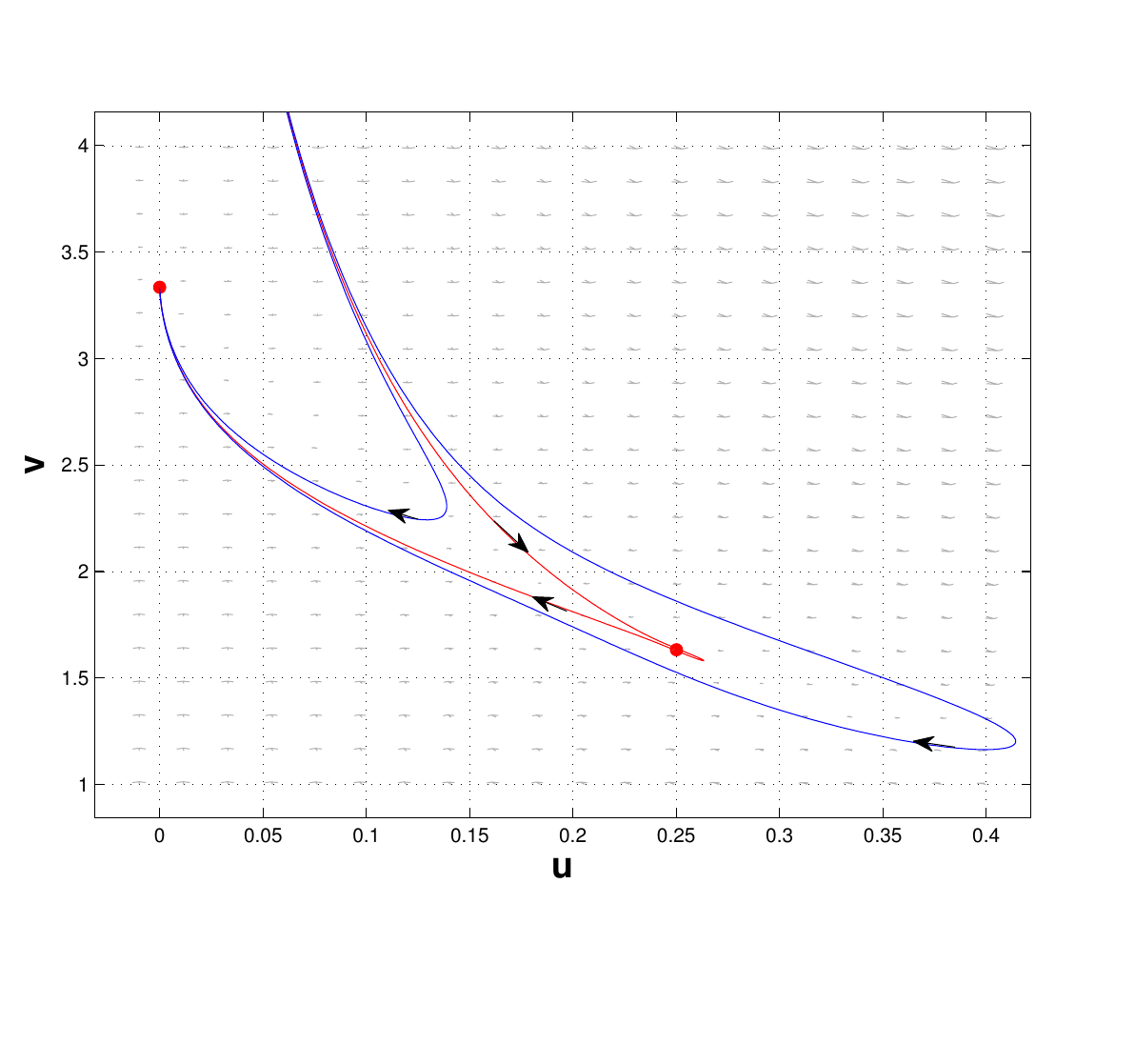}
		}
	\end{minipage}
	\caption{Taking $(c,d,m)=(1,1,2)$, then we have $h(d)=47$ and $a_{*}=\frac{\sqrt{6}}{4}-\frac{7}{16}$. $(a)$ For $a=\frac{1}{2} \textgreater a_{*}$ and $b=\frac{5}{8}$, $E_{13}(\frac{1}{4},\frac{16}{15})$ is an attracting saddle-node. $(b)$ For $a=a_{*}=\frac{\sqrt{6}}{4}-\frac{7}{16}$ and $b=\frac{\sqrt{6}}{4}-\frac{5}{16}$, we have $a_{\#}=-\frac{5}{48}$, then $E_{13}(\frac{1}{4},\frac{2\sqrt{6}}{3})$ is a cusp of codimension 2.}\label{E13 saddle-node and cusp}
\end{figure}
Next we explore the existence of the limit cycle and global stability in $\mathbb{R}^{+}_{2}$. Simple analysis claims that the solution of \eqref{local system} whose initial value is positive also remains positive. Whereupon the following series of results are gotten. 
	\begin{theorem}\label{no limit cycle}
		If $m \leq 0$ or $a \geq \frac{1}{4}$, then there is no limit cycle for system \eqref{local system}.
	\end{theorem}
\noindent\textbf{Proof.} Let $\Upsilon_{1}(u,v)=(av-1)u+mu^{2}v-u^{3}v$ and $\Upsilon_{2}(u,v)=1-(b+cu+du^{2})v$.
Choosing a Dulac function $X_{1}(u,v)=\frac{1}{uv}$, then we gain
\begin{equation}
	\frac{\partial (\Upsilon_{1}X_{1})}{\partial u}+\frac{\partial (\Upsilon_{2}X_{1})}{\partial v}=m-2u-\frac{1}{uv^{2}} \textless 0 \ \ \text{for} \ m\leq0.
\end{equation}
	Choosing another  Dulac function $X_{2}(u,v)=\frac{1}{u^{2}v}$, then we gain
	\begin{equation}
	\frac{\partial (\Upsilon_{1}X_{2})}{\partial u}+\frac{\partial (\Upsilon_{2}X_{2})}{\partial v}=\frac{-av^{2}+v-1}{u^{2}v^{2}}-1 \textless 0 \ \ \text{for} \ a \geq \frac{1}{4}.
	\end{equation}
	Accordingly, the conclusion holds.
	\begin{theorem}\label{limit cycle exist}
		If  $c\geq 0, \ b \textless a$ and $u_{10}v^{2}_{10}(m-2u_{10})\textgreater1$, then at least one stable limit cycle exists in system \eqref{local system}.
	\end{theorem}
	\noindent\textbf{Proof.} Set $L_{1}=\{(u,v): 0 \leq u \leq \frac{m+\sqrt{m^{2}+4a}}{2}, v=0\}$, $L_{2}=\{(u,v): u=0, 0\leq v \leq \frac{1}{b}\}$, $L_{3}=\{(u,v): 0 \leq u \leq \frac{m+\sqrt{m^{2}+4a}}{2}, v=\frac{1}{b}\}$ and $L_{4}=\{(u,v): u= \frac{m+\sqrt{m^{2}+4a}}{2},  0\leq v \leq \frac{1}{b}\}$. Then we get 
	\begin{flalign*}
	 	&\frac{\text{d}v}{\text{d}t}\Big|_{L_{1}}=1 \textgreater0, \ \
		\frac{\text{d}u}{\text{d}t}\Big|_{L_{2}}=0,\\
		&\frac{\text{d}v}{\text{d}t}\Big|_{L_{3}}=-\frac{cu+du^{2}}{b} \textless 0 \  \text{for} \ c\geq0, \ \ 
		\frac{\text{d}u}{\text{d}t}\Big|_{L_{4}}=u((a+mu-u^{2})v-1) \textless 0,
	\end{flalign*}
	Combining the instability of $E_{10}$ under condition $u_{10}v^{2}_{10}(m-2u_{10})\textgreater1$ and the instability of $E_{0}$ under condition $b\textless a$, then the conclusion holds.
	
	\begin{theorem}
	In addition to  $c\geq 0$, $b\geq a$,	if $m \leq c$ or $b\textgreater a+\frac{(c-m)^{2}}{4(1+d)}$, then $E_{0}$ is globally asymptotically stable.
	\end{theorem}
	\noindent\textbf{Proof.} It follows from the above proof procedure that $\Omega=\{0 \leq u \leq \frac{m+\sqrt{m^{2}+4a}}{2}, \ 0\leq v \leq \frac{1}{b}\}$ is the positively invariant domain of \eqref{local system} for $c\geq0$.  Based on that $b\geq a$ and $m \leq c$ or $b\textgreater a+\frac{(c-m)^{2}}{4(1+d)}$, $E_{0}$ is stable and becomes the unique equilibrium of \eqref{local system}, which intimates the global stability of $E_{0}$.
	 
	 \begin{theorem}
	 In addition to $c\geq 0$, $a\textgreater a^{*}$,	if $a\geq \frac{1}{4}$ or $m\leq \sqrt{\frac{a^{3}}{1+a}}$,  then $E_{11}$ is globally asymptotically stable.
	 \end{theorem}
	\noindent\textbf{Proof.} 
	By Theorem \ref{no limit cycle}, we claim that no closed orbit appears for $a\geq \frac{1}{4}$. Selecting the same Dulac function $X_{1}(u,v)$ in Theorem \ref{no limit cycle}, we gain that if $b^{2} \geq \frac{m(m+\sqrt{m^{2}+4a})}{2}$ and $c\geq 0$, then 
	\begin{equation}
		\begin{aligned}
				\frac{\partial (\Upsilon_{1}X_{1})}{\partial u}+\frac{\partial (\Upsilon_{2}X_{1})}{\partial v}&=m-2u-\frac{1}{uv^{2}} \textless m-\frac{1}{uv^{2}} \leq m-\frac{1}{\frac{m+\sqrt{m^{2}+4a}}{2b^{2}}}=m-\frac{2b^{2}}{m+\sqrt{m^{2}+4a}}\\
			&\leq 0 \ \ \text{for} \ (u,v) \in \Omega=\{0 \leq u \leq \frac{m+\sqrt{m^{2}+4a}}{2}, \ 0\leq v \leq \frac{1}{b}\}.
		\end{aligned}
	\end{equation}
	This suggests that no periodic orbit emerges in $\Omega$. Notice that under the existence condition of $E_{11}$, $b^{2} \geq \frac{m(m+\sqrt{m^{2}+4a})}{2}$ is equivalent to $m\leq \sqrt{\frac{a^{3}}{1+a}}$.
	In the light of the positive invariance of $\Omega$ for $c\geq0$ and local  stability of $E_{11}$ for $a\textgreater a^{*}$, th global stability is established.
	
	Analogously, the following outcome stands.
	 \begin{theorem}
	 In addition to $b \textless a, \ c\geq 0$, $u_{10}v^{2}_{10}(m-2u_{10})\textless 1$, if $a\geq \frac{1}{4}$ or $m\leq \sqrt{\frac{a^{3}}{1+a}}$, then $E_{10}$ is globally asymptotically stable.
	\end{theorem}
	Next the transcritical bifurcation and saddle-node bifurcation of \eqref{local system} are researched by the means in \cite{Perko}, and a short proof is given.
	\begin{theorem}
		If $a=b, \ m\neq c$, then system \eqref{local system} has the transcritical bifurcation near $E_{0}$, where $a_{TC}=b$ is the bifurcation threshold.
	\end{theorem}
	\noindent\textbf{Proof.} The following two eigenvectors are first given.
		\begin{equation*}
		W_{1}=
		\left(
		\begin{array}{c}
			1 \\ -\frac{c}{a^{2}}
		\end{array}
		\right), \ 
		W_{2}=
		\left(
		\begin{array}{c}
			1 \\ 0
		\end{array}
		\right).
	\end{equation*}
	Then we obtain	
\begin{equation*}	
\begin{aligned}
	&W^{T}_{2}F_{a}(E_{0},a_{TC})=0, \ 	W^{T}_{2}[DF_{a}(E_{0},a_{TC})]=\frac{1}{b}\neq0, \\
	&W^{T}_{2}[D^{2}F(E_{0},a_{TC})(W_{1},W_{1})]=\frac{2(m-c)}{b}\neq0.
\end{aligned}
\end{equation*}
Then the conclusion is tenable.
	
		\begin{theorem}
		For $b=a+\frac{(c-m)^{2}}{4(1+d)}$, $m+c+3md\neq cd$ and $a \neq a_{*}$, system \eqref{local system} has the saddle-node bifurcation near $E_{13}$, where $b_{SN}=a+\frac{(c-m)^{2}}{4(1+d)}$ is the bifurcation threshold.
	\end{theorem}
	\noindent\textbf{Proof.} The following two eigenvectors are first given.
	\begin{equation*}
	\widetilde{W}_{1}=
		\left(
		\begin{array}{c}
			1 \\ -\frac{m_{10}}{m_{01}}
		\end{array}
		\right), \ 
	\widetilde{W}_{2}=
		\left(
		\begin{array}{c}
			1 \\ -\frac{m_{10}}{n_{10}}
		\end{array}
		\right).
	\end{equation*}
	Then we obtain	
	\begin{equation*}	
		\begin{aligned}	
			\widetilde{W}^{T}_{2}F_{a}(E_{0},a_{TC})&=\frac{m_{10}}{n_{10}}v_{13}=\frac{2(1+d)(c-m)}{4ad^{2}+(8a+2m^{2}-2mc)d+4a+m^{2}-c^{2}}\neq0, \\
			\widetilde{W}^{T}_{2}[D^{2}F(E_{0},a_{TC})(	\widetilde{W}_{1},	\widetilde{W}_{1})]&=m_{20}-\frac{2m_{10}m_{11}}{m_{01}}-\frac{m_{10}}{n_{10}}\left( n_{20}-\frac{2m_{10}n_{11}}{m_{01}} \right)\\
			&=\frac{2(1+d)(cd-m-c-3md)}{4ad^{2}+(8a+2m^{2}-2mc)d+4a+m^{2}-c^{2}}\neq0.
		\end{aligned}	
	\end{equation*}
	Then the conclusion is tenable. 
\section{Analysis for the space-time system}	
It follows from the analysis in Section 2 that only equilibria $E_{10}$ and $E_{11}$ may experience Turing instability. This sections considers the bifurcations of system \eqref{diffusion system} at $E_{11}$. The case at $E_{10}$ could be similarly processing. Rewritig $E_{11}$ as $E_{*}(u_{*},v_{*})$ and
linearizing  \eqref{diffusion system} at $E_{*}(u_{*},v_{*})$ generates
\begin{equation}\label{linearized system}
	\partial_{t}\textbf{U}
	=J_{1}
	\textbf{U}
	+J_{2}\textbf{U},
\end{equation}
where
$\textbf{U}=(u \ v)^{T}$,
\begin{equation}\label{J1_2_3}
	J_{1}=\text{diag}\{d_{1}\Delta,d_{2}\Delta \} \ \text{and} \
	J_{2}=	\left(
	\begin{array}{cc}
		r_{11}           &  r_{12} \\
		r_{21}           &  r_{22}
	\end{array}
	\right),
\end{equation}
with
$r_{11}=u_{*}v_{*}(m-2u_{*}), \ r_{12}=\frac{u_{*}}{v_{*}}, \ r_{21}=-(c+2du_{*})v_{*}, \ r_{22}= -\frac{1}{v_{*}}$.\\
\indent Denoting the eigenvalues of
\begin{equation}
	\Delta \varepsilon(x)+\theta \varepsilon(x)=0, \ x \in (0,\mathcal{l}\pi), \\
	\ \varepsilon_{x}|_{x=0,\mathcal{l}\pi}=0,
\end{equation}
by $\theta_{n}$,
then the corresponding eigenfunctions of  $\theta_{n}=\frac{n^{2}}{\mathcal{l}^{2}}$ are $\varepsilon_{n}(x)=\cos\frac{n}{\mathcal{l}}x$.\\
Plugging
\begin{equation}
	\left(
	\begin{array}{c}
		u(t,x) \\ v(t,x)
	\end{array}
	\right)=
	\sum\limits_{n=0}^{\infty}	
	\left(
	\begin{array}{c}
		\mathcal{P}_{n} \\ \mathcal{Q}_{n}
	\end{array}
	\right)e^{\lambda_{n}t}\varepsilon_{n}(x)
\end{equation}
into \eqref{diffusion system} produces  the characteristic equation $\Gamma(\lambda)$:
\begin{flalign}\label{diffusion system characteristic equation}
	\lambda^{2}-(r_{11}+r_{22}-(d_{1}+d_{2})\theta_{n})\lambda+d_{1}d_{2}\theta^{2}_{n}-(d_{1}r_{22}+d_{2}r_{11})\theta_{n}+r_{11}r_{22}-r_{12}r_{21}=0.
\end{flalign}
Solving $r_{11}+r_{22}=0$ obtains the spatially homogeneous Hopf curve
\begin{equation}\label{Hopf bifurcation curve}
	H: \ a_{H}=a^{*}=\frac{(c-m)(md+c)+(1+d)\sqrt{(m-c)(md+2c-m)}}{(1+d)^{2}}.
\end{equation} 
Solving $\Gamma(0)=0$ obtains the Turing curve
\begin{equation}\label{Turing bifurcation curve}
	T_{n}: \ d^{(n)}_{1,T}=d^{(n)}_{1,T}(a) \triangleq \frac{d_{2}r_{11}\theta_{n}-(r_{11}r_{22}-r_{12}r_{21})}{\theta_{n}(d_{2}\theta_{n}-r_{22})} \ \text{for} \ n \textgreater \mathcal{l} \sqrt{\frac{r_{11}r_{22}-r_{12}r_{21}}{d_{2}r_{11}}}.
\end{equation}
Note that Hopf and Turing bifurcations are only possible if $r_{11}\textgreater 0 \ (\Leftrightarrow m\textgreater 2u_{*})$.
Based on $\frac{\partial r_{11}}{\partial a}=-u_{*}v^{2}_{*}(m-2u_{*})\textless 0$ and $\frac{\partial r_{22}}{\partial a}=-1\textless 0$, we gain
\begin{equation}
	\frac{\partial 	d^{(n)}_{1,T}}{\partial a}=\frac{d_{2}\theta^{2}_{n}(d_{2}\theta^{2}_{n}-r_{22})\frac{\partial r_{11}}{\partial a}+(d_{2}r_{11}\theta^{2}_{n}-(r_{11}r_{22}-r_{12}r_{21}))\frac{\partial r_{22}}{\partial a}}{\theta^{2}_{n}(d_{2}\theta^{2}_{n}-r_{22})^{2}}\textless 0,
\end{equation}
which means that $d^{(n)}_{1,T}$  decreases monotonically with respect to $a$ for fixed $n$.\\
Letting
\begin{equation}
	f(y)=\frac{d_{2}r_{11}y-(r_{11}r_{22}-r_{12}r_{21})}{y(d_{2}y-r_{22})},
\end{equation}
then 
\begin{equation}
	f^{\prime}(y)=\frac{-r_{11}d^{2}_{2}y^{2}+2d_{2}(r_{11}r_{22}-r_{12}r_{21})y-r_{22}(r_{11}r_{22}-r_{12}r_{21})}{y^{2}(d_{2}y-r_{22})^{2}}.
\end{equation}
Then $f(y)$ reaches the maximum at $y=y_{*}$, where
\begin{equation}
	y_{*}=\frac{r_{11}r_{22}-r_{12}r_{21}+\sqrt{-r_{12}r_{21}(r_{11}r_{22}-r_{12}r_{21})}}{d_{2}r_{11}}.
\end{equation} 
That is, for fixed $a$, $d^{(n)}_{1,T}$ reaches the maximum at $n=n^{\#}$,
where 
\begin{equation}
	n^{\#}=
	\begin{cases}
		\mathcal{l}\left[\sqrt{y_{*}}\right] \ \ \ \ \ \ \ \ \ \ \text{for} \ d^{(	\mathcal{l}\left[\sqrt{y_{*}}\right] )}_{1,T} \geq d^{(	\mathcal{l}\left[\sqrt{y_{*}}\right] )+1}_{1,T}, \\	1+\mathcal{l}\left[\mathcal{l}\sqrt{y_{*}}\right] \ \ \ \ \text{for} \ d^{(	\mathcal{l}\left[\sqrt{y_{*}}\right] )}_{1,T} \textless d^{(	\mathcal{l}\left[\sqrt{y_{*}}\right])+1}_{1,T}.
	\end{cases}
\end{equation}
To study Turing-Turing bifurcation, we further consider \eqref{Turing bifurcation curve}. By calculation, the equivalent expression of $d_{2}r_{11}\theta_{n}-(r_{11}r_{22}-r_{12}r_{21})\textgreater0$ is $a\textless a^{(n)}$,
where
\begin{equation}
 a^{(n)}=\frac{1}{(1+d)^2}\left( \frac{d_{2}(1+d)(md+2c-m)}{m-c}\theta_{n} -(m-c)(md+c)\right).
\end{equation}
Let the roots of $d^{(n)}_{1,T}(a)=d^{(n+1)}_{1,T}(a)$ be $a_{n} \ (\textless a^{(n)})$ with $a_{0}=+\infty$ and $a_{+\infty}=0$. Then we gain a series of segmented curves $T_{n}:d^{(n)}_{1,T}(a)$ for $a \in (a_{n},a_{n+1})$. \\
The transversal conditions about Hopf and Turing bifurcations are established below.
\begin{flalign*}\label{diffusion transversal conditions}
	&\frac{\text{d} \lambda (d_{1})}{\text{d} d_{1}} \Big|_{d_{1}=d^{(n)}_{1,T}}=\frac{d_{2}\theta^{2}_{n}-r_{22}\theta_{n}}{r_{11}+r_{22}-(d_{1}+d_{2})\theta_{n}}\textless 0,\\
	&\frac{\text{d} \text{Re} \lambda (a)}{\text{d} a} \Big|_{a=a^{*}}=\frac{1}{2}\left( (mv_{*}-4u_{*}v_{*})\frac{\partial u_{*}}{\partial a}+(mu_{*}-2u^{2}_{*}+\frac{1}{v^{2}_{*}})\frac{\partial v_{*}}{\partial a}\right)\bigg|_{a =a^{*}}
	=-1.
\end{flalign*}
In consequence, we get the following summary.
\begin{theorem}
	For system \eqref{diffusion system}, suppose that $a_{H}$ and $d^{(n)}_{1,T}$ are defined by \eqref{Hopf bifurcation curve} and \eqref{Turing bifurcation curve}, respectively.\\  
	$(i)$ Spatial homogeneous Hopf bifurcation emerges at $a=a^{*}$ and Turing  bifurcation  emerges at  $d_{1}=d^{(n)}_{1,T}$.\\
	$(ii)$ The $(0,n^{\#})$-mode Hopf-Turing  bifurcation  emerges at $(a,d_{1})=(a^{*},d^{(n^{\#})}_{1,T}(a^{*}))$.\\
	$(iii)$ The $(n,n+1)$-mode Turing-Turing bifurcation emerges at $(a,d_{1})=(a_{n},d^{(n)}_{1,T}(a_{n}))$.
\end{theorem}
	
	\begin{figure}[H]
		\centering
		\begin{minipage}[c]{0.33\textwidth}
			\centering
			\includegraphics[height=6cm,width=7cm]{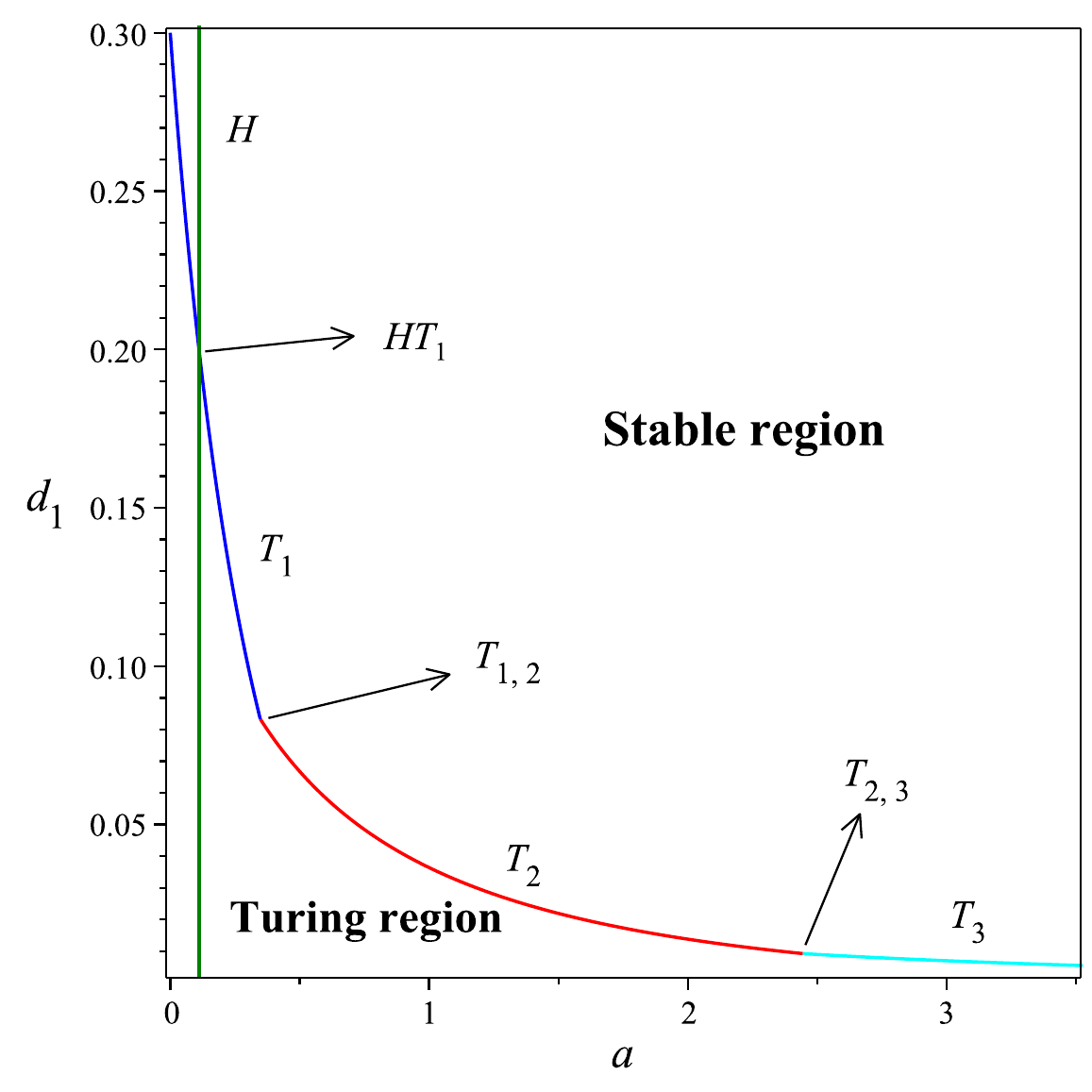}
		\end{minipage}
		\caption{Stability domain and bifurcations. Taking $(c,d,m,d_{2},\mathcal{l})=(1,2,2,1,1)$, then the Hopf-Turing bifurcation point, mode-(1,2) and mode-(2,3) Turing-Turing bifurcation points are $HT_{1}(\frac{1}{9},\frac{1}{5})$, \ $T_{1,2}(\frac{\sqrt{417}}{6}-\frac{55}{18},\frac{1}{12})$ and $T_{2,3}(\frac{22}{9},\frac{1}{108})$, respectively.}\label{Stability and bifurcation region}
	\end{figure}
	
Next we focus on the stability of periodic solutions caused by spatial Hopf bifurcation. By the foregoing analysis, we assert that if $a=a^{*}$, then $J_{2}$ admits a pair of pure imaginary eigenvalues $\pm \text{i}w_{0}(=\pm \text{i}\sqrt{r_{11}r_{22}-r_{12}r_{21}})$.	Next we employ the technique in {\cite{Hassard}} to deduce the normal form. 
	\begin{equation}
\partial_{t}\textbf{U}
=J_{1}
\textbf{U}
+J_{2}\textbf{U}
	\end{equation}
	where $J^{*}_{2}=J_{2}$ with $X$. If $L=J_{1}+J_{2}$, 
	\begin{equation*}
		Q=
		\left(
		\begin{array}{c}
			1\\  \frac{-r_{11}+\text{i}w_{0}}{r_{12}}
		\end{array}
		\right)
		\triangleq
		\left(
		\begin{array}{c}
			Q_{1}\\ Q_{2}
		\end{array}
		\right),
		Q^{*}=
		\left(
		\begin{array}{c}
			\frac{w_{0}+\text{i}r_{11}}{2\pi w_{0}} \\  \frac{\text{i}r_{12}}{2\pi w_{0}}
		\end{array}
		\right)
		\triangleq
		\left(
		\begin{array}{c}
			Q^{*}_{1}\\ Q^{*}_{2}
		\end{array}
		\right),
	\end{equation*}
	then we see $LQ=\text{i}w_{0}Q, L^{*}Q^{*}=-\text{i}w_{0}Q^{*}, <Q^{*},Q>=1,<Q^{*},\overline{Q}>=0$, where the inner product is measured by $<z_{1},z_{2}>=\int_{0}^{\pi}\overline{z_{1}}^{T}z_{2} \ d\tau$. As per {\cite{Hassard}}, we obtain
	\begin{equation}
	c_{1}(\theta_{0})=\frac{\text{i}}{2w_{0}}(\delta_{20}\delta_{11}-2|\delta_{11}|^{2}-\frac{1}{3}|\delta_{02}|^{2})+\frac{1}{2}\delta_{21}.
\end{equation}
Now let's calculate $\delta_{02}, \ \delta_{20}, \ \delta_{11}, \ \delta_{21}$. Setting  $\widetilde{F}=(F_{1},F_{2})^{T}$, where $F_{j} \ (j=1,2)$ denote the right-hand function of \eqref{local system}. Shifting $E_{*}$ to $(0,0)$ and completing Taylor expansion of \eqref{local system} about $(0,0)$ yields
\begin{equation}
	\begin{cases}
	u_{t}=\kappa_{10} u+\kappa_{10} v+\kappa_{20} u^{2}+\kappa_{11}uv+\kappa_{30}u^{3}+\kappa_{21}u^{2}v+O(|(u,v)|^{4}), \\
	u_{t}=\rho_{10} u+\rho_{10} v+\rho_{20} u^{2}+\rho_{11}uv+\rho_{30}u^{3}+\rho_{21}u^{2}v+O(|(u,v)|^{4}), 
	\end{cases}
\end{equation} 
where 
\begin{flalign*}
	&\kappa_{10}=av_{13}-1+2mu_{13}v_{13}-3u^{2}_{13}v_{13}, \  \kappa_{11}=a+2mu_{13}-3u^{2}_{13}, \ \kappa_{20}=mv_{13}-3u_{13}v_{13}, \\
	&\rho_{10}=-(c+2du_{13})v_{13}, \ \rho_{01}=-(b+cu_{13}+du^{2}_{13}), \  \rho_{11}=-(c+2du_{13}), \ \rho_{20}=-dv_{13}, \ \rho_{30}=0.
\end{flalign*}
Besides, we have
\begin{equation*}
	\begin{aligned}
		<Q^{*},\widetilde{F}>&=\frac{1}{2w_{0}}(w_{0}F_{1}-\text{i}(r_{11}F_{1}+r_{12}F_{2})),\\
		<\overline{Q^{*}},\widetilde{F}>&=\frac{1}{2w_{0}}(w_{0}F_{1}+\text{i}(r_{11}F_{1}+r_{12}F_{2})),\\
		<Q^{*},\widetilde{F}>Q&=\frac{1}{2w_{0}}
		\left(
		\begin{array}{c}
			w_{0}F_{1}-\text{i}(r_{11}F_{1}+r_{12}F_{2}) \\ (w_{0}F_{1}-\text{i}(r_{11}F_{1}+r_{12}F_{2}))\frac{-r_{11}+\text{i}w_{0}}{r_{12}}
		\end{array}
		\right),\\
		<\overline{Q^{*}},\widetilde{F}>\overline{Q}&=\frac{1}{2w_{0}}
		\left(
		\begin{array}{c}
			w_{0}F_{1}+\text{i}(r_{11}F_{1}+r_{12}F_{2}) \\ (w_{0}F_{1}+\text{i}(r_{11}F_{1}+r_{12}F_{2}))\frac{-r_{11}-\text{i}w_{0}}{r_{12}}
		\end{array}
		\right).
	\end{aligned}
\end{equation*}
In addition,
\begin{equation*}
	\delta_{02}=<\overline{Q^{*}},(c_{0},d_{0})^{T}>,\delta_{20}=<Q^{*},(c_{0},d_{0})^{T}>,\delta_{11}=<Q^{*},(e_{0},f_{0})^{T}>,\delta_{21}=<Q^{*},(g_{0},h_{0})^{T}>,
\end{equation*}
where
\begin{equation*}
	\begin{aligned}
		c_{0}&=F_{1uu}Q^{2}_{1}+2F_{1uv}Q_{1}Q_{2}+F_{1vv}Q^{2}_{2}=2\kappa_{20}+2\kappa_{11}Q_{2},\\
		d_{0}&=F_{2uu}Q^{2}_{1}+2F_{2uv}Q_{1}Q_{2}+F_{2vv}Q^{2}_{2}=2\rho_{20}+2\rho_{11}Q_{2}+2\rho_{02}Q^{2}_{2},\\
		e_{0}&=F_{1uu}|Q_{1}|^{2}+F_{1uv}(Q_{1}\overline{Q_{2}}+\overline{Q_{1}}Q_{2})+F_{1vv}|Q_{2}|^{2}=2\kappa_{20}+\kappa_{11}(\overline{Q_{2}}+Q_{2}),\\
		f_{0}&=F_{2uu}|Q_{1}|^{2}+F_{2uv}(Q_{1}\overline{Q_{2}}+\overline{Q_{1}}Q_{2})+F_{2vv}|Q_{2}|^{2}=2\rho_{20}+\rho_{11}(\overline{Q_{2}}+Q_{2})+2\rho_{02}|Q_{2}|^{2},\\
		g_{0}&=F_{1uuu}|Q_{1}|^{2}Q_{1}+F_{1uuv}(2|Q_{1}|^{2}Q_{2}+Q^{2}_{1}\overline{Q_{2}})+F_{1uvv}(2|Q_{2}|^{2}Q_{1}+Q^{2}_{2}\overline{Q_{1}}+F_{1vvv}|Q_{2}|^{2}Q_{2})\\
		&=6\kappa_{30}+2\kappa_{21}(2Q_{2}+\overline{Q_{2}}),\\
		h_{0}&=F_{2uuu}|Q_{1}|^{2}Q_{1}+F_{2uuv}(2|Q_{1}|^{2}Q_{2}+Q^{2}_{1}\overline{Q_{2}})+F_{2uvv}(2|Q_{2}|^{2}Q_{1}+Q^{2}_{2}\overline{Q_{1}}+F_{2vvv}|Q_{2}|^{2}Q_{2})\\
		&=6\rho_{30}+2\rho_{21}(2Q_{2}+\overline{Q_{2}}),
	\end{aligned}
\end{equation*}
and the partial derivatives above are all at point $(u,v,a)=(0,0,a^{*})$. Thereupon, we gain
\begin{flalign*}
	\delta_{02}&=\kappa_{20}-\rho_{11}+\frac{2\kappa_{10}(\rho_{02}-\kappa_{11})}{\kappa_{01}}+\frac{\text{i}}{w_{0}}\left(\kappa_{10}(\kappa_{20}-\rho_{11})+\kappa_{01}\rho_{20}+\frac{(\kappa^{2}_{10}-w^{2}_{0})(\rho_{02}-\kappa_{11})}{\kappa_{01}}\right),\\
	\delta_{20}&=\kappa_{20}+\rho_{11}+\frac{2\kappa_{10}\rho_{02}}{\kappa_{01}}-\frac{\text{i}}{w_{0}}\left(\kappa_{10}(\kappa_{20}-\rho_{11})+\kappa_{01}\rho_{20}+\frac{\rho_{02}(\kappa^{2}_{10}-w^{2}_{0})-\kappa_{11}(\kappa^{2}_{10}+w^{2}_{0})}{\kappa_{01}}\right),\\
	\delta_{11}&=\kappa_{20}-\frac{\kappa_{10}\kappa_{11}}{\kappa_{01}}-\frac{\text{i}}{w_{0}}\left(\kappa_{10}(\kappa_{20}-\rho_{11})+\kappa_{01}\rho_{20}+\frac{\rho_{02}(\kappa^{2}_{10}+w^{2}_{0})-\kappa_{11}\kappa^{2}_{10}}{\kappa_{01}}\right),\\
	\delta_{21}&=3\kappa_{30}+\rho_{21}-\frac{2\kappa_{21}\kappa_{10}}{\kappa_{01}}+\frac{\text{i}}{w_{0}}
	\left(3(\kappa_{10}\rho_{21}-\kappa_{01}\rho_{30}-\kappa_{10}\kappa_{30})+\frac{\kappa_{21}(3\kappa^{2}_{10}+w^{2}_{0})}{\kappa_{01}} \right).
\end{flalign*}
\indent Evidently, we have $\text{Re}(\lambda^{\prime}(a^{*}))\textless 0$. 
Then we claim that for the spatial homogeneous Hopf bifurcation of \eqref{diffusion system} around $E_{*}$, its direction and stability depend on $\text{Re}(c_{1}(a^{*}))$. Specifically, if $\text{Re}(c_{1}(a^{*}))\textgreater 0$, then it's subcritical and unstable. If $\text{Re}(c_{1}(a^{*})) \textless0$, then it's supercritical and stable.
	
\section{Numerical computation and simulation for pattern formation}	
\begin{figure}[H] 
	\centering
	\begin{minipage}[b]{.22\linewidth}
		\centering
		\subfigure[]
		{
			\includegraphics[scale=0.25]{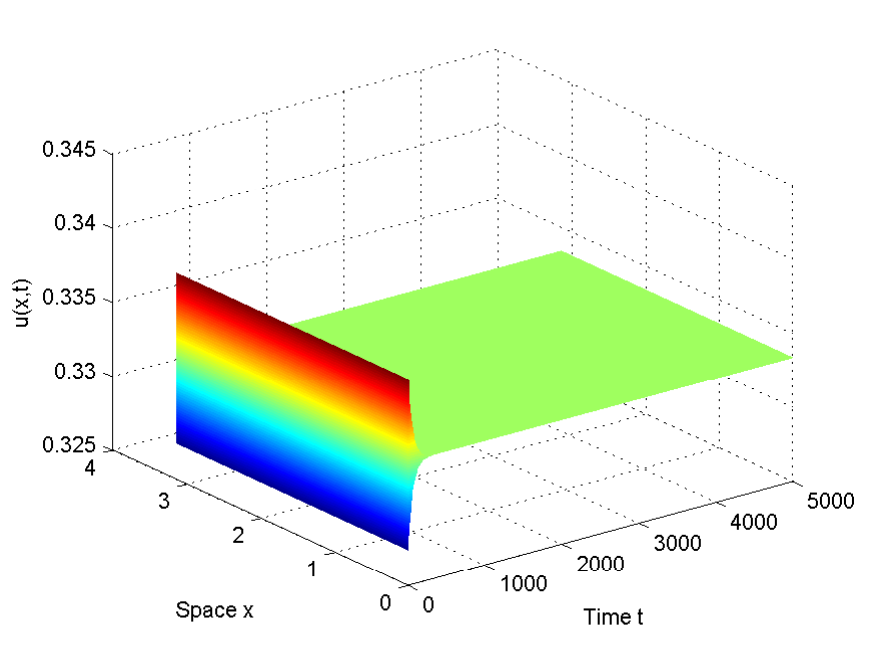}
		}
	\end{minipage}
	\begin{minipage}[b]{.22\linewidth}
		\subfigure[]
		{
			\includegraphics[scale=0.25]{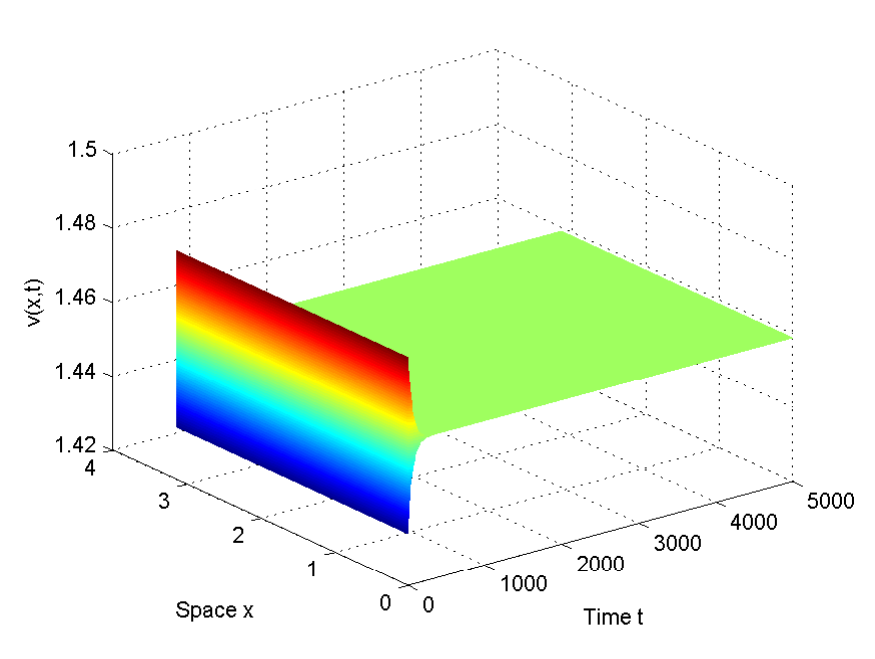}
		}
	\end{minipage}
	\begin{minipage}[b]{.22\linewidth}
		\centering
		\subfigure[]
		{
			\includegraphics[scale=0.25]{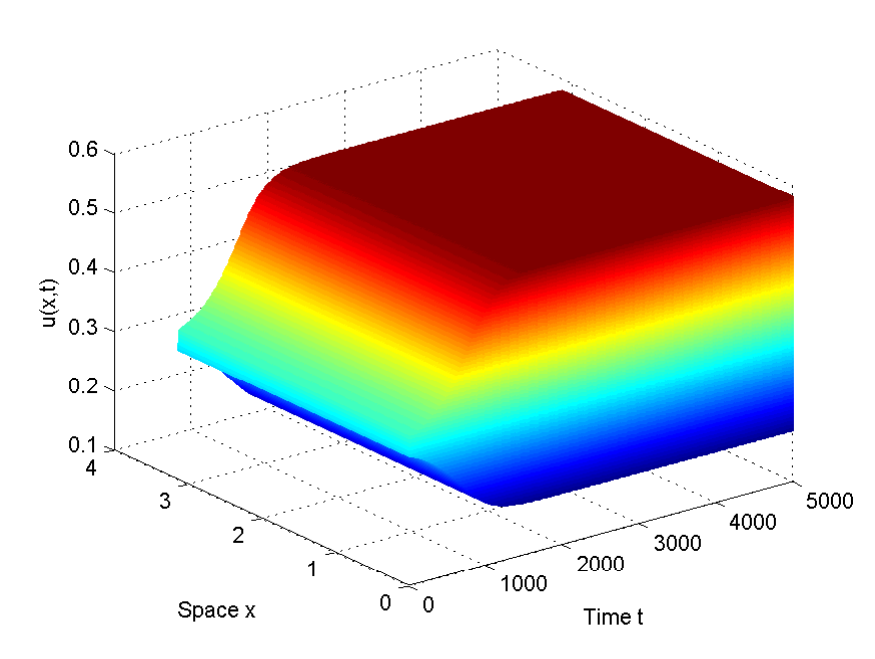}
		}
	\end{minipage}
	\begin{minipage}[b]{.22\linewidth}
		\subfigure[]
		{
			\includegraphics[scale=0.25]{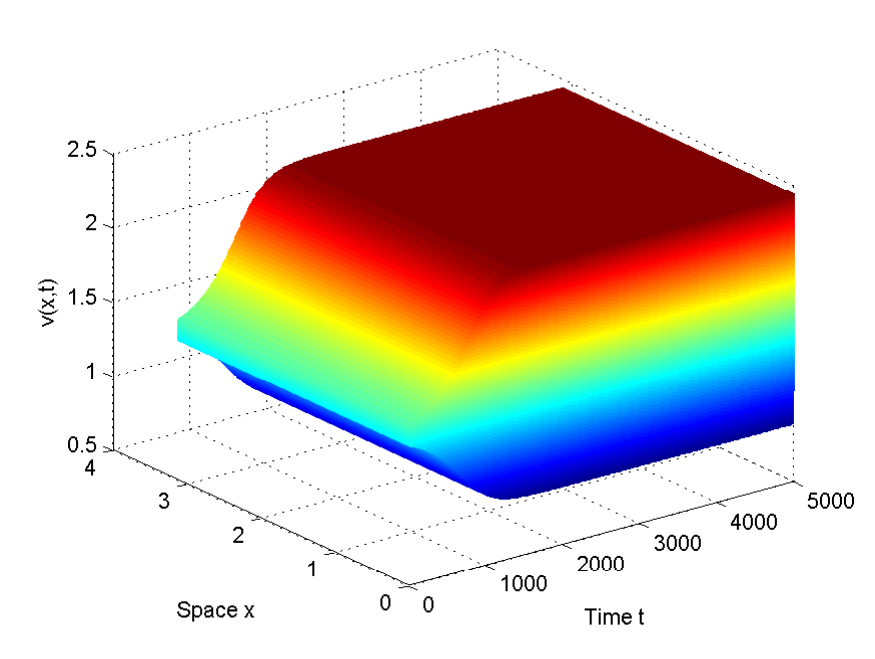}
		}
	\end{minipage}
	\caption{$(a)-(b)$ Taking $(b,c,d,m,d_{1},d_{2},\mathcal{l})=(0.13,1,2,2,0.2,2,1)$ and $a=0.13 \textgreater a_{H}=\frac{1}{9}$, then $E_{*}(\frac{1}{3},1.4377)$ is stable. $(c)-(d)$ Taking  $(b,c,d,m,d_{1},d_{2},\mathcal{l})=(0.11,1,2,2,0.17,2,1)$ and $a=0.11\textless a_{H}$, then the stable periodic solutions appear. The initial values are all  $(0.34,1.44)$.}\label{Hopf0 and stable}
\end{figure}

\begin{figure}[H]
	\centering
	\begin{minipage}[b]{.22\linewidth}
		\centering
		\subfigure[]
		{
			\includegraphics[scale=0.25]{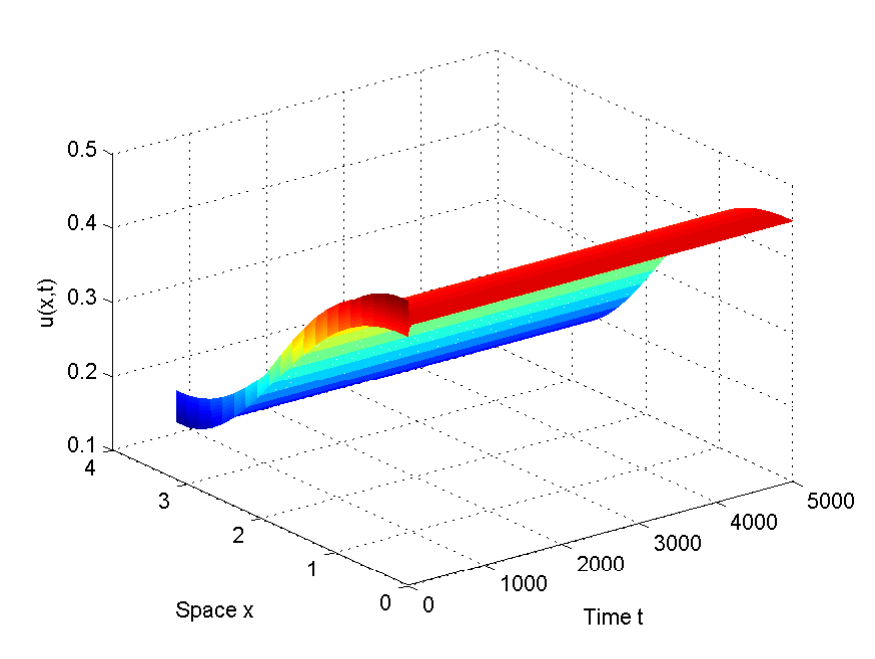}
		}
	\end{minipage}
	\begin{minipage}[b]{.22\linewidth}
		\subfigure[]
		{
			\includegraphics[scale=0.25]{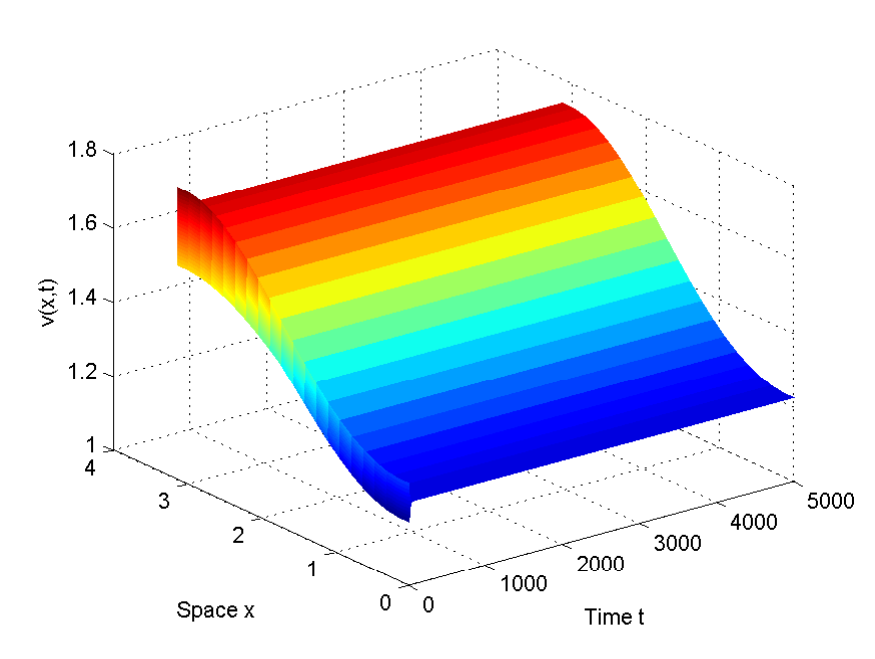}
		}
	\end{minipage}
	\begin{minipage}[b]{.22\linewidth}
		\centering
		\subfigure[]
		{
			\includegraphics[scale=0.25]{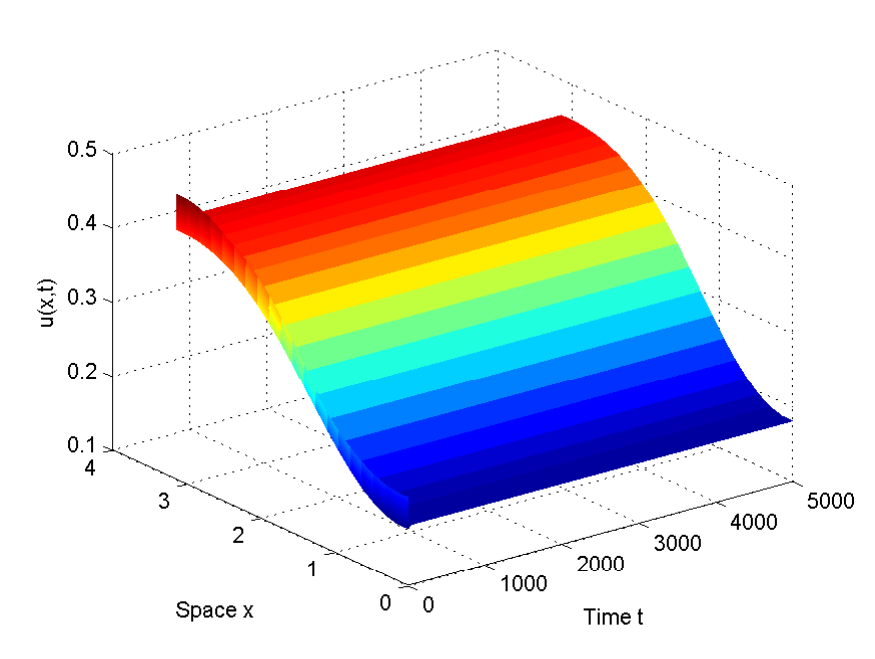}
		}
	\end{minipage}
	\begin{minipage}[b]{.22\linewidth}
		\subfigure[]
		{
			\includegraphics[scale=0.25]{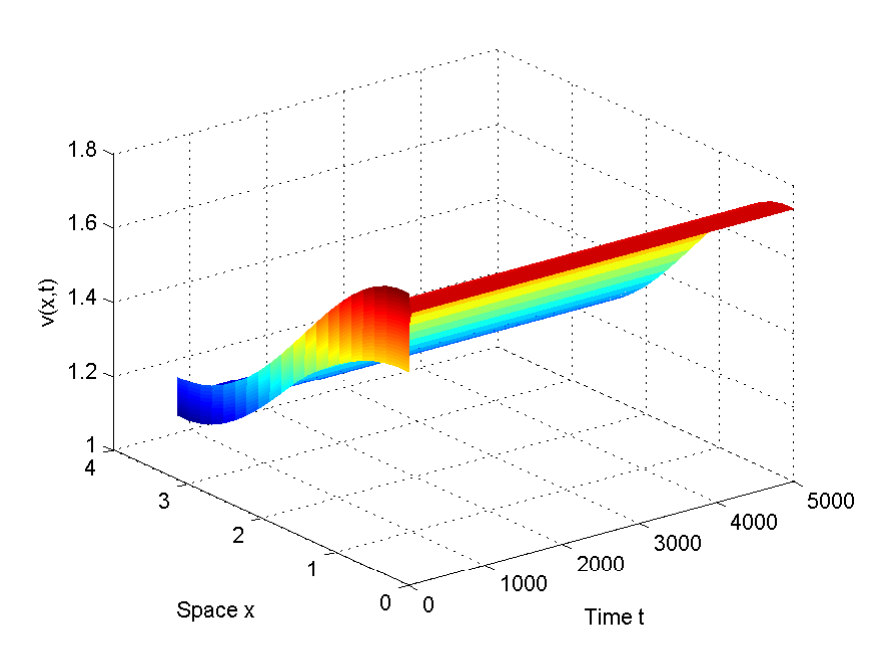}
		}
	\end{minipage}
	\caption{Taking $(a,b,c,d,m,d_{2},\mathcal{l})=(0.14,0.14,1,2,2,1,1)$ and $d_{1}=0.17\textless d^{(1)}_{1,T}=0.1803$, then stable spatially inhomogeneous steady states of $\cos x$-type appear. The initial values of  $(a)-(b)$ and $(c)-(d)$ take respectively  $(0.34+0.1 \cos x, \ 1.44+0.1 \cos x)$ and  $(0.34-0.1 \cos x, \ 1.44-0.1 \cos x)$.}\label{Turing1}
\end{figure}

\begin{figure}[H]
	\centering
	\begin{minipage}[b]{.22\linewidth}
		\centering
		\subfigure[]
		{
			\includegraphics[scale=0.25]{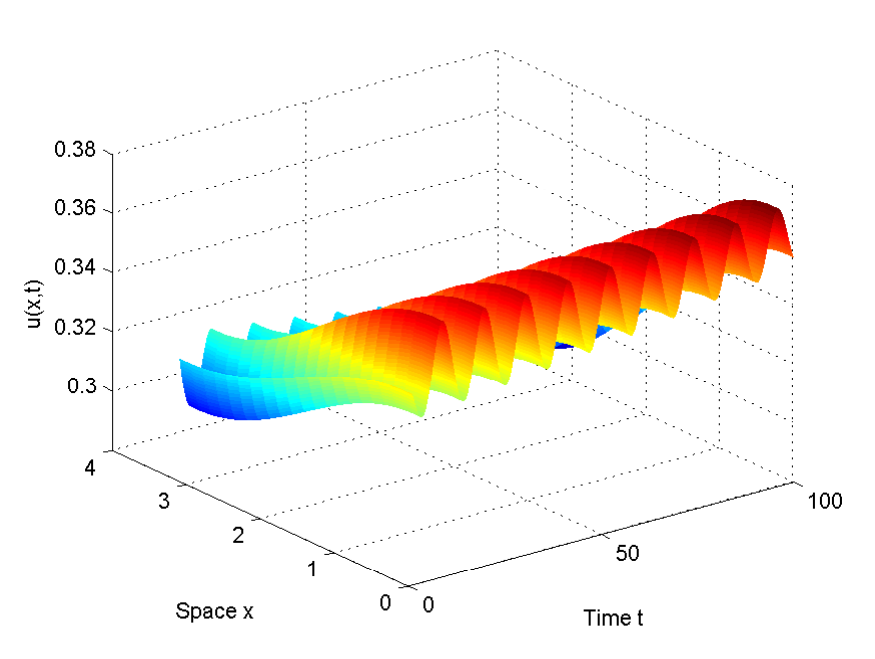}
		}
	\end{minipage}
	\begin{minipage}[b]{.22\linewidth}
		\subfigure[]
		{
			\includegraphics[scale=0.25]{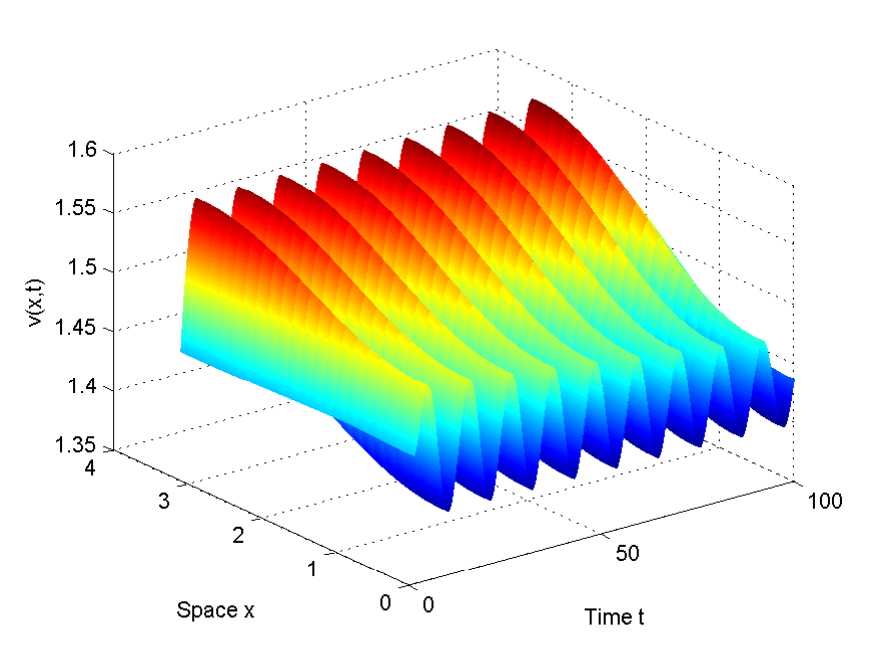}
		}
	\end{minipage}
	\begin{minipage}[b]{.22\linewidth}
		\centering
		\subfigure[]
		{
			\includegraphics[scale=0.25]{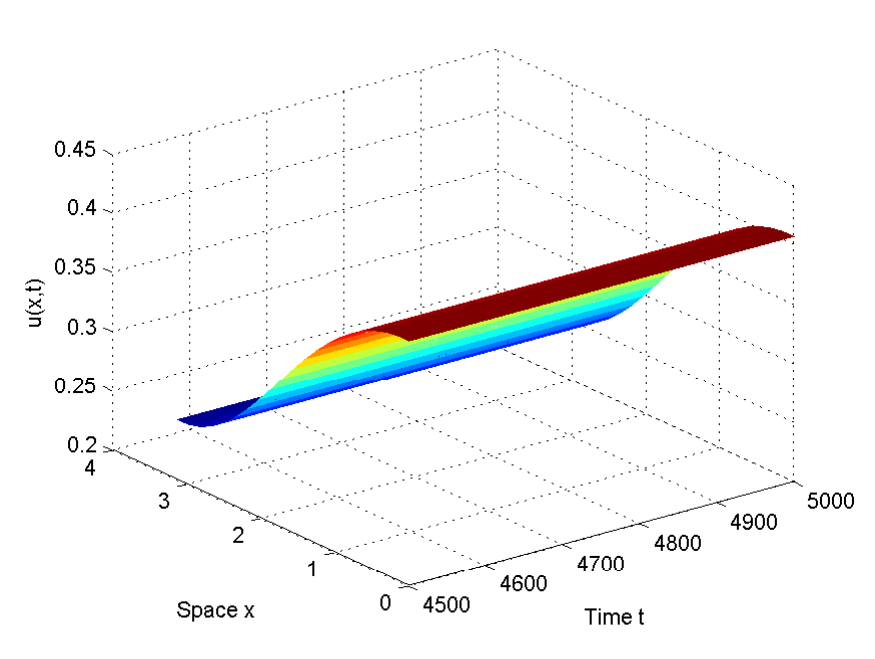}
		}
	\end{minipage}
	\begin{minipage}[b]{.22\linewidth}
		\subfigure[]
		{
			\includegraphics[scale=0.25]{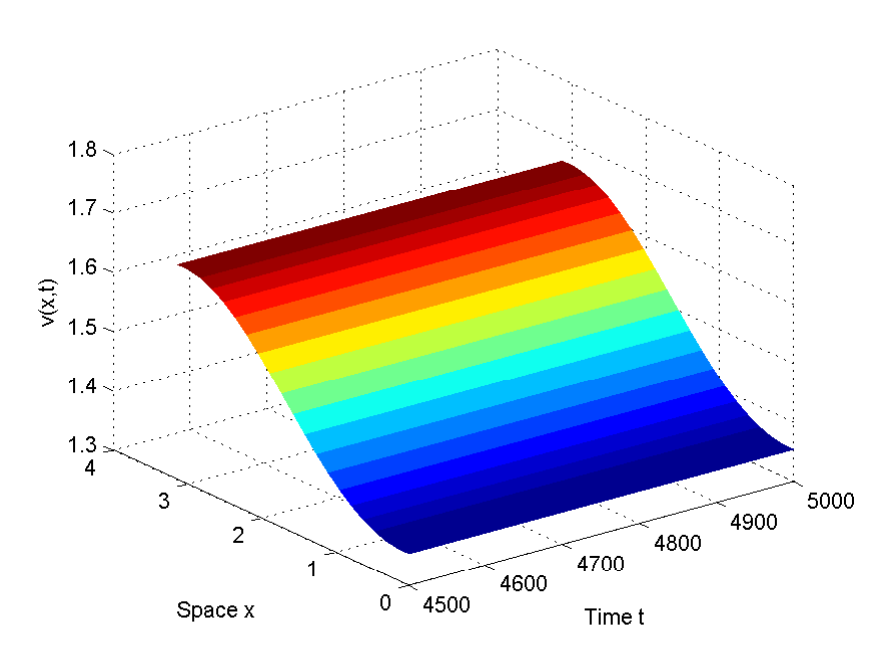}
		}
	\end{minipage}
	\caption{Taking $(b,c,d,m,d_{2},\mathcal{l})=(0.115,1,2,2,1,1)$ and $a=0.115, \ d_{1}=0.199$, then there are heteroclinic solutions connecting unstable periodic solutions to stable spatially inhomogeneous steady states of $\cos x$-type.
		The initial value is $(0.34, 1.44+0.1\cos x)$.}\label{Hopf-Turing}
\end{figure}

\begin{figure}[H]
	\centering
	\begin{minipage}[b]{.22\linewidth}
		\centering
		\subfigure[]
		{
			\includegraphics[scale=0.25]{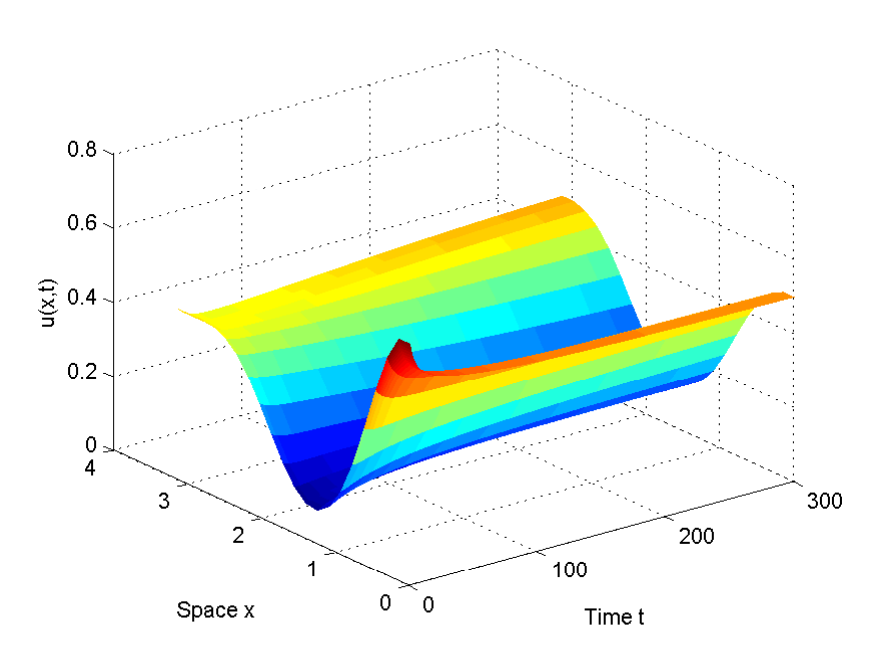}
		}
	\end{minipage}
	\begin{minipage}[b]{.22\linewidth}
		\subfigure[]
		{
			\includegraphics[scale=0.25]{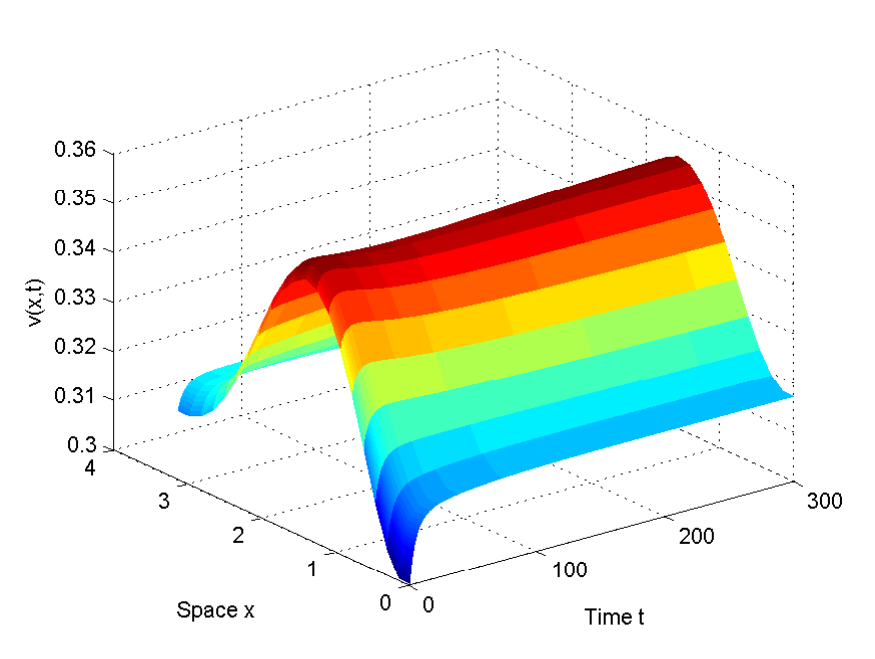}
		}
	\end{minipage}
	\begin{minipage}[b]{.22\linewidth}
		\centering
		\subfigure[]
		{
			\includegraphics[scale=0.25]{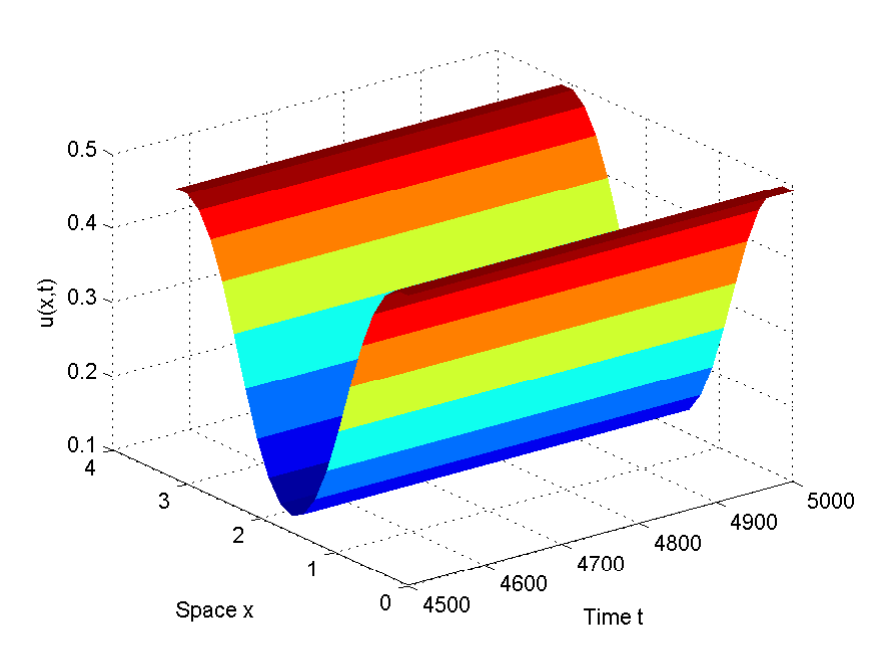}
		}
	\end{minipage}
	\begin{minipage}[b]{.22\linewidth}
		\subfigure[]
		{
			\includegraphics[scale=0.25]{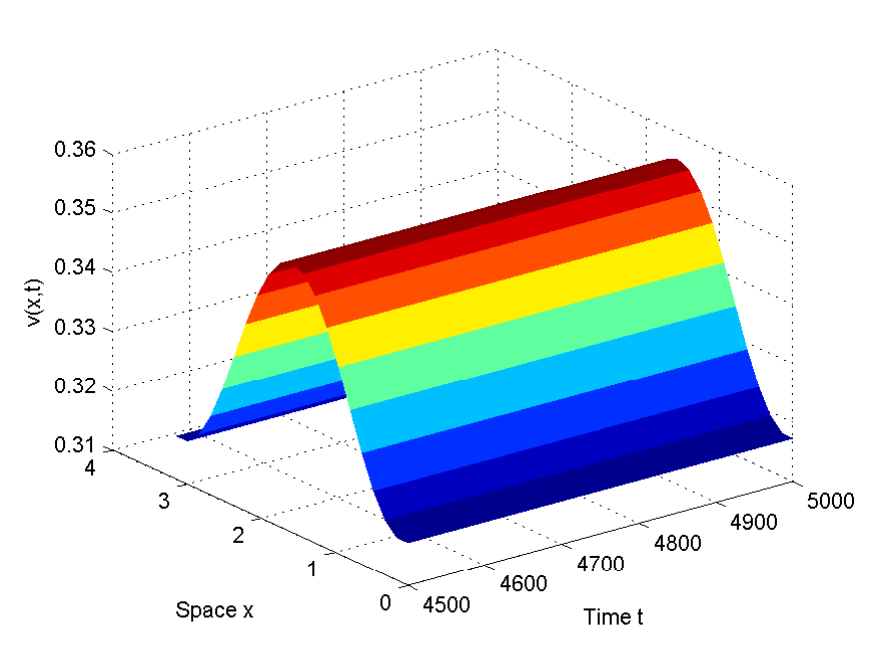}
		}
	\end{minipage}
	\caption{Taking $(b,c,d,m,d_{2},\mathcal{l})=(\frac{22}{9},1,2,2,1,1)$ and $a=\frac{22}{9}, \ d_{1}=0.008259\textless d^{(2)}_{1,T}(a^{\#})=d^{(3)}_{1,T}(a^{\#})=\frac{1}{108}$, then 
		stable spatially heterogeneous steady states of $\cos 2x$-type emerge.
		The initial value is $(\frac{1}{3}+0.2 \cos 2x+0.1\cos 3x, \frac{1}{3}+0.2 \cos 2x+0.1\cos 3x)$.}\label{1 Turing-Turing}
\end{figure}

\begin{figure}[H]
	\centering
	\begin{minipage}[b]{.22\linewidth}
		\centering
		\subfigure[]
		{
			\includegraphics[scale=0.25]{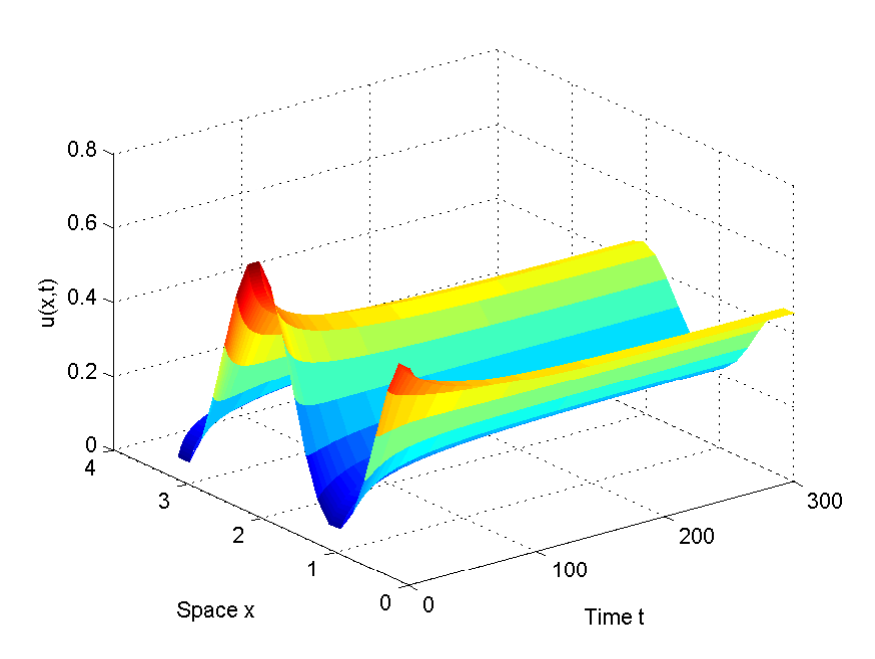}
		}
	\end{minipage}
	\begin{minipage}[b]{.22\linewidth}
		\subfigure[]
		{
			\includegraphics[scale=0.25]{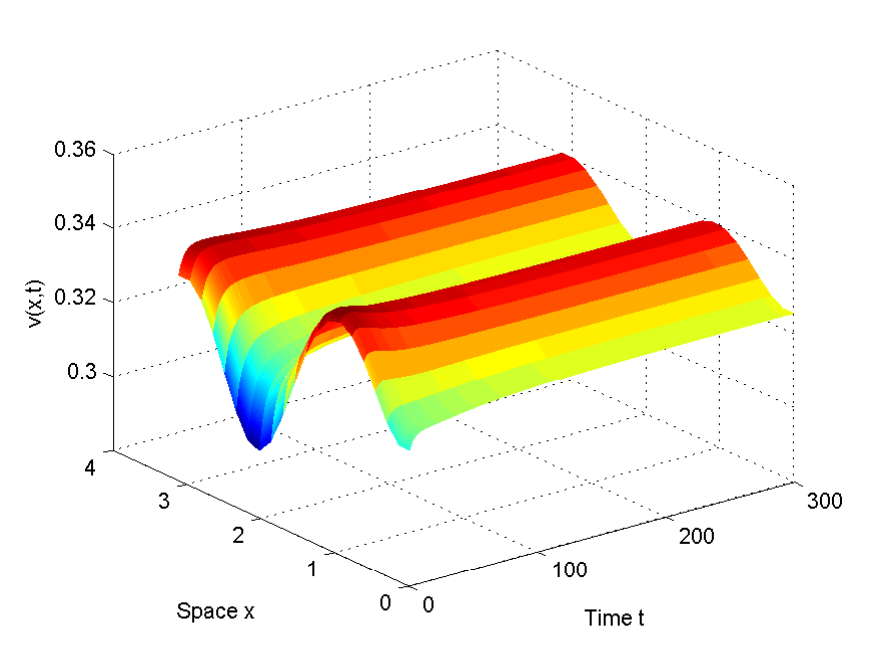}
		}
	\end{minipage}
	\begin{minipage}[b]{.22\linewidth}
		\centering
		\subfigure[]
		{
			\includegraphics[scale=0.25]{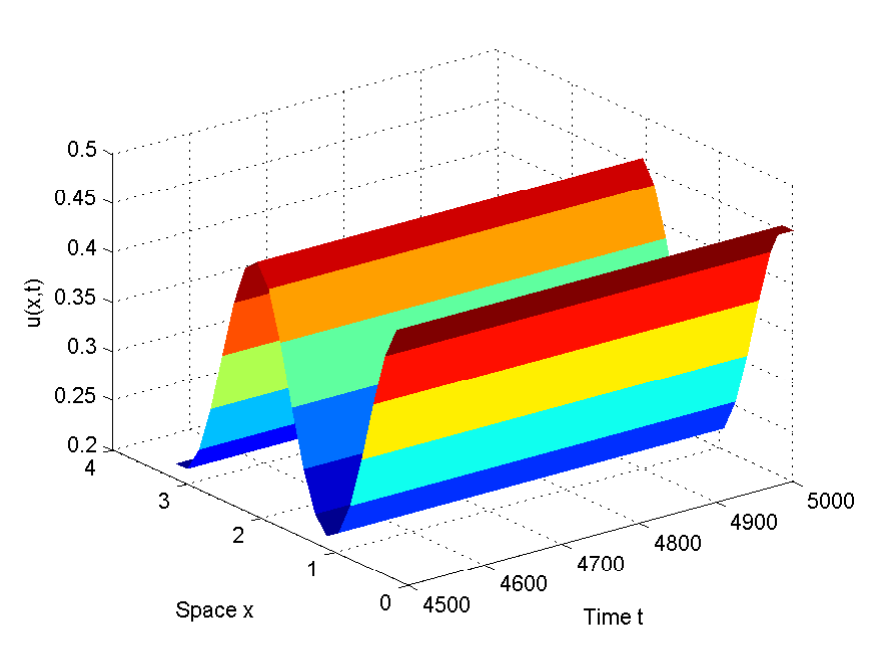}
		}
	\end{minipage}
	\begin{minipage}[b]{.22\linewidth}
		\subfigure[]
		{
			\includegraphics[scale=0.25]{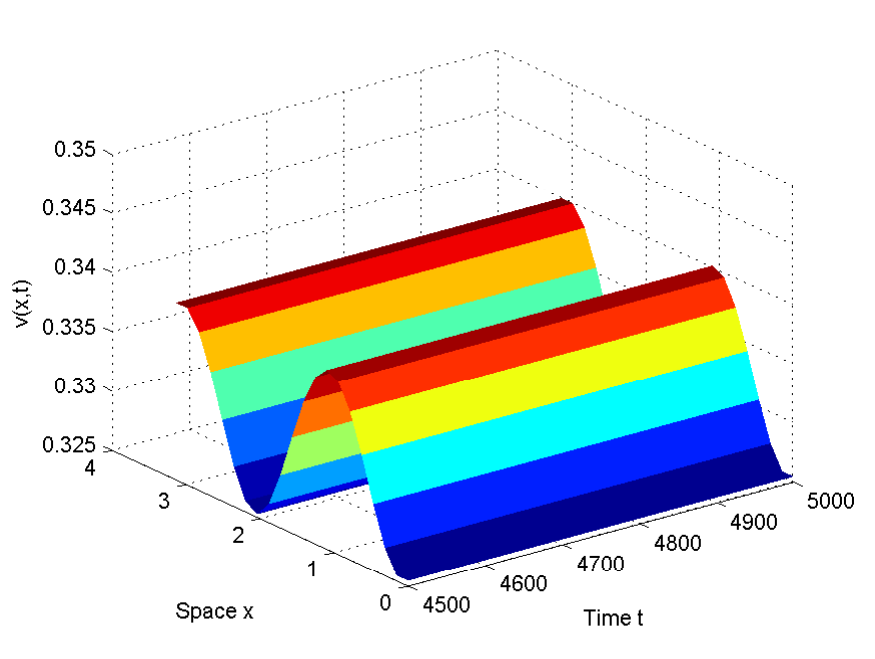}
		}
	\end{minipage}
	\caption{Here the parameter values keep up with Fig. \ref{1 Turing-Turing} and the different initial value takes $(\frac{1}{3}-0.03 \cos 2x+0.3\cos 3x, \frac{1}{3}-0.03 \cos 2x+0.3\cos 3x)$. Then stable spatially heterogeneous steady states of $\cos 3x$-type emerge.}\label{2 Turing-Turing}
\end{figure}

This section employs numerical simulations to validate theoretical analysis. For temporal system \eqref{local system}, Figs. \ref{E0}-\ref{E13 saddle-node and cusp} illustrate the type and stability of equilibria. In Fig. \ref{E0}, the type of $E_{0}(0,1)$ is discussed by fixing $(b,c,d)=(1,2,2)$ and varying $a$ and $m$. In Fig. \ref{E11 limit cycle}, we take $(b,c,d,m)=(\frac{\sqrt{2}}{2}-\frac{9}{16},2,1,\frac{5}{2})$, then we have  $g(d)=\frac{47}{8}$ and $a^{*}=\frac{\sqrt{2}}{2}-\frac{9}{16}$. By taking $a=0.15\textgreater a^{*}$, $E_{11}(\frac{1}{4},\sqrt{2})$ is stable; by taking $a=0.144 \textless a^{*}$, a limit cycle appears.  In Fig. \ref{E13 saddle-node and cusp}, we take $(c,d,m)=(1,1,2)$, then $h(d)=47$ and $a_{*}=\frac{\sqrt{6}}{4}-\frac{7}{16}$. By taking $a=\frac{1}{2} \textgreater a_{*}$ and $b=\frac{5}{8}$, $E_{13}(\frac{1}{4},\frac{16}{15})$ is an attracting saddle-node; by taking $a=a_{*}=\frac{\sqrt{6}}{4}-\frac{7}{16}$ and $b=\frac{\sqrt{6}}{4}-\frac{5}{16}$, we have $a_{\#}=-\frac{5}{48}$, then $E_{13}(\frac{1}{4},\frac{2\sqrt{6}}{3})$ is a cusp of codimension 2.\\
\indent For space-time system \eqref{diffusion system}, we fix $(c,d,m,d_{2},\mathcal{l})=(1,2,2,1,1)$, then we obatin the spatial homogeneous Hopf line $a_{H}=\frac{1}{9}$. In Fig. \ref{Turing1}, we take $b=a=0.14$, then $y_{*}=1.1483$ and $d^{(	\mathcal{l}\left[\sqrt{y_{*}}\right] )}_{1,T}=0.1803 \textgreater d^{(	\mathcal{l}\left[\sqrt{y_{*}}\right] )+1}_{1,T}=0.1183$. That is, $n^{\#}=1$ and then mode-(0,1) Hopf-Turing bifurcation arises. Besides, several Turing curves with different modes are shown below.
\begin{equation}
	\begin{cases}
		T_{1}: d^{(1)}_{1,T}(a)=\frac{21-27a}{81a^{2}+171a+70} \ \ \ \ \ \ \text{for} \ 0 \textless a \textless \frac{\sqrt{417}}{6}-\frac{55}{18}, \\
		T_{2}: d^{(2)}_{1,T}(a)=\frac{129-27a}{324a^{2}+1656a+820} \ \ \ \text{for} \ \frac{\sqrt{417}}{6}-\frac{55}{18} \textless a \textless \frac{22}{9}, \\
		T_{3}: d^{(3)}_{1,T}(a)=\frac{103-9a}{243a^{2}+2457a+1290} \ \ \text{for} \ \frac{22}{9} \textless a \textless \frac{\sqrt{1393}}{2}-\frac{235}{18}.
	\end{cases}
\end{equation}
Thus, the Hopf-Turing bifurcation point, mode-(1,2) and mode-(2,3) Turing-Turing bifurcation points are $HT_{1}(\frac{1}{9},\frac{1}{5})$, \ $T_{1,2}(\frac{\sqrt{417}}{6}-\frac{55}{18},\frac{1}{12})$ and $T_{2,3}(\frac{22}{9},\frac{1}{108})$, respectively.
Fig. \ref{Stability and bifurcation region} illustrates the stability domain and bifurcation curves in the $d_{1}-a$ plane.
Fig. \ref{Hopf0 and stable} displays the local stability $E_{*}(\frac{1}{3},1.4377)$ by taking $a=0.13 \textgreater a_{H}=\frac{1}{9}$ and periodic solutions by taking $a=0.11\textless a_{H}$. For this set of parameters  $(b,c,d,m,d_{1},d_{2},\mathcal{l})=(0.11,1,2,2,0.17,2,1)$, we work out $\text{Re}(c_{1}(a^{*}))=-\frac{1}{3}\textless0$. Thereupon, the periodic solutions are stable. Fig. \ref{Turing1} depicts stable nonconstant steady states by taking $d_{1}=0.17\textless d^{(1)}_{1,T}=0.1803$. Fig. \ref{Hopf-Turing} reflects the key feature around  Hopf-Turing bifurcation point $(\frac{1}{9},0.2)$. 
Specifically, we found the evolution of solutions in system \eqref{diffusion system}, transitioning from unstable periodic solutions to stable spatially inhomogeneous steady states. Figs. \ref{1 Turing-Turing}-\ref{2 Turing-Turing} display that different initial values result in vastly divergent spatial patterns. For instance, we here focus on point $T_{2,3}$, then the profiles of the ultimate behaviors in Figs. \ref{1 Turing-Turing} and \ref{2 Turing-Turing} are $\pm \cos 2x$ and $\pm \cos 3x$, respectively. 

\begin{figure}[H]
	\centering
	\begin{minipage}[b]{.45\linewidth}
		\centering
		\subfigure[]
		{
			\includegraphics[scale=0.15]{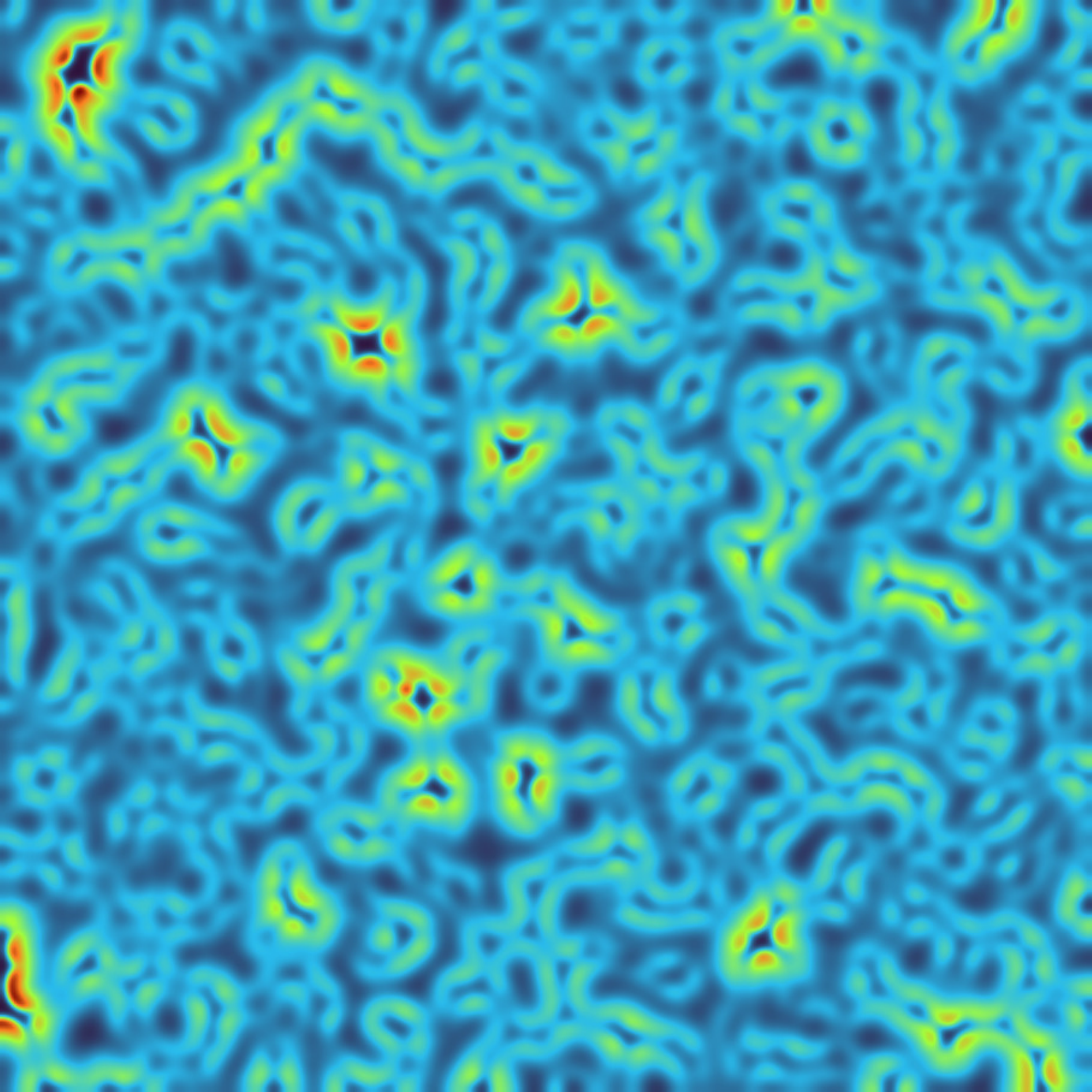}
		}
	\end{minipage}
	\hspace{0.05\linewidth} 
	\begin{minipage}[b]{.45\linewidth}
		\centering
		\subfigure[]
		{
			\includegraphics[scale=0.15]{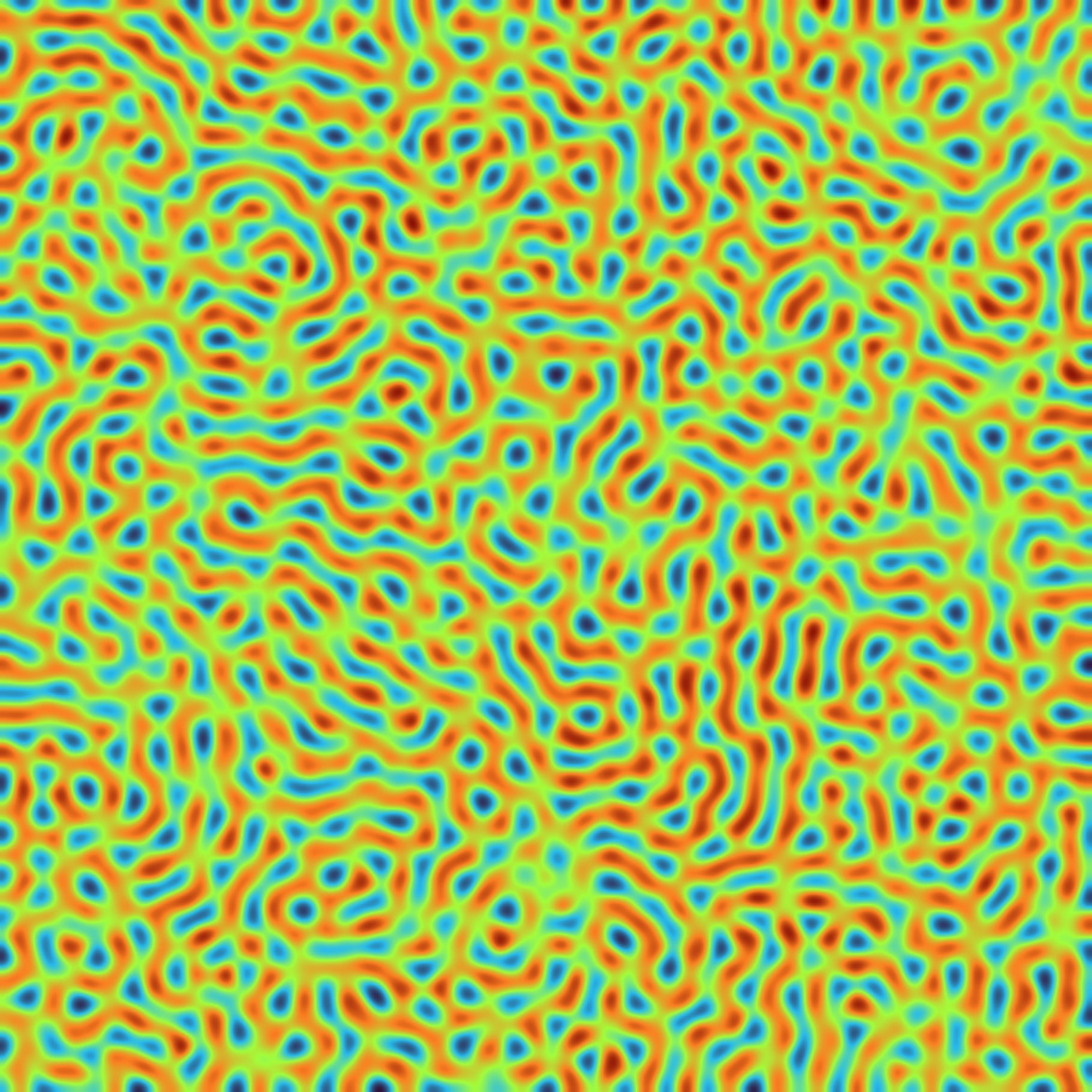}
		}
	\end{minipage}
	\begin{minipage}[b]{.45\linewidth}
		\centering
		\subfigure[]
		{
			\includegraphics[scale=0.15]{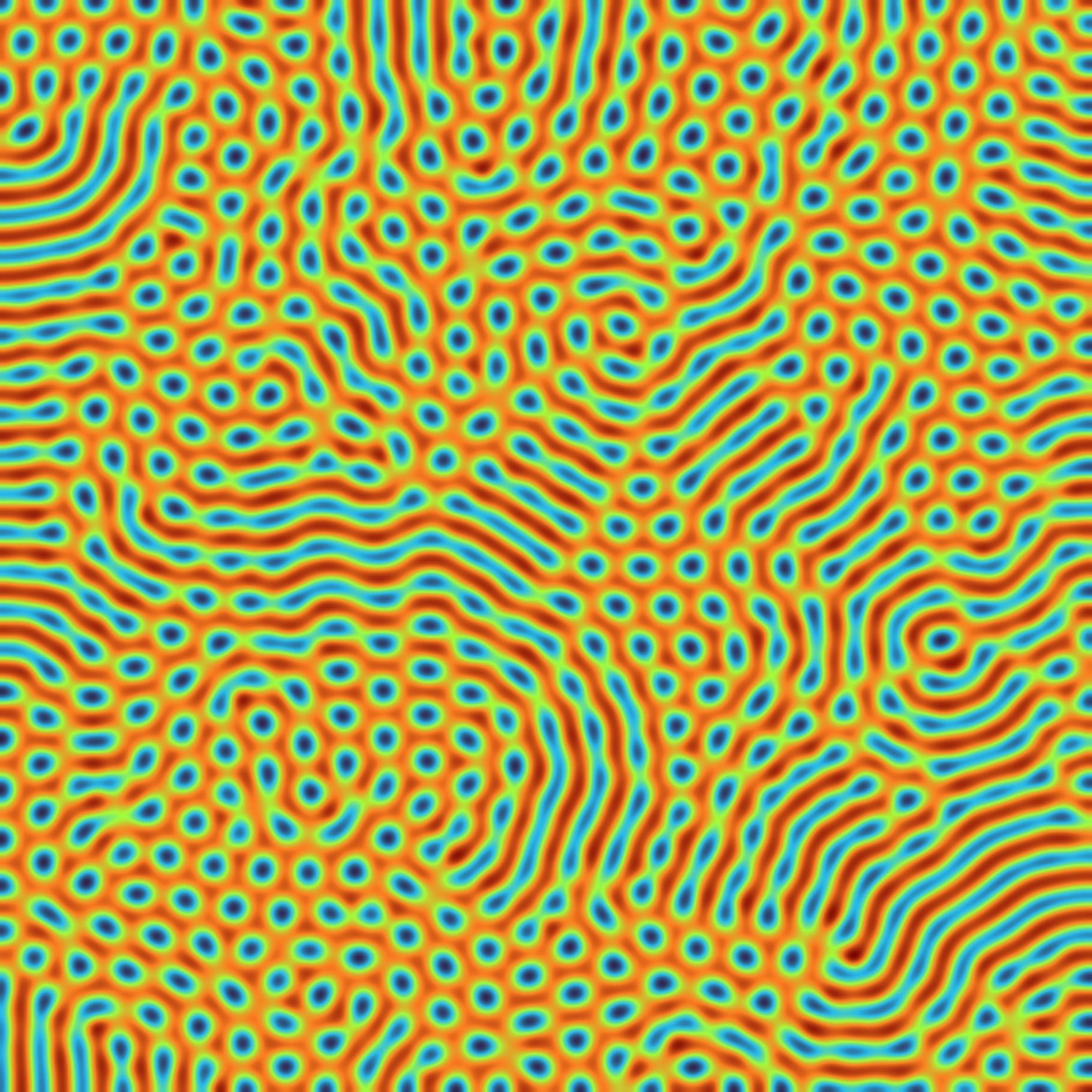}
		}
	\end{minipage}
	\hspace{0.05\linewidth} 
	\begin{minipage}[b]{.45\linewidth}
		\centering
		\subfigure[]
		{
			\includegraphics[scale=0.15]{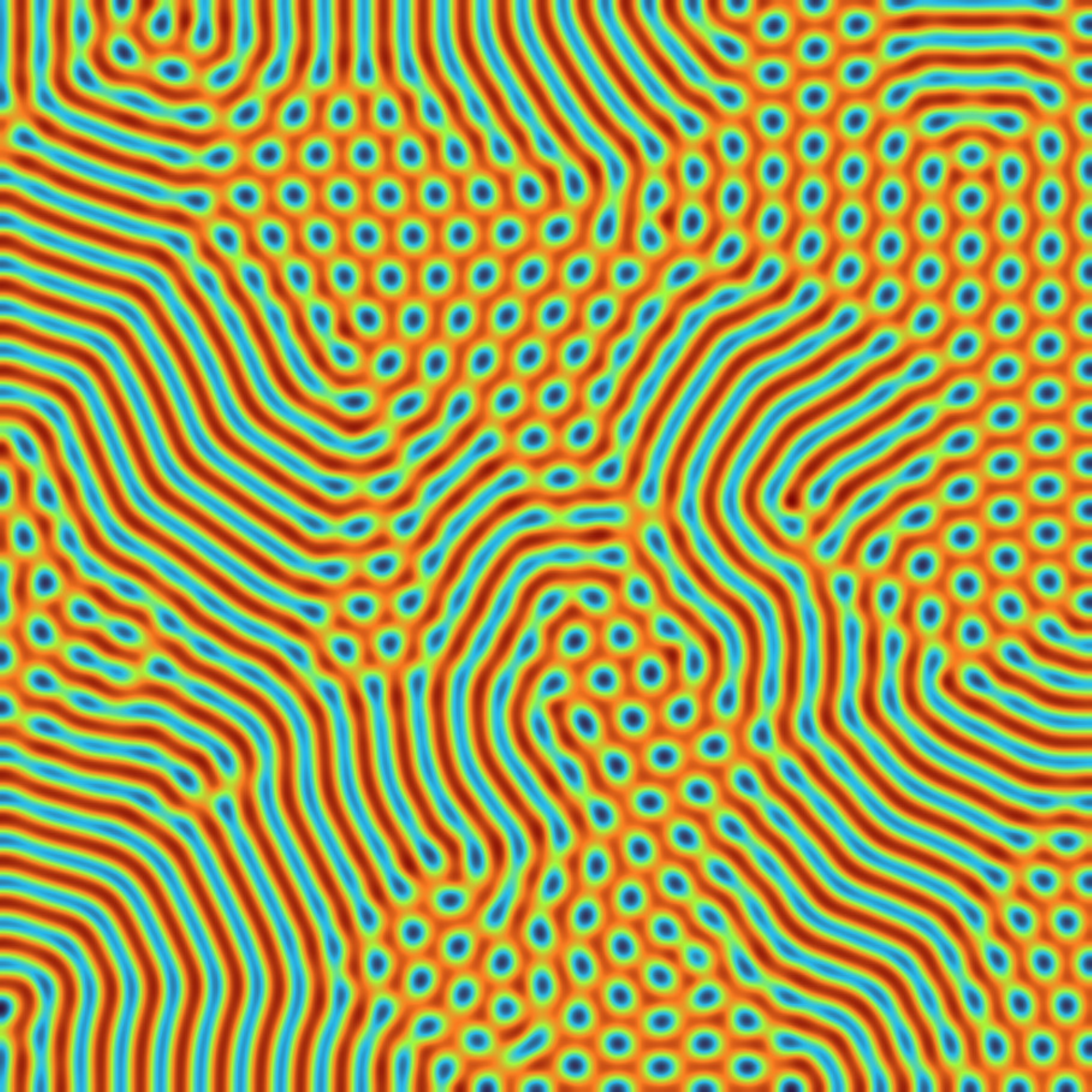}
		}
	\end{minipage}
	\caption{Patterned vegetation distribution at different moments. 
			$(a)$: $t=20$. 	$(b)$: $t=60$. 	$(c)$: $t=1000$ 	$(d)$: $t=5000$.}\label{spot_stripe}
\end{figure}

\begin{figure}[H]
	\centering
	\begin{minipage}[b]{.45\linewidth}
		\centering
		\subfigure[]
		{
			\includegraphics[scale=0.15]{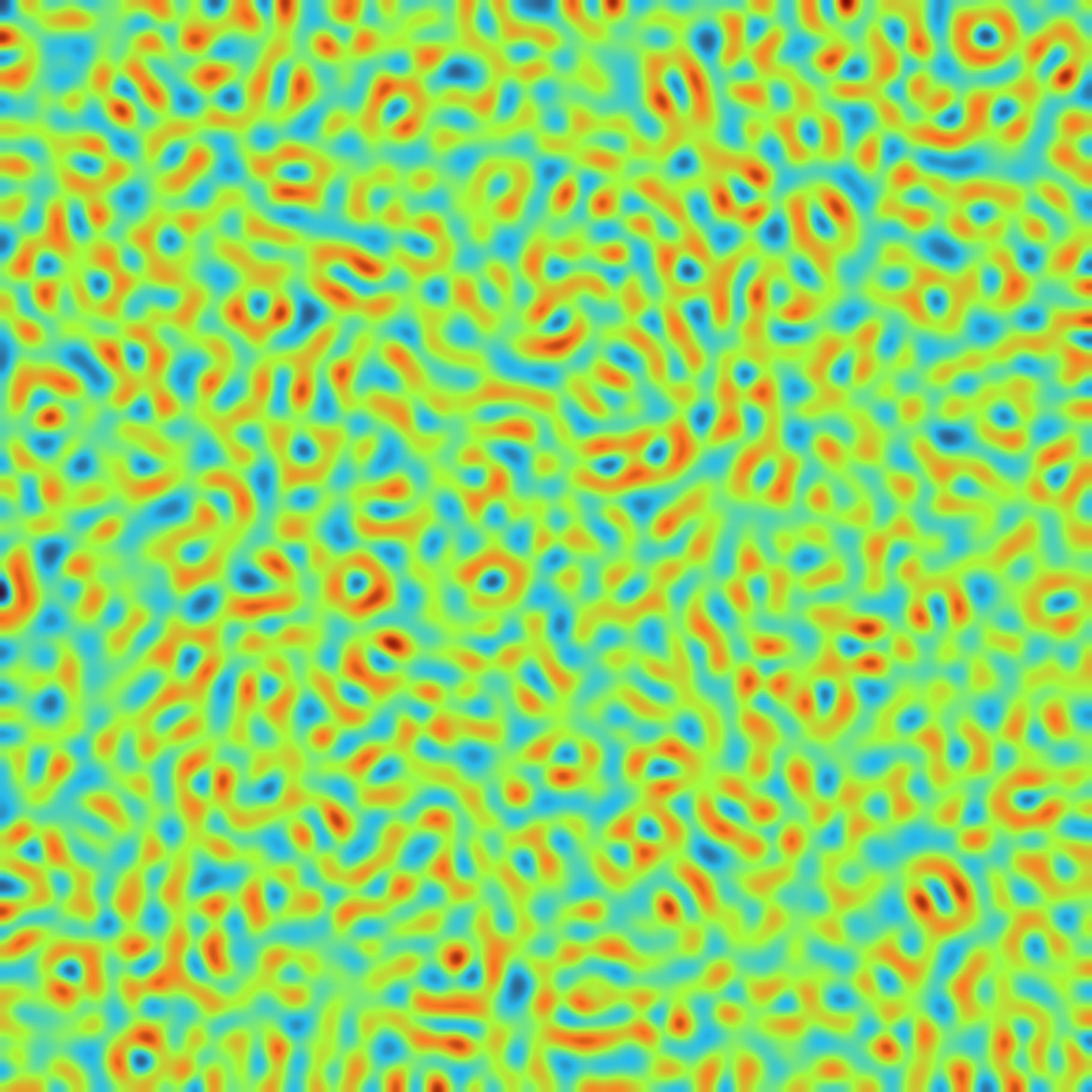}
		}
	\end{minipage}
	\hspace{0.05\linewidth} 
	\begin{minipage}[b]{.45\linewidth}
		\centering
		\subfigure[]
		{
			\includegraphics[scale=0.15]{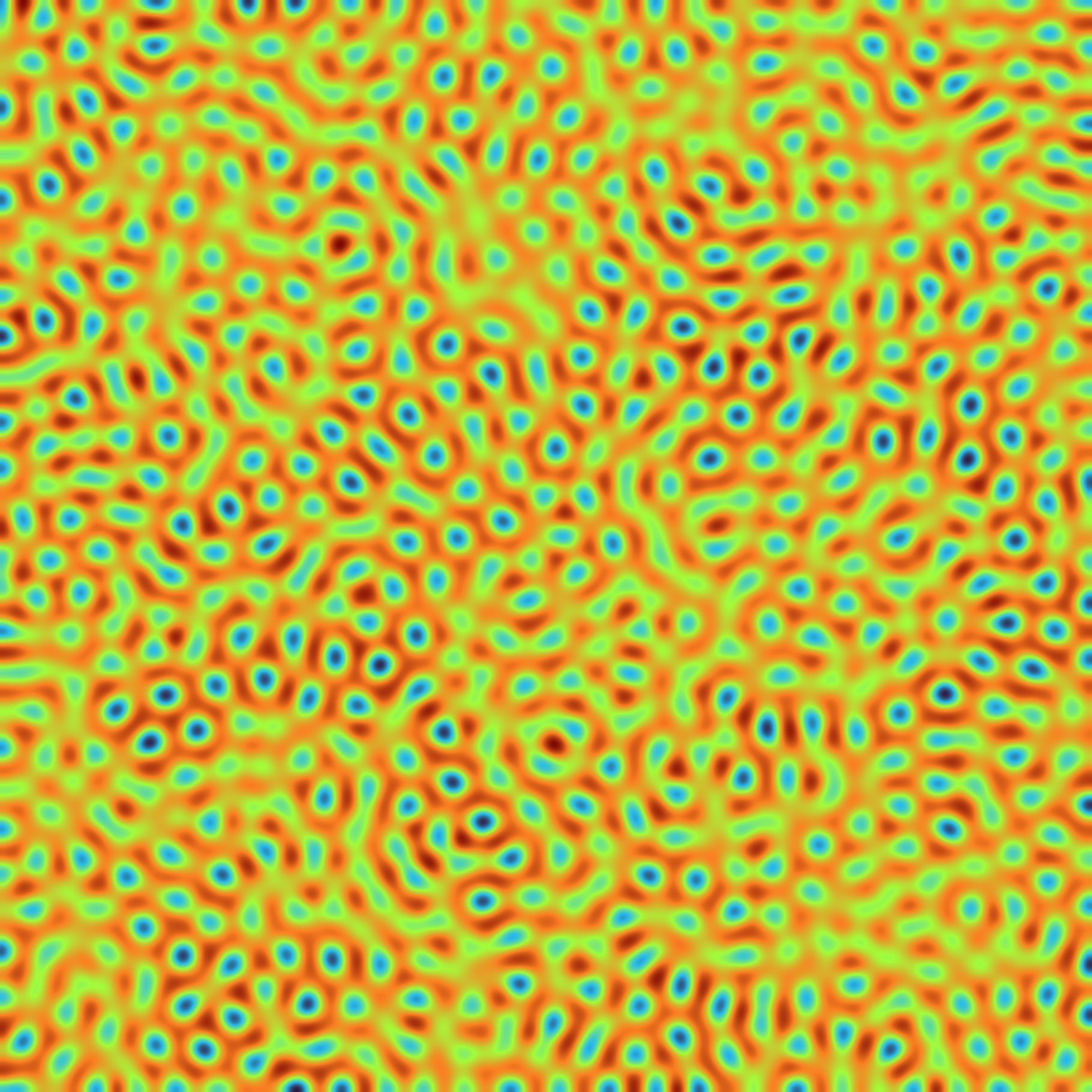}
		}
	\end{minipage}
	
	\vspace{0.5\baselineskip} 
	
	\begin{minipage}[b]{.45\linewidth}
		\centering
		\subfigure[]
		{
			\includegraphics[scale=0.15]{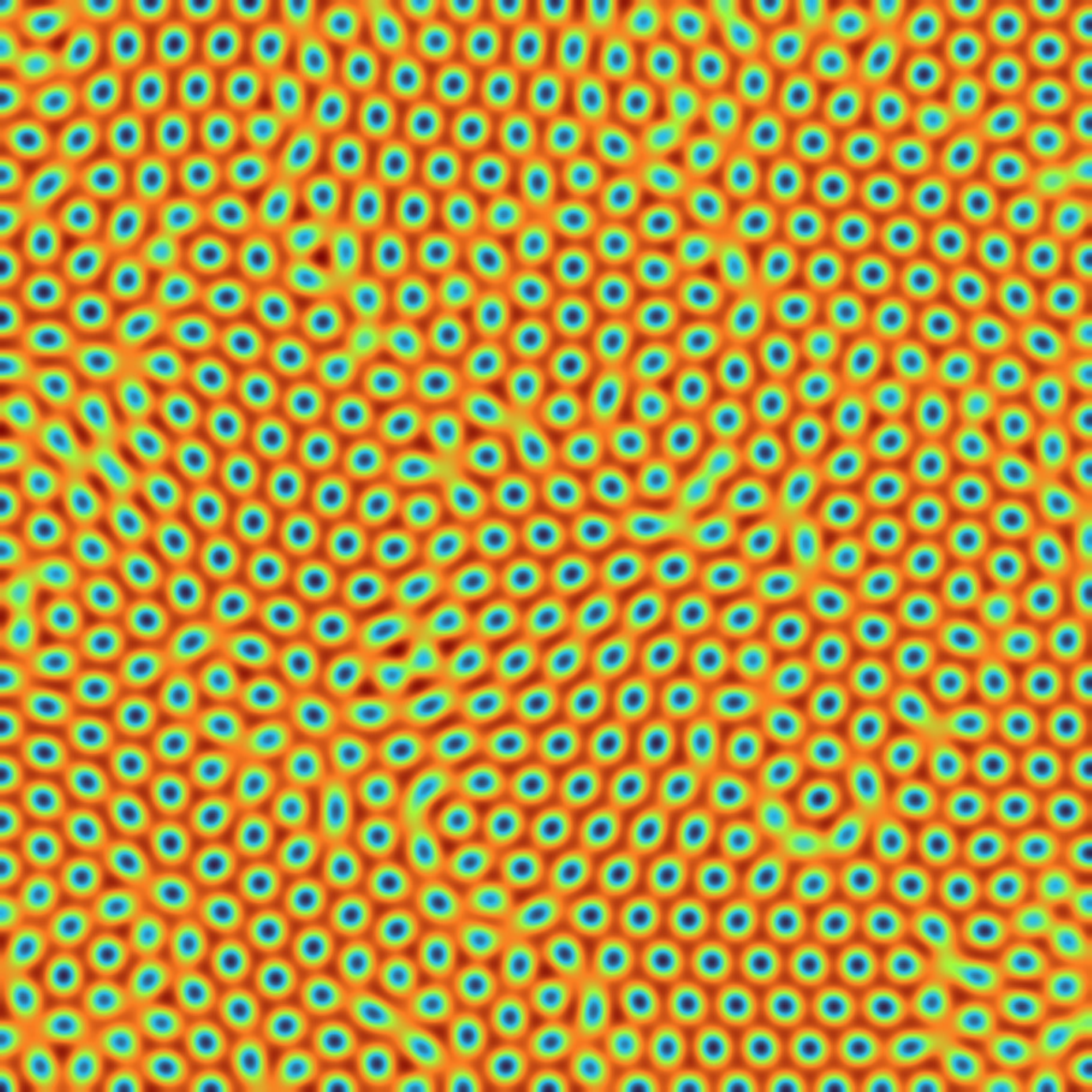}
		}
	\end{minipage}
	\hspace{0.05\linewidth} 
	\begin{minipage}[b]{.45\linewidth}
		\centering
		\subfigure[]
		{
			\includegraphics[scale=0.15]{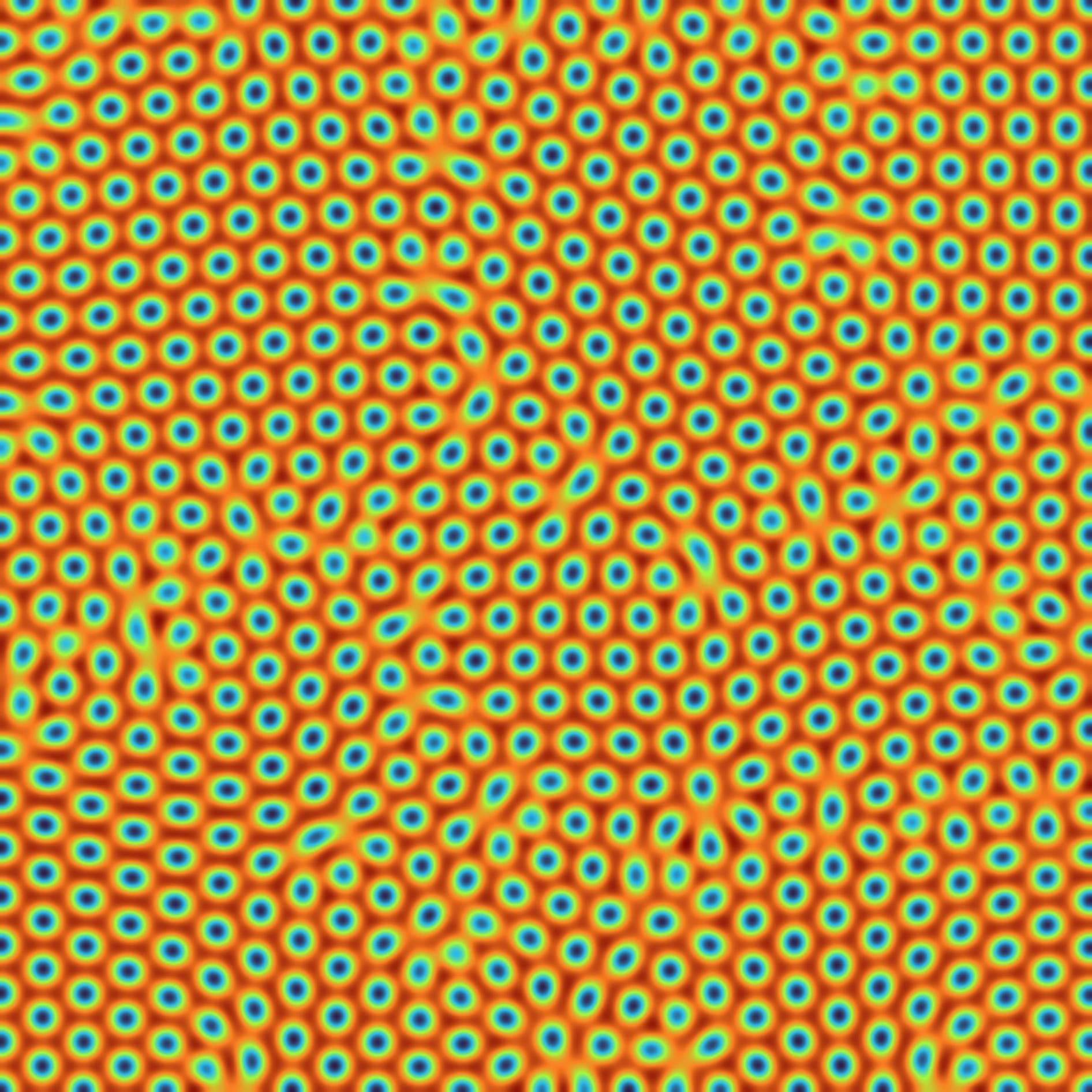}
		}
	\end{minipage}
	\caption{Patterned vegetation distribution at different moments. 
		$(a)$: $t=20$. 	$(b)$: $t=60$. 	$(c)$: $t=1000$ $(d)$: $t=5000$.}\label{spot}
\end{figure}

\begin{figure}[H]
	\centering
	\begin{minipage}[b]{.32\linewidth}
			\centering
		\subfigure[]
		{
			\includegraphics[scale=0.15]{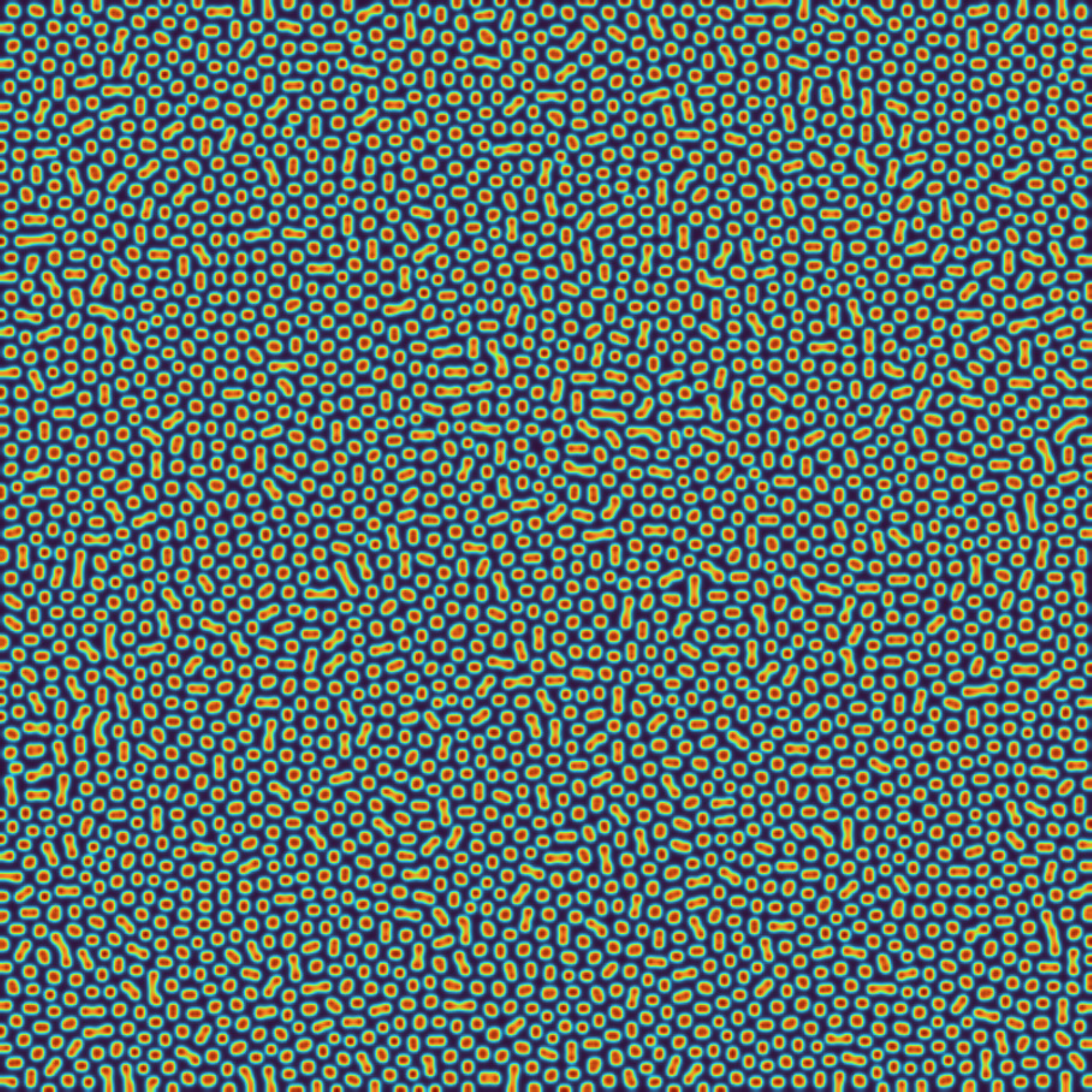}
		}
	\end{minipage}
	\begin{minipage}[b]{.32\linewidth}
			\centering
		\subfigure[]
		{
			\includegraphics[scale=0.15]{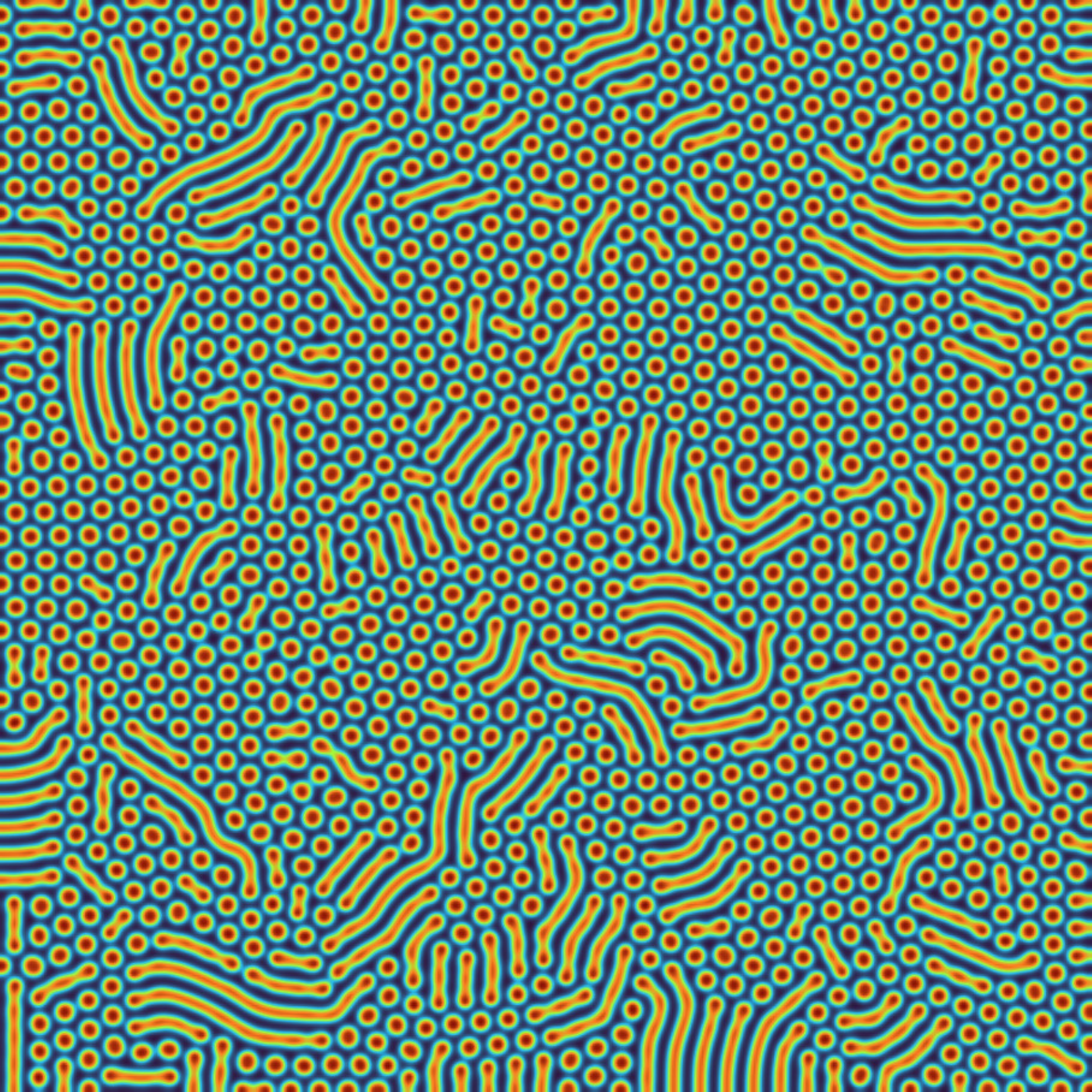}
		}
	\end{minipage}
	\begin{minipage}[b]{.32\linewidth}
			\centering
		\subfigure[]
		{
			\includegraphics[scale=0.15]{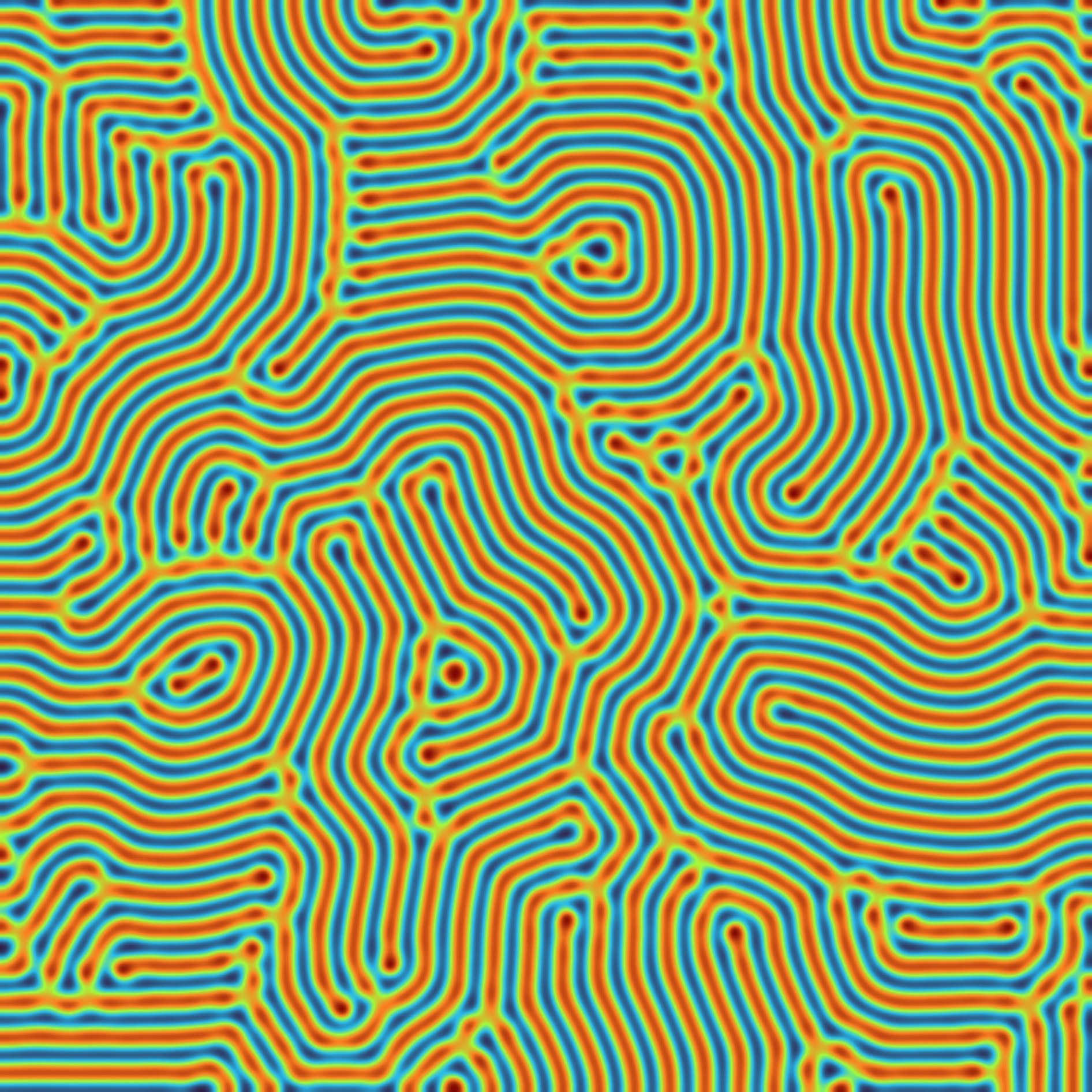}
		}
	\end{minipage}
	
	\begin{minipage}[b]{.32\linewidth}
			\centering
		\subfigure[]
		{
			\includegraphics[scale=0.15]{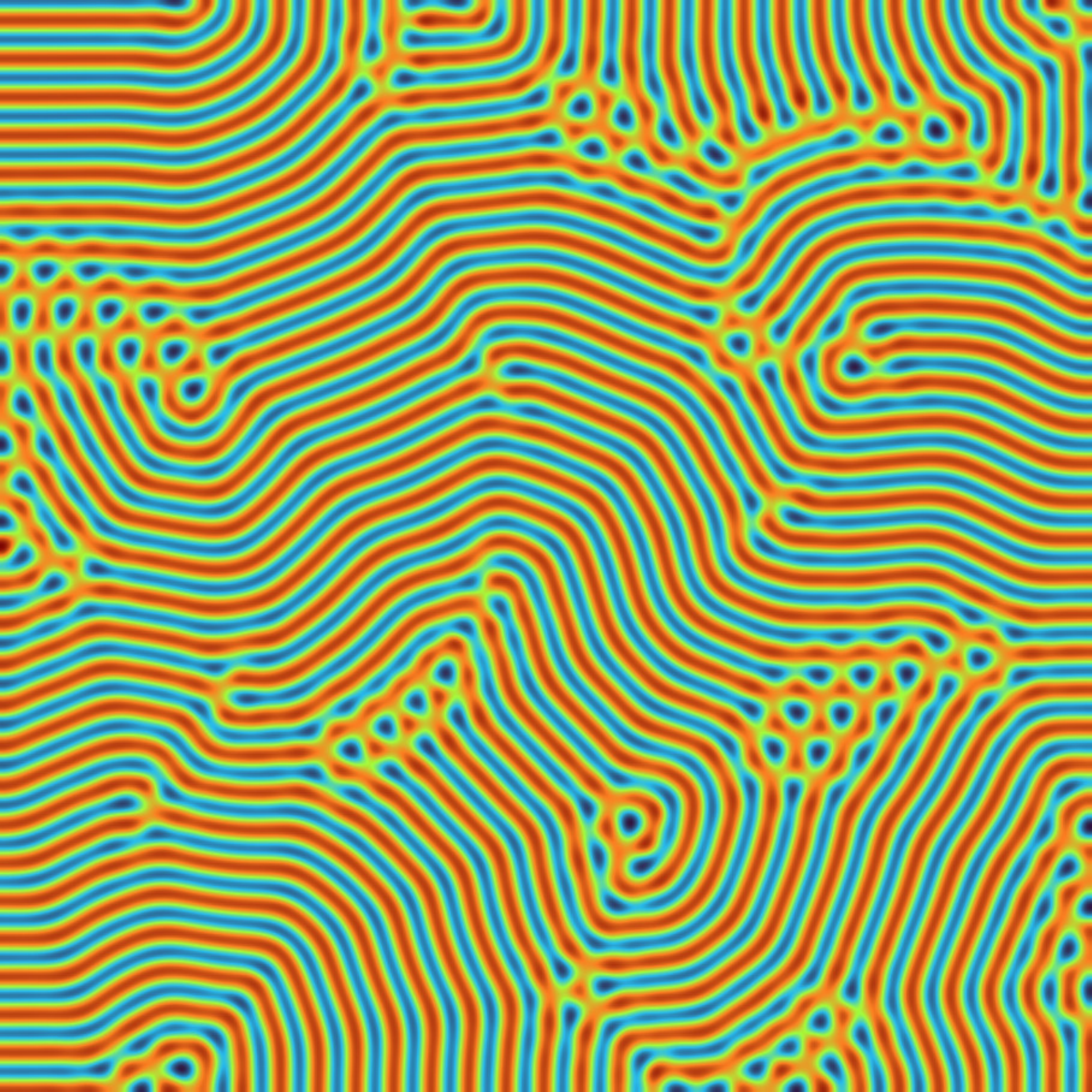}
		}
	\end{minipage}
	\begin{minipage}[b]{.32\linewidth}
			\centering
		\subfigure[]
		{
			\includegraphics[scale=0.15]{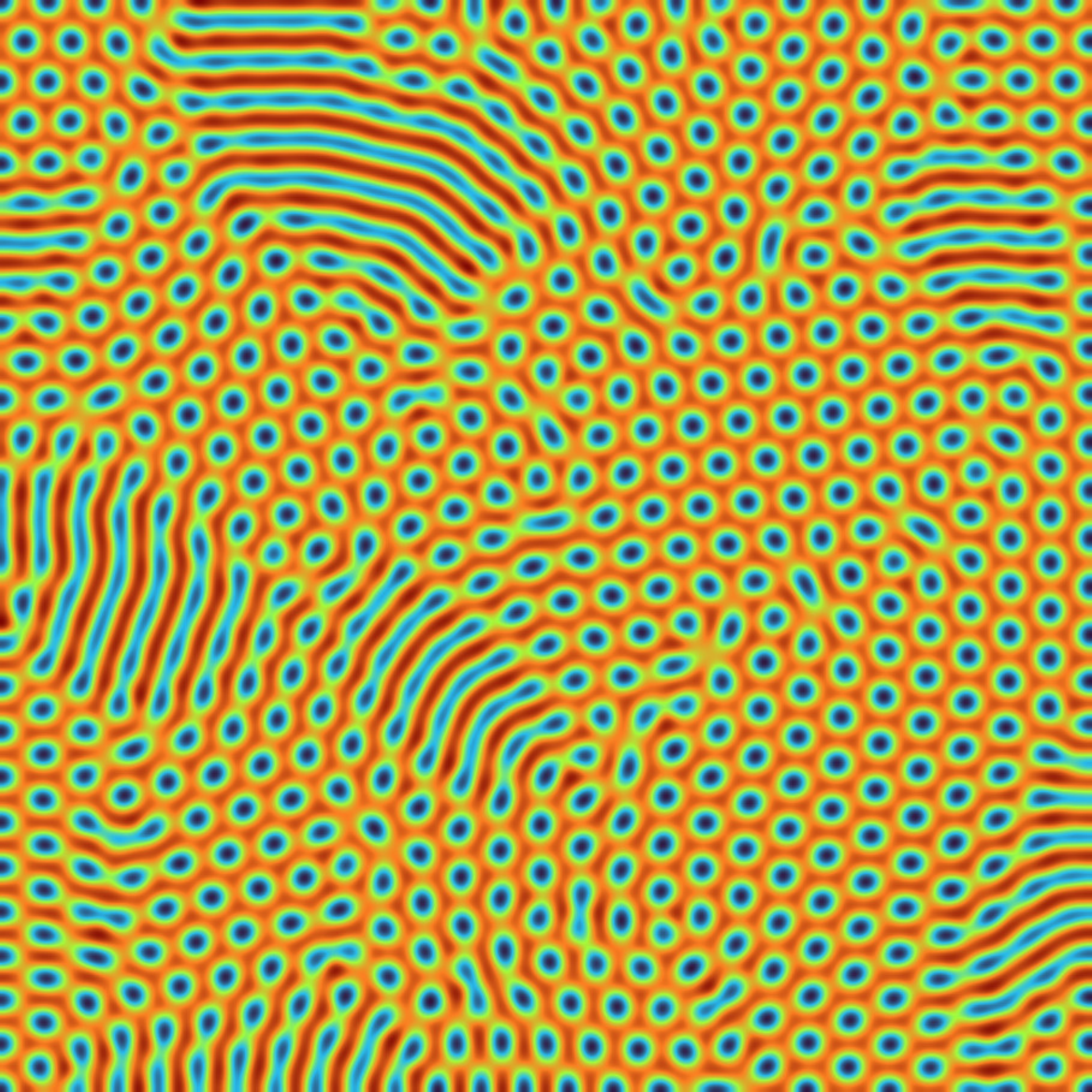}
		}
	\end{minipage}
	\begin{minipage}[b]{.32\linewidth}
			\centering
		\subfigure[]
		{
			\includegraphics[scale=0.15]{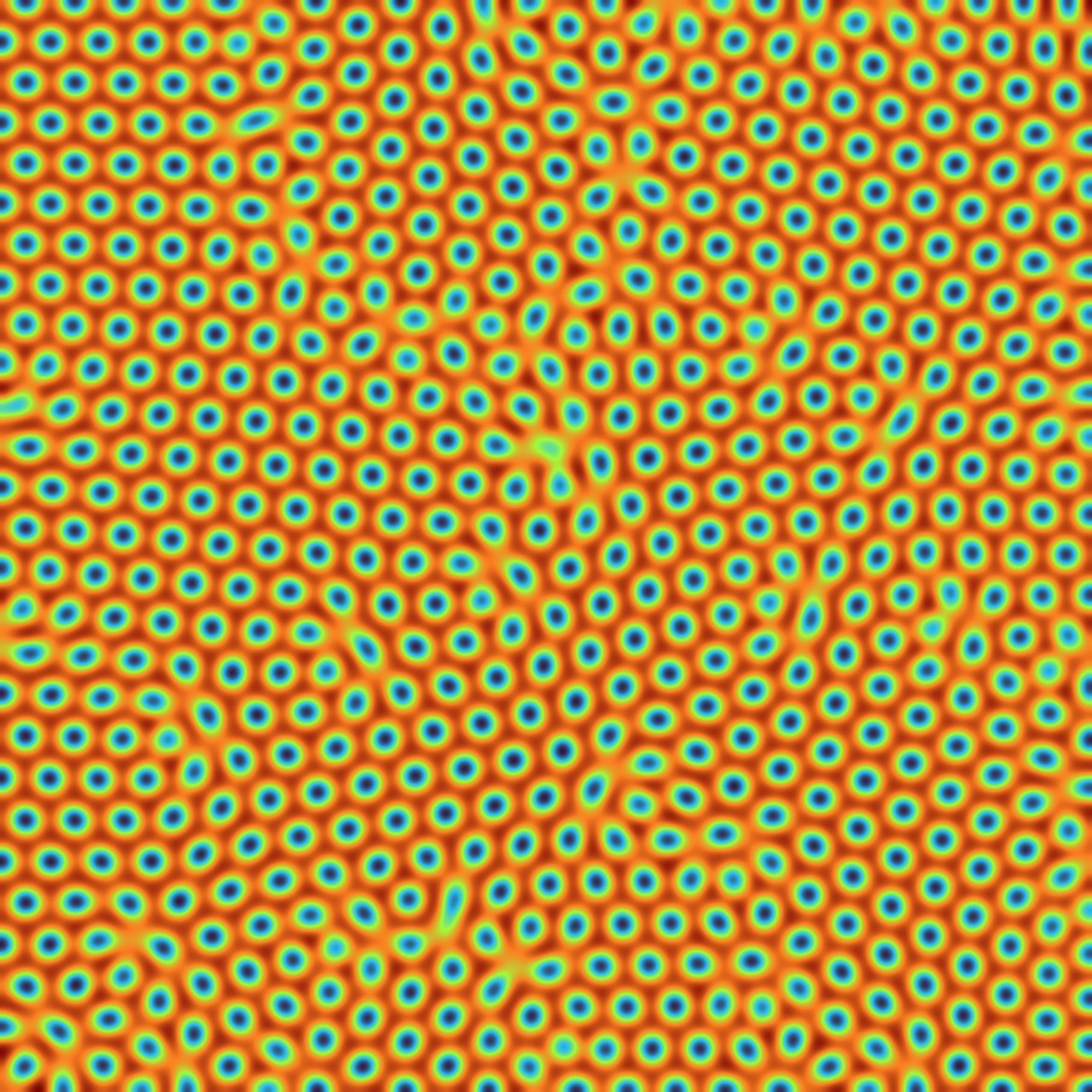}
		}
	\end{minipage}
	\caption{Pattern structure of vegetation with different parameter value $d_{1}$.
	$(a)$: $d_{1}=0.05$.  $(b)$: $d_{1}=0.12$. $(c)$: $d_{1}=0.25$. $(d)$: $d_{1}=0.3$. $(e)$: $d_{1}=0.32$. $(f)$: $d_{1}=0.35$ }\label{gap1_6}
\end{figure}

\begin{figure}[H]
	\centering
	\begin{minipage}[b]{.32\linewidth}
			\centering
		\subfigure[]
		{
			\includegraphics[scale=0.15]{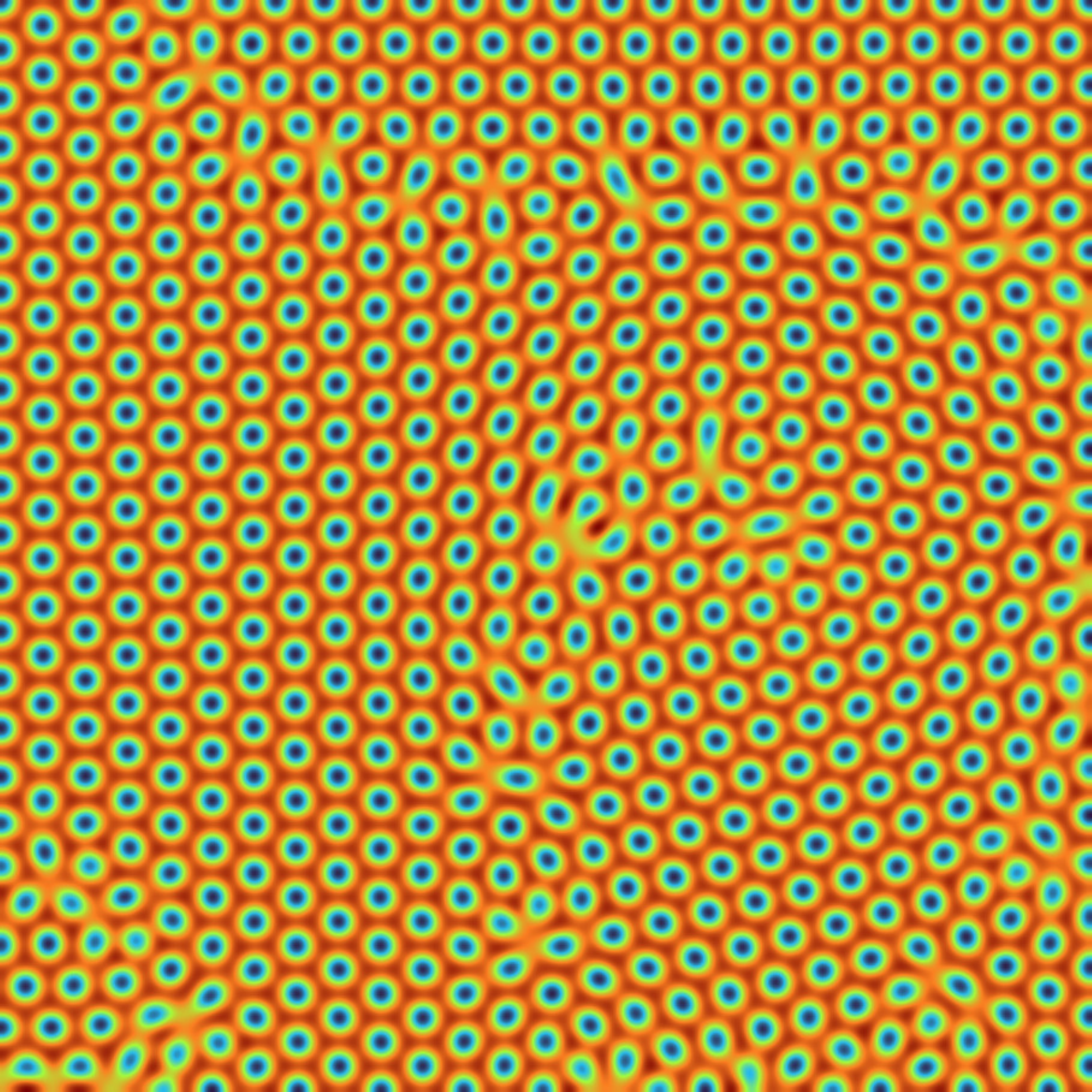}
		}
	\end{minipage}
	\begin{minipage}[b]{.32\linewidth}
			\centering
		\subfigure[]
		{
			\includegraphics[scale=0.15]{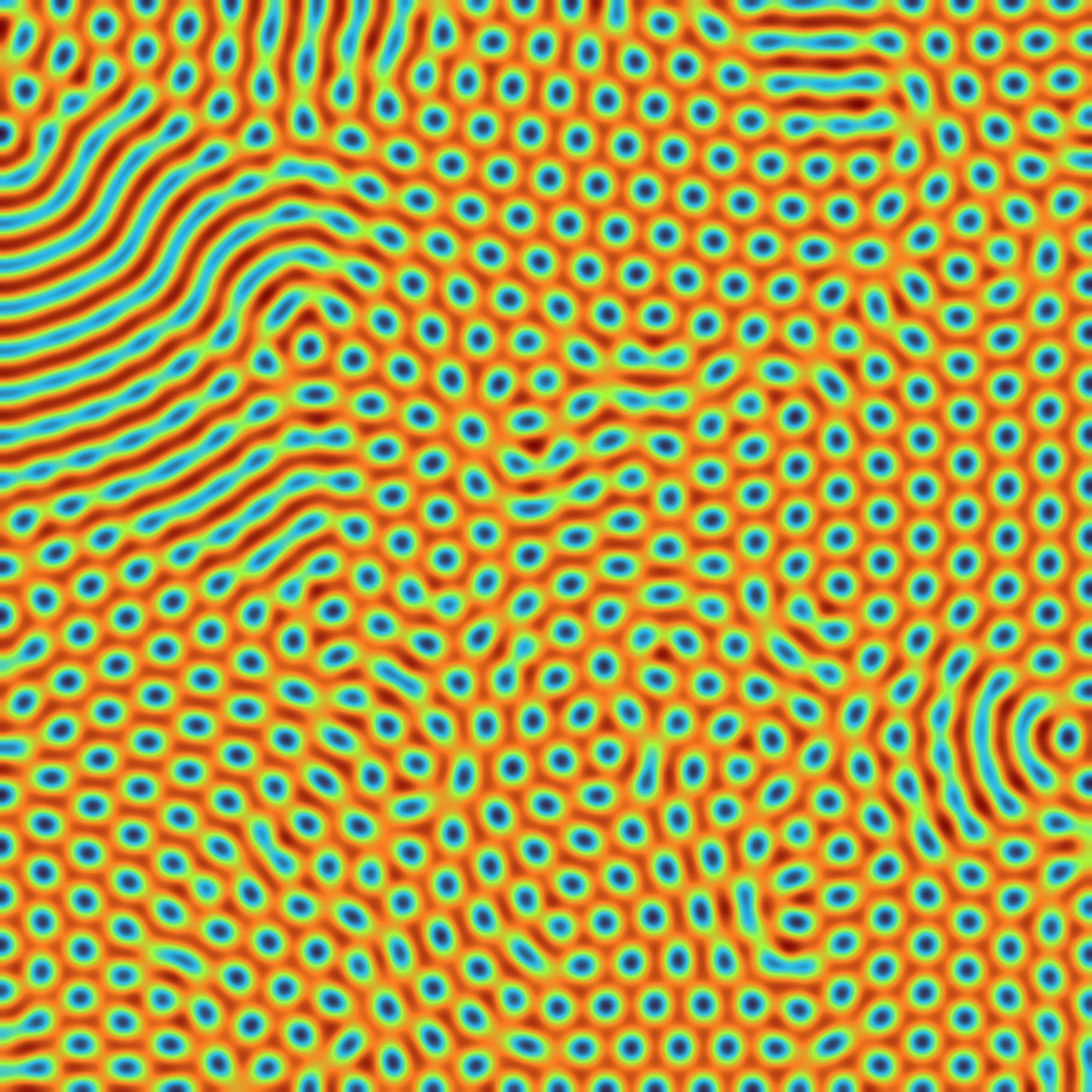}
		}
	\end{minipage}
	\begin{minipage}[b]{.32\linewidth}
			\centering
		\subfigure[]
		{
			\includegraphics[scale=0.15]{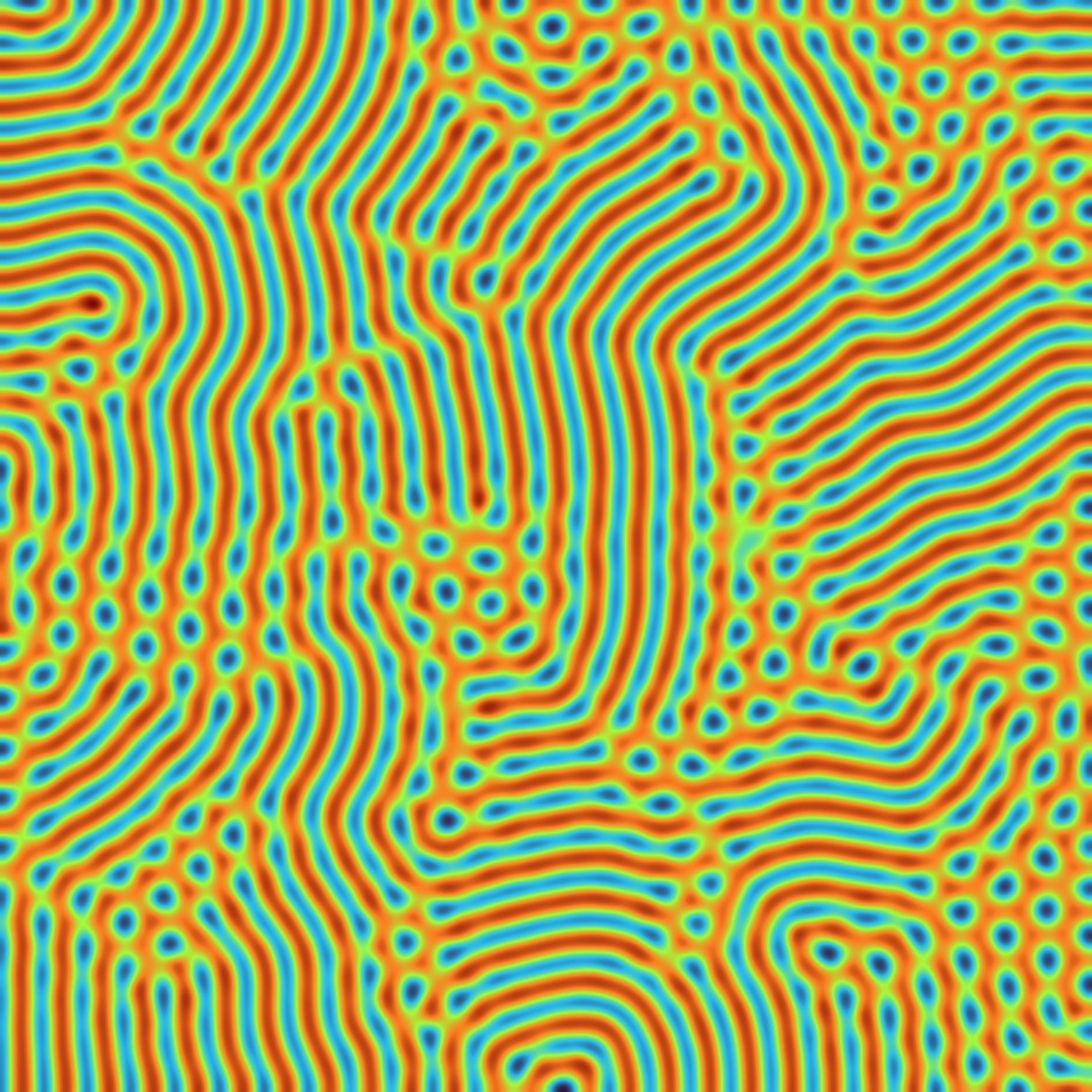}
		}
	\end{minipage}
	
	\begin{minipage}[b]{.32\linewidth}
			\centering
		\subfigure[]
		{
			\includegraphics[scale=0.15]{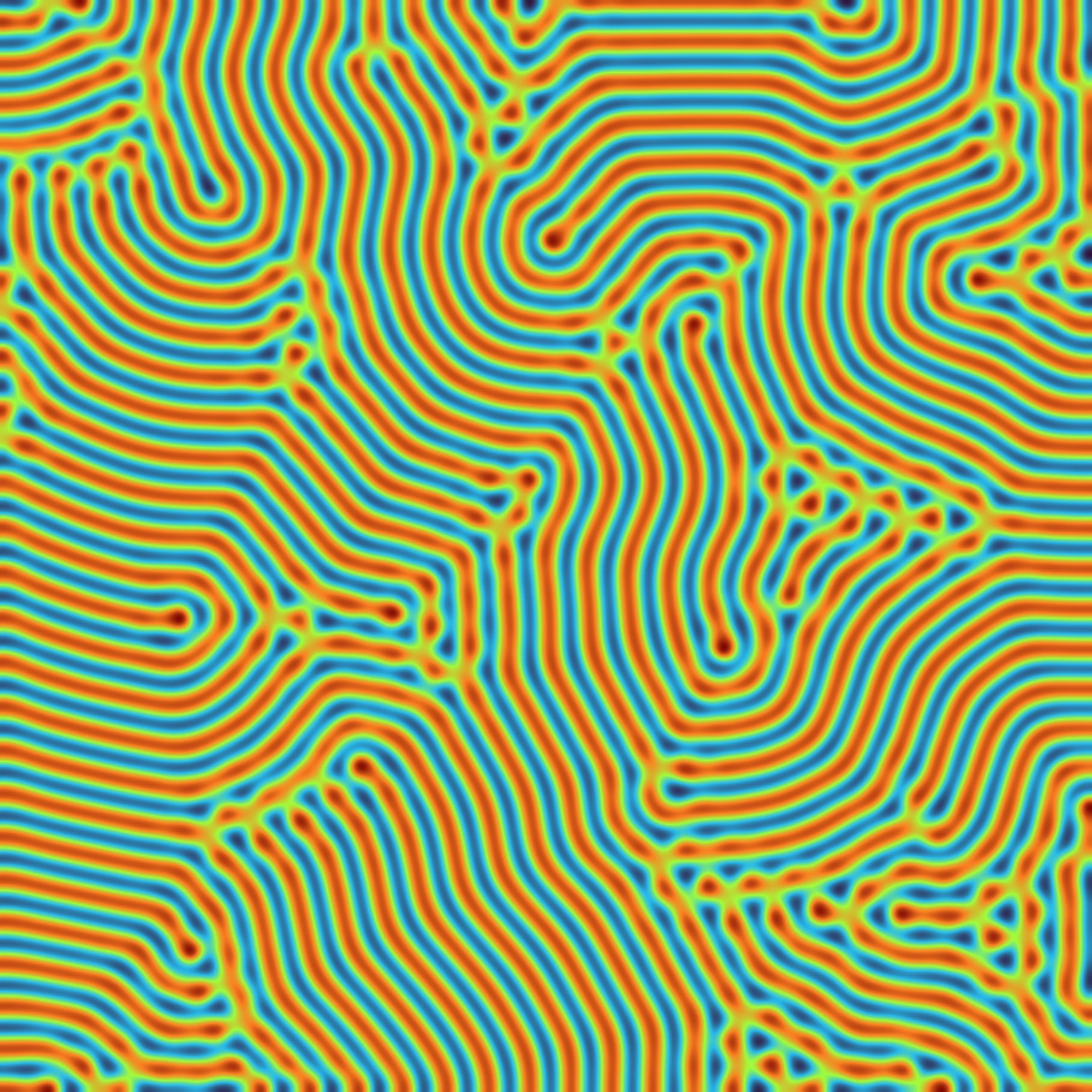}
		}
	\end{minipage}
	\begin{minipage}[b]{.32\linewidth}
			\centering
		\subfigure[]
		{
			\includegraphics[scale=0.15]{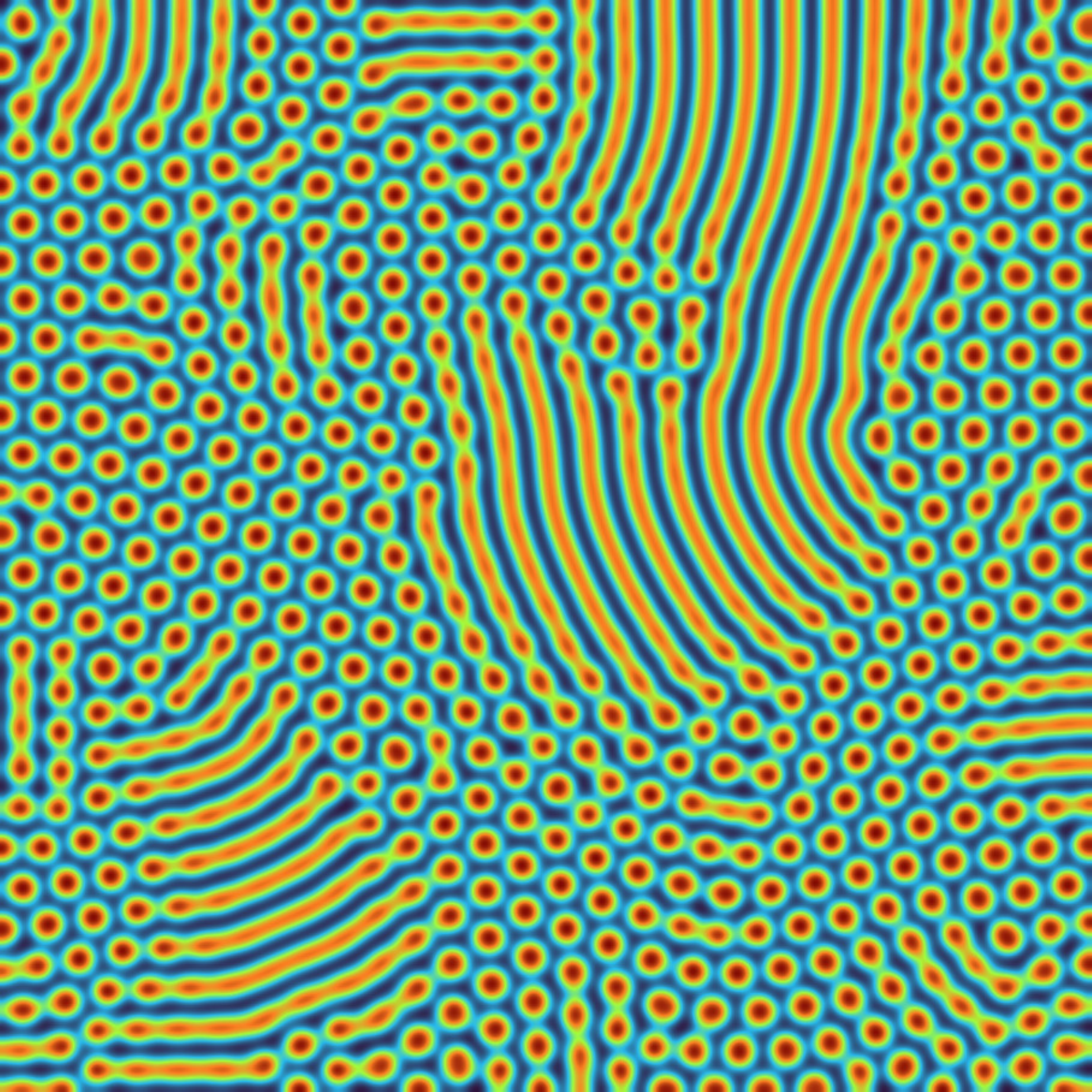}
		}
	\end{minipage}
	\begin{minipage}[b]{.32\linewidth}
			\centering
		\subfigure[]
		{
			\includegraphics[scale=0.15]{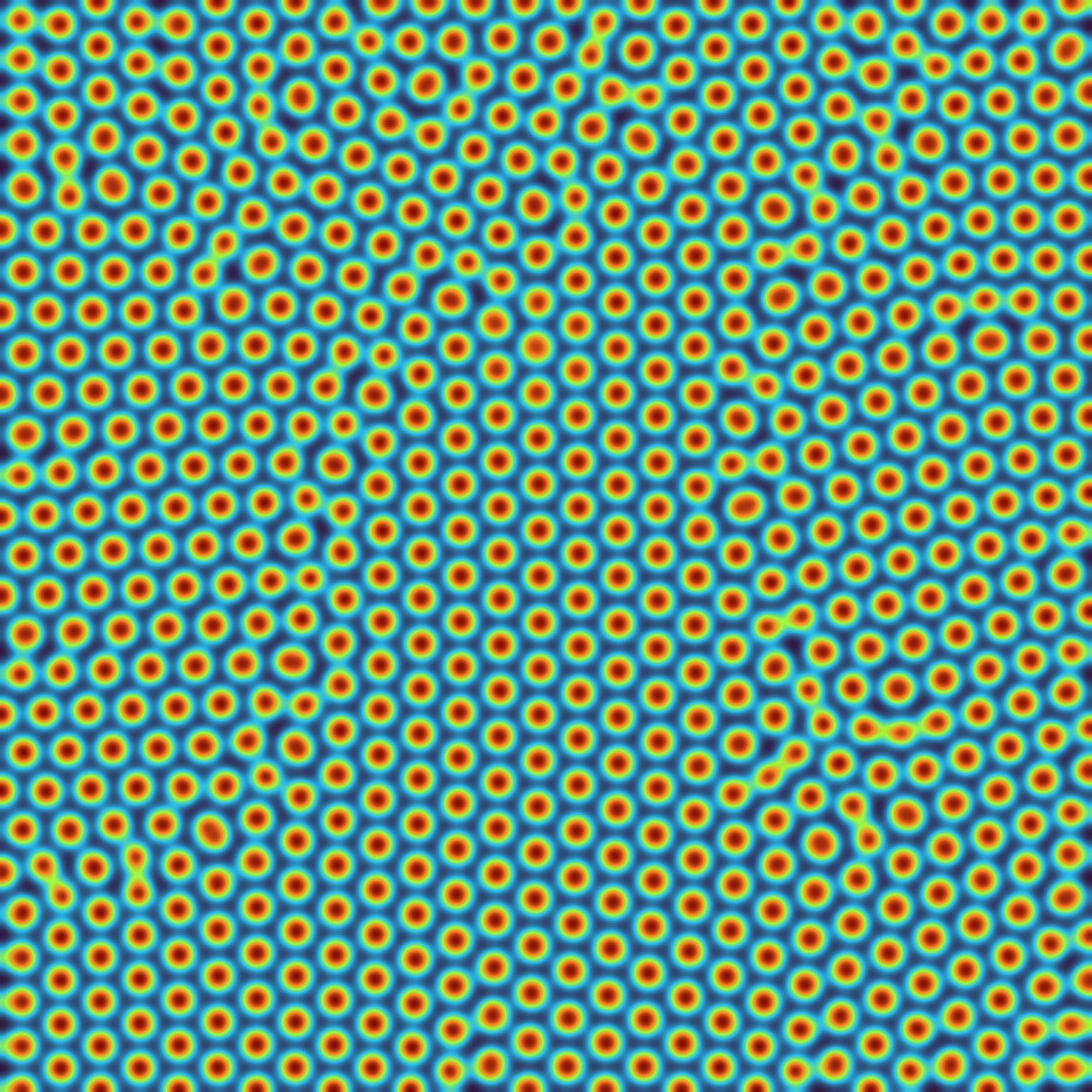}
		}
	\end{minipage}
	\caption{Pattern structure of vegetation with different parameter value $d_{1}$.
		$(a)$: $d=2$.  $(b)$: $d=2.2$. $(c)$: $d=2.3$. $(d)$: $d=2.7$. $(e)$: $d=4$. $(f)$: $d=4.6$ }\label{gap11_66}
\end{figure}

\begin{figure}[H]
	\centering
	\begin{minipage}[b]{.45\linewidth}
			\centering
		\subfigure[]
		{
			\includegraphics[scale=0.25]{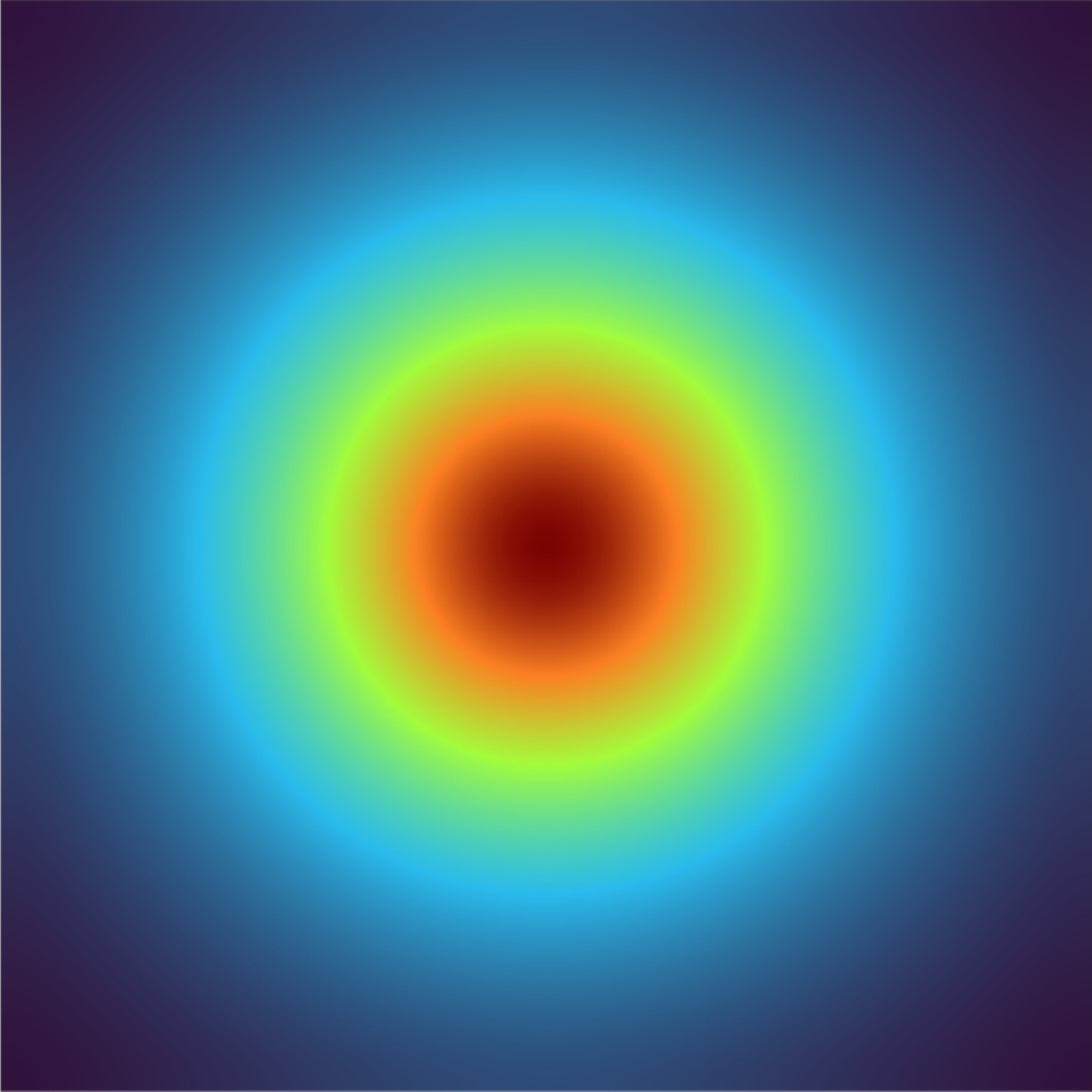}
		}
	\end{minipage}
	\hspace{0.05\linewidth} 
	\begin{minipage}[b]{.45\linewidth}
			\centering
		\subfigure[]
		{
			\includegraphics[scale=0.15]{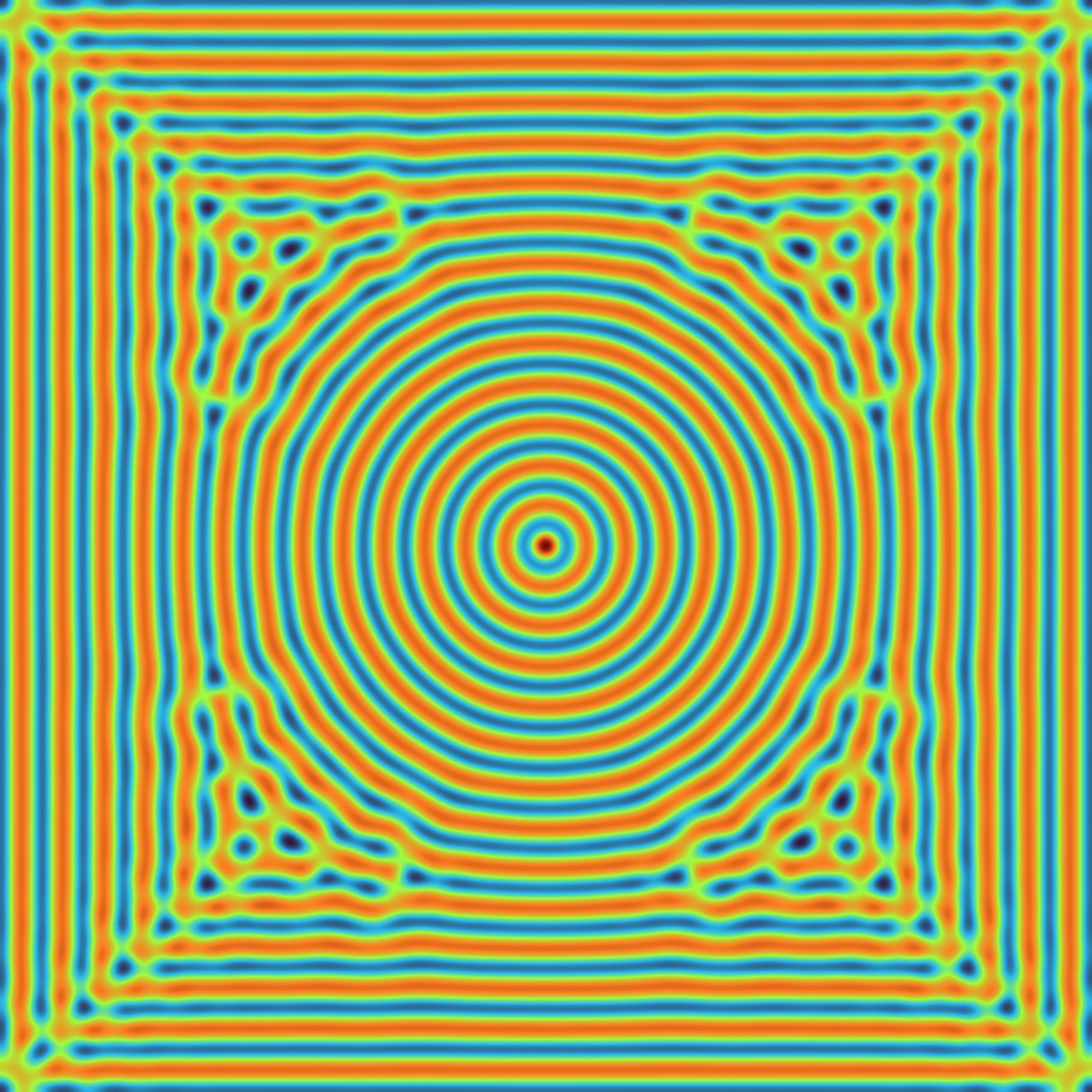}
		}
	\end{minipage}
	
	\vspace{0.5\baselineskip} 
	
	\begin{minipage}[b]{.45\linewidth}
			\centering
		\subfigure[]
		{
			\includegraphics[scale=0.15]{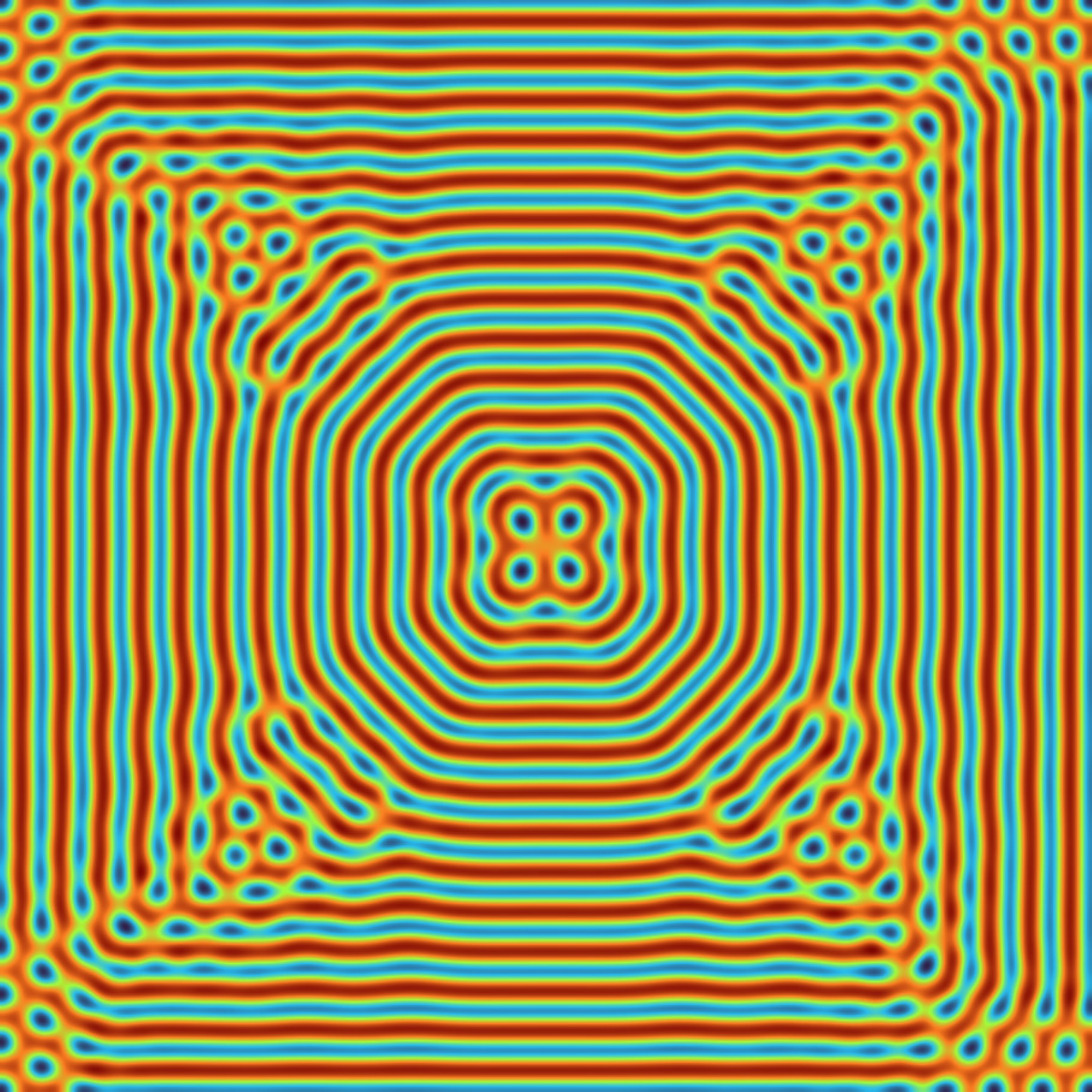}
		}
	\end{minipage}
	\hspace{0.05\linewidth} 
	\begin{minipage}[b]{.45\linewidth}
			\centering
		\subfigure[]
		{
			\includegraphics[scale=0.15]{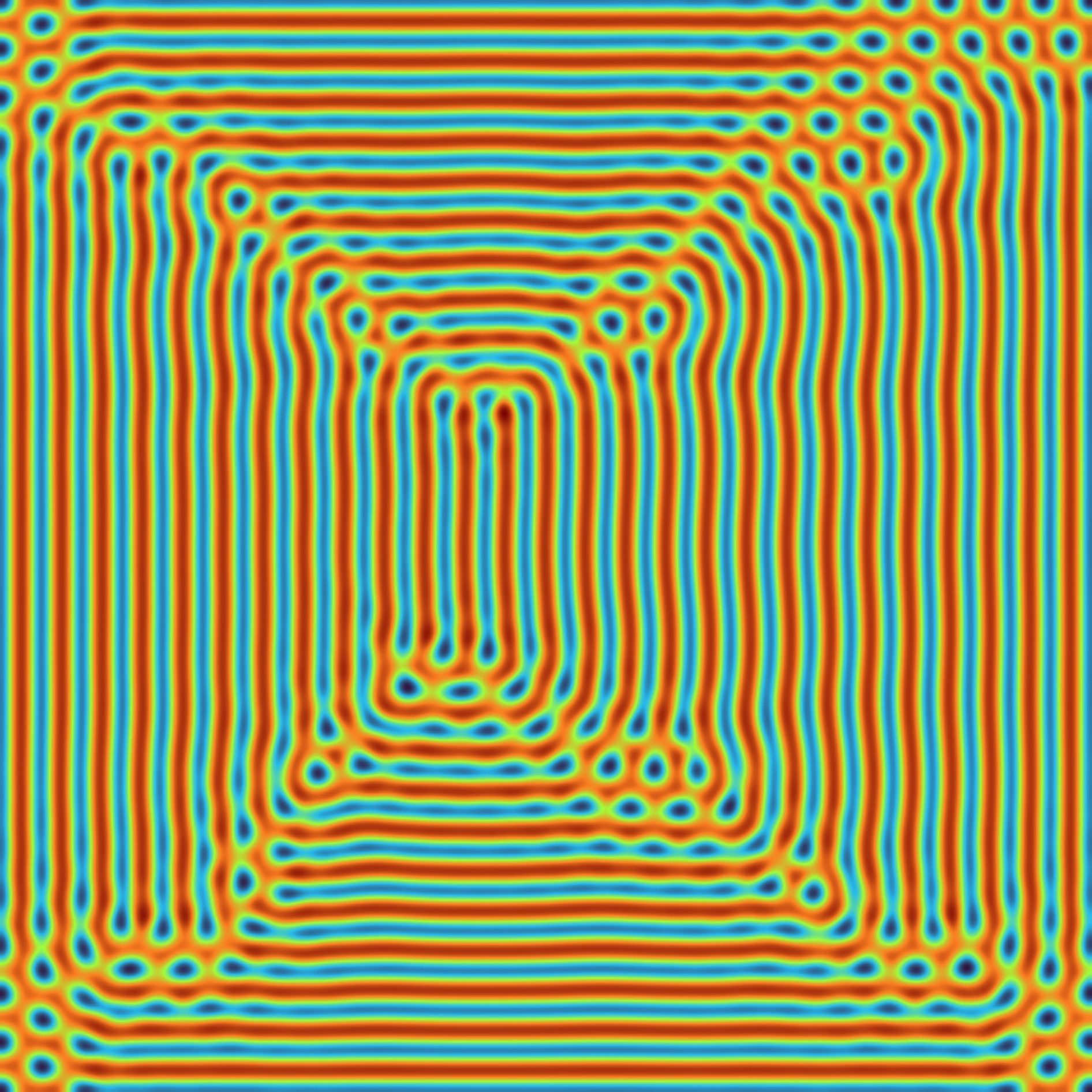}
		}
	\end{minipage}
	\caption{Pattern structure of vegetation at different moments. 
		$(a)$: $t=60$. 	$(b)$: $t=500$. 	$(c)$: $t=5000$ 	$(d)$: $t=20000$.}\label{wave}
\end{figure}

Next we focus on the effect of parameters $a, \ d$ and $d_{1}$ on the diversity of plant patterns. Fix $(c,m,d_{2},\mathcal{l})=(1,2,2,1)$. In a 2D space 200 $\times$ 200, we observe a variety of plant patterns by using computer technique, where the spatial step and time step are 0.5 and 0.01, respectively.\\
$\textbf{(i)}$ Taking $(a,d,d_{1})=(0.12,2,0.35)$, 
 then the evolutionary process of spot-stripe patterns are observed in Fig. \ref{spot_stripe}. The spot represents a high vegetation density.\\
 $\textbf{(ii)}$ Taking $(a,d,d_{1})=(0.16,2,0.35)$, then the evolutionary process of spot patterns is observed in Fig. \ref{spot}.\\
$\textbf{(iii)}$  For fixed $(a,d)=(0.16,2)$, we vary $d_{1}$ in Fig. \ref{gap1_6}. When $d_{1}$ is small, gap patterns appear in $(a)$. When we increase $d_{1}$, stripe patterns begin to  appear in $(b)$. When $d_{1}$ is further increased, stripe patterns fill almost the space in $(c)$. When $d_{1}$ continues to be increased, the space is filled with spot-stripe patterns with more stripe patterns in $(d)$, spot-stripe patterns with more spot patterns in $(e)$. When $d_{1}$ is larger $(d_{1}=0.35)$, spot patterns cover the entire space in $(f)$.\\
$\textbf{(iv)}$ For fixed $(a,d_{1})=(0.16,0.35)$, we vary $d$ in Fig. \ref{gap11_66}. For $d=2$, spot parameter appear in $(a)$. If $d$ is increased slowly, then mixed patterns appear in $(b), \ (c)$. 
If $d$ is increased further, then strip patterns take up almost the entire space in $(d)$. If $d$ continues to be increased, then gap patterns emerge in $(e)$. When $d$ is larger, gap patterns cover the entire space in $(f)$.\\
  \indent In addition,  we take the following initial value for fixed $(a,d,d_{1})=(0.16,2,0.35)$.
\begin{equation}
	u_{0}(x)=\frac{1}{3}-0.0001RAND, \ 	v_{0}(x)=\frac{9}{14}-0.0001((x-100)^{2}+(y-100)^{2}),
\end{equation}
then the initial target wave is observed and finally evolves into spot-stripe patterns with a few spots in Fig. \ref{wave}.

%
%

\section{Conclusion}
This paper deals with a reduced plant-water system in a drought-prone area. Initially, we contemplate a non-diffusive scenario for the model, examine the local and global stability of equilibria. It was observed that under suitable system parameters, plant-extinction equilibrium $E_{0}$ becomes a globally asymptotically stable state. Moreover, suitable parameter conditions also lead to the global asymptotic stability of two coexisting plant-water equilibria $E_{10}$ and $E_{11}$. This underscores the pivotal role of controlling system parameters in sustaining plant growth. We analyzed the conditions for the existence and non-existence of the limit cycle. Furthermore, we explored bifurcation behaviors, and find that appropriate parameter conditions result in the emergence of transcritical bifurcations near $E_{0}$, saddle-node bifurcation, Hopf bifurcation, and even bifurcations of higher codimension near the coexisting equilibria. This illustrates the significant implications of controlling system parameters for maintaining coexistence between plants and water.\\
	\indent For the space-time system, we investigate Hopf, Turing, Hopf-Turing and Turing-Turing bifurcations. The stability domain and bifurcation diagrams are visually depicted in the $d_{1}-a$ plane. It's notable to mention that mixed nonconstant steady states around the Turing-Turing bifurcation point are observed.
   This suggest that different initial values are likely to bring about vastly divergent spatial patterns. The heteroclinic solutions connecting unstable periodic solutions to stable spatially inhomogeneous steady states of $\cos x$-type are observed around the Hopf-Turing bifurcation point. By deducing the normal form, we obtain stable and supercritical spatial homogeneous periodic solutions. Additionally, we primarily explores the influence of three key parameters $a, \ d, \ d_{1}$ on the pattern structure of plants. Various patterns are observed, including spot, stripe and coexisting patterns. Interestingly, we observe  transient wave patterns evolving into cycle patterns, ultimately transitioning into spot-stripe patterns. Hence, the seed diffusivity enables the transformation of its own pattern structures. And  controlling system parameters may prevent desertification from occurring.

			\section*{Conflict of Interest}
		The authors declare that they have no conflict of interest.
		
		\section*{Author Contribution Statement}
		We declare that the authors are ranked in alphabetic order of their names and all of them have the same contributions to this paper.

			\section*{Data Availability Statement}
		No data was used for the research in this article.

			\section*{Ethics statement}
		
		 This article does not present research with ethical considerations.

\end{document}